\documentclass[referee]{raa}            

\usepackage{geometry}
\usepackage{graphicx}             
\usepackage{subfigure}
\usepackage{paralist}
\usepackage{amssymb}
\usepackage{bbding}
\usepackage{amsmath}
\usepackage{natbib}
\usepackage{ulem}
\usepackage{color}
\usepackage{cancel}
\usepackage{bm}
\usepackage[a4paper=true,dvipdfm=true,pagebackref=true]{hyperref}
\hypersetup{pdftitle = The title of my PDF, pdfauthor = My name, pdfsubject= The subject, pdfkeywords = keyword1 keyword2 keyword3}
\hypersetup{colorlinks = true, linkcolor = green, anchorcolor = red, citecolor = blue, filecolor = red, pagecolor = red, urlcolor = red}

\begin{document}

\title{A new expansion of planetary disturbing function and applications to interior, co-orbital and exterior resonances with planets
}

\volnopage{Vol.0 (2002x) No.0, 000--000}      
\setcounter{page}{1}          

   \author{Hanlun Lei\inst{}
      \inst{}
   }

\institute{School of Astronomy and Space Science, Nanjing University, Nanjing 210023, China; {\it leihl@nju.edu.cn}\\
\and
{Key Laboratory of Modern Astronomy and Astrophysics in Ministry of Education, Nanjing University, Nanjing 210023, China}\\
   }

\date{Received~~2021 month day; accepted~~2021~~month day}

\abstract{In this study, a new expansion of planetary disturbing function is developed for describing the resonant dynamics of minor bodies with arbitrary inclinations and semimajor axis ratios. In practice, the disturbing function is expanded around circular orbits in the first step and then, in the second step, the resulting mutual interaction between circular orbits is expanded around a reference point. As usual, the resulting expansion is presented in the Fourier series form, where the force amplitudes are dependent on the semimajor axis, eccentricity and inclination and the harmonic arguments are linear combinations of the mean longitude, longitude of pericenter and longitude of ascending node of each mass. The resulting new expansion is valid for arbitrary inclinations and semimajor axis ratios. In the case of mean motion resonant configuration, the disturbing function can be easily averaged to produce the analytical expansion of resonant disturbing function. Based on the analytical expansion, the Hamiltonian model of mean motion resonances is formulated, and the resulting analytical developments are applied to Jupiter's inner and co-orbital resonances and Neptune's exterior resonances. Analytical expansion is validated by comparing the analytical results with the associated numerical outcomes.
}
   \authorrunning{H.L. Lei}            
   \titlerunning{Disturbing function}  
   \maketitle

%
%
\section{Introduction}
\label{Section1}

Expanding the planetary disturbing function as a power series in orbital elements is an essential and fundamental issue in the development of perturbation theories in celestial mechanics. To understand the dynamics of resonances and secular evolutions, it is a key step to study the properties of the planetary disturbing function experienced by the object of interest \citep{murray1999solar}. In history, there are various types of expansions for planetary disturbing function: (a) Laplace-type expansions where the force amplitudes are expressed in terms of Laplace coefficients \citep{brouwer1961methods, ellis2000disturbing, murray1999solar, morais2013retrograde, namouni2017disturbinga, namouni2017disturbingb}, (b) Legendre-type expansions where Legendre polynomials are used in the derivation \citep{kaula1962development}, (c) asymmetric and local expansions where the disturbing function is expanded around a specific resonant center \citep{ferraz1987expansion, ferraz1989very, yokoyama1994expansion, roig1998high}, (d) Taylor-type expansions where the disturbing function is expanded around an artificially chosen reference point in Taylor series \citep{paez2015trojan, morais1999secular, petrovskaya1970expansions, petrovskaya1972expansions, beauge1996global}, and (e) semianalytical expansions where the force amplitudes are determined by numerical Fourier analysis \citep{klioner2000numerical}. Among these expansions, the Laplace- and Legendre-type expansions are classical and they are widely used in previous studies.

The Laplace-type expansion of planetary disturbing function starts from \citet{peirce1849development}, who derived a sixth-order expansion. As an extension, \citet{newcomb1895development} performed a seventh-order expansion. \citet{brown1933planetary} provided a clear and detailed derivation for the expansion up to the second-order in terms of the classical orbital elements. \citet{brouwer1961methods} formulated a third-order expansion for the planetary disturbing function, which has been widely adopted as a standard and low-order expansion. In more recent times, higher-order explicit expansions of the planetary disturbing function can be found in \citet{ellis2000disturbing} and \citet{murray1999solar}, who expanded the disturbing function as a power series in eccentricities and inclinations of the objects involved. Particularly, \citet{ellis2000disturbing} presented a new algorithm for the determination of the terms associated with a certain argument in the expansion of planetary disturbing function. Recently, \citet{morais2013retrograde} and \citet{namouni2017disturbinga, namouni2017disturbingb} made further extensions of the classical expansions to retrograde orbits, polar orbits and orbits with arbitrary inclinations. Due to the choice of the reference orbit as a coplanar circular orbit, the classical expansions are valid in low-inclination and low-eccentricity configurations. However, the Laplace coefficients and their derivatives with respect to the semimajor axis ratio arising in the Laplace-type expansions are divergent when the semimajor axis ratio is approaching unity. This results in the fact that the Laplace-type expansions are divergent for co-orbital configurations.

Concerning the Legendre-type expansion, \citet{kaula1962development} expanded the disturbing function as a power series of the semimajor axis ratio ($\alpha$) between two objects in triple systems. Please refer to \citet{murray1999solar} for explicit expansions of such an expansion. Obviously, the Legendre-type expansion is applicable for objects with arbitrary inclinations but with semimajor axis ratio much smaller than unity. It has been widely applied to varieties of triple systems with hierarchical configurations \citep{kozai1962secular, beauge2006high, naoz2011hot, naoz2013secular, naoz2016eccentric, lei2018modified, lei2019semi}. However, when the configuration of system is not highly hierarchical, the disturbing function needs to be truncated at a high enough order in the semimajor axis ratio in order to achieve a certain precision and, in this case, the convergence becomes very slow. In particular, when the semimajor axis ratio is close to unity, the Legendre-type expansion fails to converge.

According to the aforementioned discussions, both the Laplace- and Legendre-type expansions of planetary disturbing function are valid for the resonant dynamics in these configurations where the semimajor axis ratio is not close to unity. Concerning this issue, we may ask: is there an expansion valid for minor bodies with arbitrary semimajor axis ratios? To this end, a new expansion is developed in this study. Based on such a `general' expansion, it becomes possible to analytically explore dynamical structures of minor bodies located inside the interior, co-orbital and exterior mean motion resonances at arbitrary inclinations.

The structure of the remaining part is organized as follows. In Section \ref{Sect2}, the disturbing function for the minor bodies in planetary triple systems is briefly introduced, and the new expansion of planetary disturbing function is developed in Section \ref{Sect3}. In Section \ref{Sect4}, the resonant disturbing function are presented under the assumption that the perturber is moving in a circular orbit and then the analytical results obtained by the new expansion is compared with numerical integration results. The Hamiltonian dynamics are discussed in Section \ref{Sect5} and the practical applications of the analytical developments to the interior, co-orbital and exterior resonances are reported in Section \ref{Sect6}. At last, conclusions are summarized in Section \ref{Sect7}.

\section{Planetary disturbing function}
\label{Sect2}

The planetary system considered in this study is composed of a central body with mass $m_0$ (i.e. a central star), a planet with mass $m_p$ and an asteroid with mass $m$. Usually, the mass of the asteroid $m$ is much smaller than $m_0$ and $m_p$, so that it is reasonable to approximate the asteroid as a test particle.

To describe the orbits, we introduce an inertial coordinate system originated at the central body, where the $x$--$y$ plane is aligned with the invariant plane of the system, the $x$-axis points towards an arbitrary direction in the invariant plane, and the $z$-axis goes along the angular momentum vector of the planet moving around the central body. Under this reference frame, the state of the asteroid (or planet) is characterized by orbital elements: the semimajor axis $a$ ($a_p$), eccentricity $e$ ($e_p$), inclination $I$ ($I_p$), longitude of the ascending node $\Omega$ ($\Omega_p$), argument of pericenter $\omega$ ($\omega_p$) and the mean anomaly $M$ ($M_p$) or the true anomaly $f$ ($f_p$). Unless otherwise stated, in the entire study we will adopt the notations without any subscript to stand for the elements of the asteroid and use the ones with subscript $p$ to represent the elements of the planet.

The dynamical model taken in this investigation can be treated as a perturbed Keplerian problem, in which the motion of the asteroid moving around the central body is perturbed by the gravitational attraction of the planet. The disturbing function, governing the evolution of the asteroid, can be written as \citep{murray1999solar}
\begin{equation}\label{Eq1}
{\cal R} = {\cal G}{m_p}\left( {\frac{1}{\Delta } - \frac{r}{{r_p^2}}\cos \psi } \right),
\end{equation}
where ${\cal G}$ is the universal gravitational constant, $r$ and $r_p$ are, respectively, the radial distances of the asteroid and planet relative to the central body. The distance between the asteroid and planet, $\Delta$, is determined by
\begin{equation*}
\Delta  =  {\left( {{r^2} + r_p^2 - 2r{r_p}\cos \psi } \right)^{1/2}}
\end{equation*}
where $\psi$ is the separation angle between the asteroid and planet relative to the central body, given by
\begin{equation}\label{Eq2}
\cos{\psi}  = {\sin ^2}\frac{I}{2}\cos \left( {f + {\theta _p} + \omega  - \Omega } \right) + {\cos ^2}\frac{I}{2}\cos \left( {f - {\theta _p} + \omega  + \Omega } \right),
\end{equation}
with ${\theta _p} = {f_p} + {\omega _p} + {\Omega _p}$ as the true longitude of the planet.

Usually, the disturbing function can be separated into the direct part standing for the direct gravitational attraction from the planet and the indirect part corresponding to the perturbation of the planet to the origin of the selected coordinate system, given by \citep{ellis2000disturbing}
\begin{equation}\label{Eq3}
{\cal R} = {\cal G}{m_p}\left( {{{\cal R}_D} + {{\cal R}_I}} \right),
\end{equation}
where ${\cal R}_D$ and ${\cal R}_I$ are
\begin{equation}\label{Eq4}
{{\cal R}_D} = \frac{1}{\Delta },\quad {{\cal R}_I} =  - \frac{r}{{r_p^2}}\cos \psi.
\end{equation}

\section{A new expansion of planetary disturbing function}
\label{Sect3}

Before expanding the planetary disturbing function, it is necessary to introduce two small parameters with the same order of magnitude as the eccentricities of the asteroid and planet, given by \citep{murray1999solar, ellis2000disturbing}
\begin{equation}\label{Eq5}
\varepsilon  = \frac{r}{a} - 1  \sim {\cal O}(e), \quad {\varepsilon _p} = \frac{{{r_p}}}{{{a_p}}} - 1 \sim {\cal O}(e_p).
\end{equation}
With these two small parameters, the direct part of disturbing function can be expanded around $\varepsilon = \varepsilon _p = 0$ (i.e. around circular orbits) as a formal series in the following form:
\begin{equation}\label{Eq6}
\begin{aligned}
{{\cal R}_D} &= \sum\limits_{n = 0}^N {\sum\limits_{m = 0}^n {\frac{{{\varepsilon ^m}\varepsilon _p^{n - m}}}{{n!}}{a^m}a_p^{n - m}\frac{{{{\rm d}^n}}}{{{\rm d}{a^m}{\rm d}a_p^{n - m}}}\left( {\frac{1}{{{\Delta _0}}}} \right)} }\\
&= \sum\limits_{n = 0}^N {\sum\limits_{m = 0}^n {\frac{{{{\left( {r - a} \right)}^m}{{\left( {{r_p} - {a_p}} \right)}^{n - m}}}}{{n!}}\frac{{{{\rm d}^n}}}{{{\rm d}{a^m}{\rm d}a_p^{n - m}}}\left( {\frac{1}{{{\Delta _0}}}} \right)}},
\end{aligned}
\end{equation}
where $N$ specifies the truncated order in terms of $\max \left(\varepsilon, {\varepsilon _p}\right)$ or $\max \left(e, {e_p}\right)$, and the term $\frac{1}{{{\Delta _0}}}$ is given by
\begin{equation}\label{Eq7}
\frac{1}{{{\Delta _0}}} = {\left[ {{a^2} + a_p^2 - 2a{a_p}\cos \psi } \right]^{ - \frac{1}{2}}}.
\end{equation}
Obviously, the term $\frac{1}{{{\Delta _0}}}$ stands for the mutual interaction between two inclined circular orbits.

It is to be noted that, in the classical Laplace-type expansions, the first step is to expand the disturbing function around a prograde or retrograde coplanar orbit \citep{murray1999solar, ellis2000disturbing, morais2013retrograde} or around an inclined reference orbit \citep{namouni2017disturbinga, namouni2017disturbingb} as a power series of inclination and then, in the second step, the disturbing function is expanded around circular orbits as power series of eccentricities. Thus, the step of series expansion represented by equation (\ref{Eq6}) is in accordance with the second step of the traditional Laplace-type expansions. Through this step of expansion, the disturbing function between two elliptic orbits transforms into a summation of disturbing functions between circular orbits.

Observing equation (\ref{Eq6}), we can find that the difficulty of expanding the direct part of disturbing function ${{\cal R}_D}$ lies in the expansion of the term $\frac{1}{{{\Delta _0}}}$. Let us remind that, in the classical Laplace-type expansions, the term $\frac{1}{{{\Delta _0}}}$ given by equation (\ref{Eq7}) is expanded by using two-dimensional Laplace coefficients as follows \citep{namouni2017disturbingb}:
\begin{equation*}
\begin{aligned}
\frac{1}{{{\Delta _0}}} &= \frac{1}{{{a_p}}}\frac{1}{{\sqrt {1 + {\alpha ^2} - 2\alpha \cos \psi } }}\\
&= \frac{1}{{{a_p}}}\sum\limits_{\scriptstyle - \infty  < j,k < \infty \hfill\atop
\scriptstyle\bmod (j + k,2) = 0\hfill} {\frac{1}{4}b_{1/2}^{jk}(\alpha ,I)\cos \left[ {j\left( {\Omega  - {\lambda _p}} \right) + k\left( {f + \omega } \right)} \right]},
\end{aligned}
\end{equation*}
where $\alpha = a/a_p$ is the semimajor axis ratio between asteroid and planet and $b_{1/2}^{jk}(\alpha ,I)$ are the two-dimensional Laplace coefficients. In practical simulations, the Laplace coefficients $b_{1/2}^{jk}(\alpha ,I)$ are approximated by series expansions in $\alpha$ up to order $N_{\alpha}$. It is known that the Laplace coefficients are divergent when $\alpha$ is approaching unity ($\alpha=1$ corresponds to co-orbital configuration). Therefore, the resulting expansions of disturbing function based on Laplace coefficients are divergent in co-orbital or nearly co-orbital configurations.

To describe all types of configurations including the interior ($\alpha<1$), exterior ($\alpha>1$) and co-orbital ($\alpha = 1$) resonances, it is required to develop a `general' expansion for the planetary disturbing function which can be used for configurations with arbitrary semimajor axis ratios. To this end, we introduce a new small parameter to make it be possible to perform Taylor expansion for the planetary disturbing function whose convergence depends on the magnitude of the introduced small parameter solely. In the entire process, we avoid to use Laplace coefficients, making the new expansion of disturbing function be easy to realize in computer code.

For convenience, we organize equation (\ref{Eq7}) in the following form:
\begin{equation}\label{Eq8}
\frac{1}{{{\Delta _0}}} = \frac{1}{{a + {a_p}}}{\left[ {1 - \frac{{2a{a_p}}}{{{{\left( {a + {a_p}} \right)}^2}}}\left( {1 + \cos \psi } \right)} \right]^{ - \frac{1}{2}}} = \frac{1}{{a + {a_p}}}{\left[ {1 - x} \right]^{ - \frac{1}{2}}},
\end{equation}
where the variable $x$ is given by
\begin{equation}\label{Eq9}
x = \frac{{2a{a_p}}}{{{{\left( {a + {a_p}} \right)}^2}}}\left( {1 + \cos \psi } \right).
\end{equation}
Evidently, the domain of definition of $x$ is $x \in \left[ {0,1} \right]$. In particular, the variable $x$ is equal to zero when the separation angle is $\psi = \pi$, and the variable $x$ is equal to unity when $\psi = 0$ and $a=a_p$ (the condition with $\psi = 0$ and $a=a_p$ corresponds to collision points for two circular inclined orbits). Obviously, equation (\ref{Eq8}) is an increasing function of $x$, so that the direct part of disturbing function, ${{\cal R}_D}$, is also an increasing function of $x$, meaning that there is a positive correlation between ${{\cal R}_D}$ and $x$. Thus, for a given pair of $(a,a_p)$, when the two objects involved are located on the opposite sides of the central star ($\psi = \pi$), the disturbing function ${{\cal R}_D}$ takes the minimum and, when the two objects are located on the same side ($\psi = 0$), the disturbing function ${{\cal R}_D}$ takes the maximum.

Considering the domain of definition of $x$, we can perform Taylor expansion for equation (\ref{Eq8}) around the origin $x=0$ as follows:
\begin{equation*}
\frac{1}{{{\Delta _0}}}  = \frac{1}{{a + {a_p}}}\sum\limits_{k = 0}^{{k_{\max }}} {\frac{{\left( {2k - 1} \right)!!}}{{\left( {2k} \right)!!}}{x^k}},
\end{equation*}
where $k_{\max}$ specifies the truncated order of the expansion in terms of $x$. Except for the collision point at which $x=1$, such a Taylor expansion is uniformly convergent. However, when the variable $x$ is relatively large or close to unity, the convergence of the Taylor expansion becomes very slow. Thus, this is not an ideal expansion.

Observing the expression given by equation (\ref{Eq9}), we find that the variable $x$ oscillates around a fixed value for two circular orbits specified by a given pair of semimajor axes $(a,a_p)$. Naturally, this fixed value, denoted by $x_c$, can be taken as a reference point for $x$ and then equation~(\ref{Eq8}) can be expanded in the vicinity of $x = x_c$ as a Taylor series in the following form:
\begin{equation}\label{Eq10}
\frac{1}{{{\Delta _0}}} = \frac{1}{{a + {a_p}}}\sum\limits_{k = 0}^{{k_{\max }}} {\frac{{\left( {2k - 1} \right)!!}}{{\left( {2k} \right)!!}}\frac{{{{\left( {x - {x_c}} \right)}^k}}}{{{{\left( {1 - {x_c}} \right)}^{1/2 + k}}}}},
\end{equation}
where $k_{\max}$ specifies the truncated order of the expansion in terms of the deviation of $x$ relative to its reference point, namely $\delta x = x - x_c$.

Obviously, the accuracy of the Taylor expansion is determined by the magnitude of $\delta x$ and the truncated order $k_{\max}$. A better choice of the reference point $x_c$ results in a smaller magnitude of $\delta x$, implying that a lower truncated order $k_{\max}$ is required to achieve a given accuracy. Thus, the current difficulty lies in the determination of the reference point. Observing the expression of equation~(\ref{Eq9}), we can find a good choice for the reference point, given by
\begin{equation}\label{Eq11}
{x_c} = \frac{{2{a^0}{a_p^0}}}{{{{\left( {a^0 + {a_p^0}} \right)}^2}}} = \frac{{2{\alpha_0}}}{{{{\left( {1 + {\alpha_0}} \right)}^2}}},
\end{equation}
where the initial semimajor axis ratio, $\alpha_0$, is defined by $\alpha_0 =a^0/a_p^0$, where $a^0$ and $a_p^0$ are, respectively, the initial values of the semimajor axes of the asteroid and planet. It is not difficult to observe that the deviation of $x$ relative to its reference point $x_c$ is
\begin{equation*}
\delta x = x - {x_c} = \frac{{2\alpha }}{{{{\left( {1 + \alpha } \right)}^2}}}\left( {1 + \cos \psi } \right) - \frac{{2{\alpha _0}}}{{{{\left( {1 + {\alpha _0}} \right)}^2}}} \approx \frac{{2\alpha }}{{{{\left( {1 + \alpha } \right)}^2}}}\cos \psi
\end{equation*}
which satisfies $\delta x \in [-0.5, 0.5]$ regardless of the values of semimajor axis ratios and inclinations, ensuring that the Taylor expansion given by equation (\ref{Eq10}) is uniformly convergent. Remember that this is a key feature of our expansion method, making the final expansion be valid for arbitrary inclinations and semimajor axis ratios.

In particular, when the asteroid is located inside the $p_0$:$q_0$ mean motion resonance with respect to the planet, the reference point for $x$ can be approximated as
\begin{equation}\label{Eq12}
{x_c} = \frac{{2{{\left( {\frac{{{q_0}}}{{{p_0}}}} \right)}^{2/3}}}}{{{{\left[ {1 + {{\left( {\frac{{{q_0}}}{{{p_0}}}} \right)}^{2/3}}} \right]}^2}}}.
\end{equation}

\begin{figure*}
\centering
\includegraphics[width=0.48\textwidth]{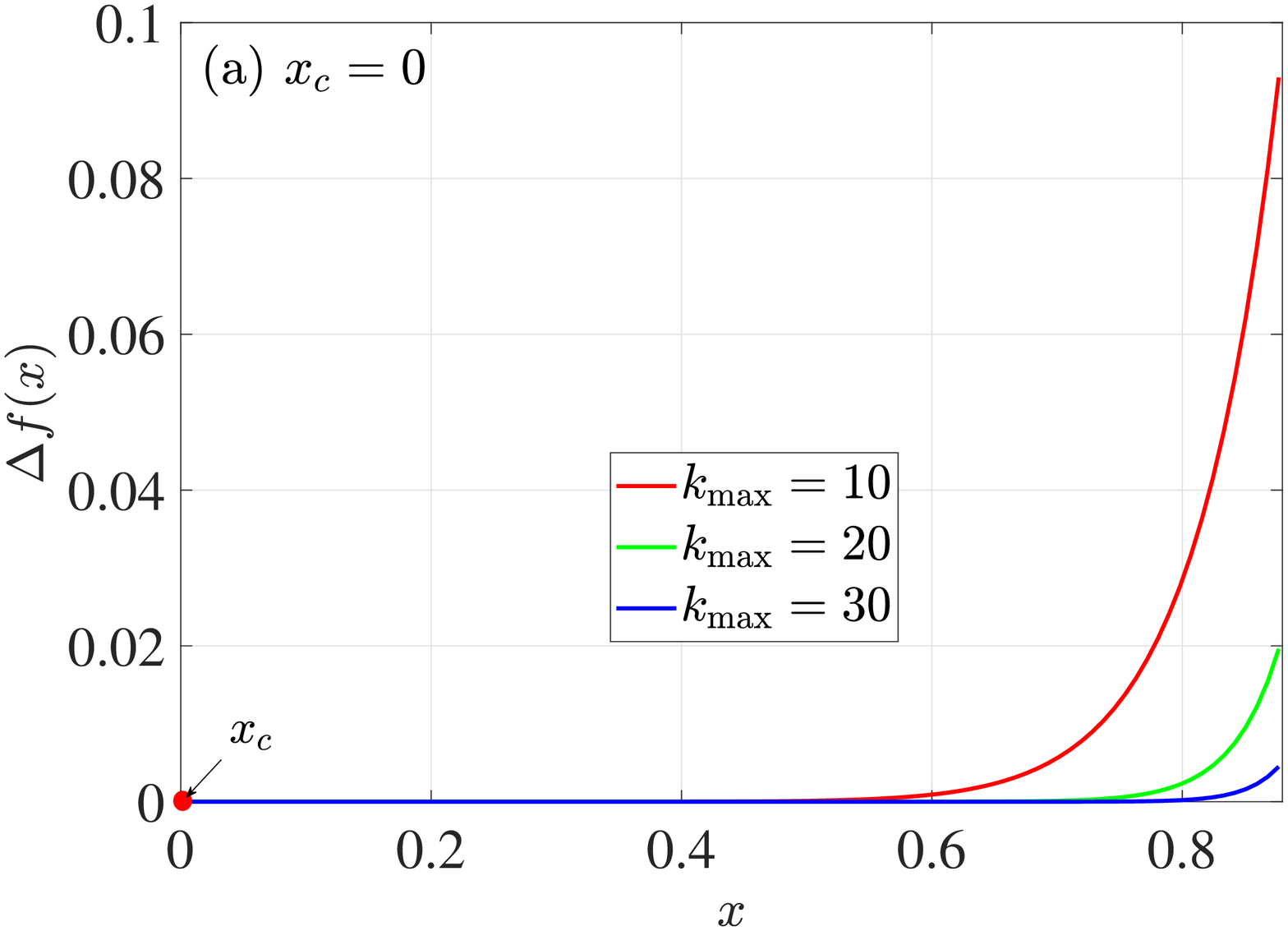}
\includegraphics[width=0.48\textwidth]{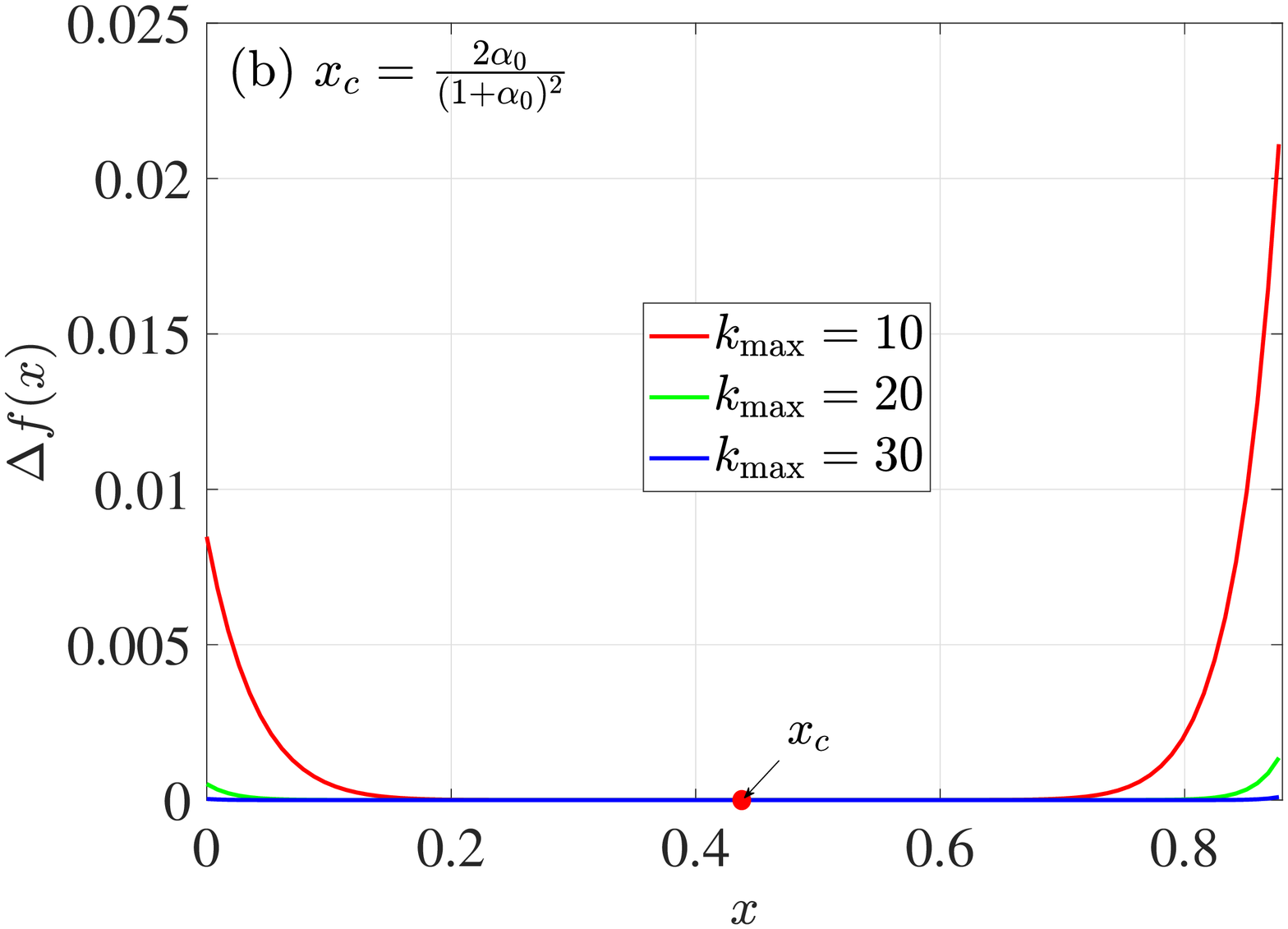}
\caption{Accuracy analysis for the Taylor expansion of the function $f(x) = \frac{1}{\sqrt{1-x}}$ in the case of the reference point at $x_c = 0$ (\emph{left panel}) and $x_c = \frac{{2{\alpha_0}}}{{{{\left( {1 + {\alpha_0}} \right)}^2}}}$ (\emph{right panel}). $\Delta f(x)$ stands for the relative deviation of $f(x)$ (please see the text for its definition). The red points stand for the location of reference points. We take the 3:1 MMR as an example ($\alpha_0 = 0.48$) to do this analysis.}
\label{Fig1_1}
\end{figure*}

In order to validate the Taylor expansion performed by equation (\ref{Eq10}), in Figure \ref{Fig1_1} we make an accuracy analysis for the Taylor expansion of the function $f(x) = \frac{1}{\sqrt{1-x}}$ (equation (\ref{Eq8}) shows $\frac{1}{\Delta_0} = \frac{1}{a + a_p} f(x)$) truncated at orders $k_{\max} = 10, 20, 30$ in the case of the reference point at $x_c = 0$ (origin) and $x_c = \frac{{2{\alpha_0}}}{{{{\left( {1 + {\alpha_0}} \right)}^2}}}$. The relative deviation is defined by
\begin{equation*}
\Delta f(x) = \frac{\left| f_{\rm approx} - f_{\rm accurate} \right|}{f_{\rm accurate}},
\end{equation*}
where $f_{\rm approx}$ is the magnitude obtained by Taylor expansion and $f_{\rm accurate}$ stands for the exact value of $f(x)$. From Figure \ref{Fig1_1}, we can see that, at the same truncated order, the expansion around $x_c = \frac{{2{\alpha_0}}}{{{{\left( {1 + {\alpha_0}} \right)}^2}}}$ has better accuracy than that around $x_c = 0$. This is in agreement with our expectation.

Substituting equation (\ref{Eq9}) into equation (\ref{Eq10}) and performing Newton's binomial expansion to the result can lead to the following expression:
\begin{equation}\label{Eq13}
\frac{1}{{{\Delta _0}}} = \sum\limits_{k = 0}^{{k_{\max }}} {\sum\limits_{q = 0}^k {\sum\limits_{l = 0}^q {\frac{{{2^q}\left( {2k - 1} \right)!!}}{{\left( {2k} \right)!!}}{{\left( { - 1} \right)}^{k - q}}{k \choose q} {q \choose l}\frac{{x_c^{k - q}}}{{{{\left( {1 - {x_c}} \right)}^{1/2 + k}}}}\frac{{{a^q}a_p^q}}{{{{\left( {a + {a_p}} \right)}^{2q + 1}}}}{{\cos }^l}\psi } } },
\end{equation}
where $i \choose j$ is the binomial coefficient, defined by ${i \choose j} = \frac{{i!}}{{j!(i - j)!}}$. For the sake of brevity, we denote the function associated with the semimajor axes $a$ and $a_p$ by
\begin{equation}\label{Eq14}
{f_q}(a,{a_p}) = \frac{{{a^q}a_p^q}}{{{{\left( {a + {a_p}} \right)}^{2q + 1}}}}
\end{equation}
and define the differential operator with respect to the semimajor axes of the asteroid and planet as
\begin{equation}\label{Eq15}
{{\rm D}_{{k_1},{k_2}}} = \frac{{{{\rm d}^{{k_1} + {k_2}}}}}{{{\rm d}{a^{{k_1}}}{\rm d}a_p^{{k_2}}}}.
\end{equation}
Through some mathematical derivation, the high-order partial derivatives of ${f_q}(a,{a_p})$ with respect to $a$ and $a_p$ can be calculated by
\begin{equation}\label{Eq16}
{{\rm D}_{{k_1},{k_2}}}{f_q} = \sum\limits_{{l_2} = 0}^{{k_2}} {\sum\limits_{{l_1} = 0}^{{k_1}} {\frac{{{{\left( { - 1} \right)}^{{l_2} + {l_1}}}\left( {2q + {l_2} + {l_1}} \right)!}}{{{{\left( {a + {a_p}} \right)}^{2q + 1 + {l_2} + {l_1}}}\left( {2q} \right)!}} {{k_2} \choose {l_2}} {{k_1} \choose {l_1}}\left[ {\frac{{{{\rm d}^{{k_1} - {l_1}}}}}{{{\rm d}{a^{{k_1} - {l_1}}}}}{a^q}} \right]\left[ {\frac{{{{\rm d}^{{k_2} - {l_2}}}}}{{{\rm d}a_p^{{k_2} - {l_2}}}}a_p^q} \right]} }.
\end{equation}

Replacing equation (\ref{Eq13}) in equation (\ref{Eq6}) and combining the differential operator defined by equation (\ref{Eq15}), we can obtain the direct part of disturbing function as follows:
\begin{equation}\label{Eq17}
\begin{aligned}
{{\cal R}_D} & =  \sum\limits_{n = 0}^N {\sum\limits_{m = 0}^n {\sum\limits_{k = 0}^{{k_{\max }}} {\sum\limits_{q = 0}^k {\sum\limits_{l = 0}^q {\frac{{{2^q}\left( {2k - 1} \right)!!}}{{\left( {2k} \right)!!}}{{\left( { - 1} \right)}^{k - q}} {k \choose q} {q \choose l} \frac{{x_c^{k - q}}}{{{{\left( {1 - {x_c}} \right)}^{1/2 + k}}}}} } } } } \\
&\times \frac{{{{\left( {r - a} \right)}^m}{{\left( {{r_p} - {a_p}} \right)}^{n - m}}}}{{n!}}\left[ {{{\rm D}_{m,n - m}}{f_q}\left( {a,{a_p}} \right)} \right]{\cos ^l}\psi ,
\end{aligned}
\end{equation}
where the term ${\cos ^l}\psi$ can be expanded to be
\begin{equation}\label{Eq18}
\begin{aligned}
{\cos ^l}\psi  &= \sum\limits_{t = 0}^l {\sum\limits_{{t_1} = 0}^t {\sum\limits_{{t_2} = 0}^{l - t} {\frac{1}{{{2^l}}} {l \choose t} {t \choose {{t_1}}} {{l - t} \choose {{t_2}}} {{\sin }^{2\left( {l - t} \right)}}({I \mathord{\left/
 {\vphantom {I 2}} \right.
 \kern-\nulldelimiterspace} 2}){\cos}^{2t}({I \mathord{\left/
 {\vphantom {I 2}} \right.
 \kern-\nulldelimiterspace} 2})} } } \\
&\times \cos \left[ {\left( {l - 2{t_1} - 2{t_2}} \right)\left( {f + \omega } \right) + \left( {l - 2t + 2{t_1} - 2{t_2}} \right)\left( {{\theta _p} - \Omega } \right)} \right]
\end{aligned}
\end{equation}
and the terms ${\left( {r - a} \right)^m}$ and ${\left( {{r_p} - {a_p}} \right)^{n - m}}$ can be expanded as follows:
\begin{equation}\label{Eq19}
\begin{aligned}
&{\left( {r - a} \right)^m} = \sum\limits_{{t_3} = 0}^m {{{\left( { - 1} \right)}^{m - {t_3}}}{a^m} {m \choose t_3} {{\left( {\frac{r}{a}} \right)}^{{t_3}}}},\\
&{\left( {{r_p} - {a_p}} \right)^{n - m}} = \sum\limits_{{t_4} = 0}^{n - m} {{{\left( { - 1} \right)}^{n - m - {t_4}}} {n-m \choose t_4} a_p^{n - m}{{\left( {\frac{{{r_p}}}{{{a_p}}}} \right)}^{{t_4}}}}.
\end{aligned}
\end{equation}
By substituting Eqs. (\ref{Eq18}) and (\ref{Eq19}) into equation (\ref{Eq17}), the direct part of disturbing function becomes
\begin{equation}\label{Eq20}
\begin{aligned}
{{\cal R}_D} & = \sum\limits_{n = 0}^N {\sum\limits_{m = 0}^n {\sum\limits_{k = 0}^{{k_{\max }}} {\sum\limits_{q = 0}^k {\sum\limits_{l = 0}^q {\sum\limits_{t = 0}^l {\sum\limits_{{t_1} = 0}^t {\sum\limits_{{t_2} = 0}^{l - t} {\sum\limits_{{t_3} = 0}^m {\sum\limits_{{t_4} = 0}^{n - m} {\kappa \left[ {{{\rm D}_{m,n - m}}{f_q}} \right]{a^m}a_p^{n - m}\frac{{{{\sin }^{2l}}\left( {{I \mathord{\left/
 {\vphantom {I 2}} \right.
 \kern-\nulldelimiterspace} 2}} \right)}}{{{{\tan }^{2t}}\left( {{I \mathord{\left/
 {\vphantom {I 2}} \right.
 \kern-\nulldelimiterspace} 2}} \right)}}} } } } } } } } } } \\
& \times {\left( {\frac{r}{a}} \right)^{{t_3}}}{\left( {\frac{{{r_p}}}{{{a_p}}}} \right)^{{t_4}}}\cos \left[ {\left( {l - 2{t_1} - 2{t_2}} \right)\left( {f + \omega } \right) + \left( {l - 2t + 2{t_1} - 2{t_2}} \right)\left( {{\theta _p} - \Omega } \right)} \right]
\end{aligned}
\end{equation}
where $\kappa$ is given by
\begin{equation*}
\kappa  = \frac{{{2^{q - l}}x_c^{k - q}\left( {2k - 1} \right)!!}}{{{{\left( {1 - {x_c}} \right)}^{1/2 + k}}n!\left( {2k} \right)!!}}{\left( { - 1} \right)^{n + k - q - {t_3} - {t_4}}} {k \choose q} {q \choose l} {t \choose {{t_1}}} {{l - t} \choose {{t_2}}} {l \choose t} {m \choose {{t_3}}} {{n - m} \choose {{t_4}}}.
\end{equation*}
The trigonometric functions related to the true anomaly arising in equation (\ref{Eq20}) can be transformed into the forms of the mean anomaly by the following relation \citep{kaula1961analysis, kaula1962development}:
\begin{equation}\label{Eq21}
\begin{aligned}
&{\left( {\frac{r}{a}} \right)^{{t_3}}}{\left( {\frac{{{r_p}}}{{{a_p}}}} \right)^{{t_4}}}\cos \left[ {\left( {l - 2{t_1} - 2{t_2}} \right)\left( {f + \omega } \right) + \left( {l - 2t + 2{t_1} - 2{t_2}} \right)\left( {{\theta _p} - \Omega } \right)} \right]\\
& = \sum\limits_{{s_1} =  - \infty }^\infty  {\sum\limits_{{s_2} =  - \infty }^\infty  {X_{{s_1}}^{{t_3},\left( {l - 2{t_1} - 2{t_2}} \right)}\left( e \right)X_{{s_2}}^{{t_4},\left( {l - 2t + 2{t_1} - 2{t_2}} \right)}\left( {{e_p}} \right)} } \\
&\times \cos \left[ {{s_1}M + {s_2}{M_p} + \left( {l - 2{t_1} - 2{t_2}} \right)\omega  + \left( {l - 2t + 2{t_1} - 2{t_2}} \right)\left( {{\varpi _p} - \Omega } \right)} \right]
\end{aligned}
\end{equation}
The mathematical expression given by equation (\ref{Eq21}) has been used in previous studies \citep{murray1999solar, ellis2000disturbing, beauge2003modelling, beauge2006high, lei2019secular, lei2019semi, lei2019three}. The Hansen coefficient $X_c^{a,b}(e)$ arising in equation (\ref{Eq21}) is a function of the eccentricity $e$, given by \citep{hughes1981computation}
\begin{equation*}
X_c^{a,b}(e) = {e^{\left| {c - b} \right|}}\sum\limits_{s = 0}^\infty  {Y_{s + t,s + u}^{a,b}{e^{2s}}},
\end{equation*}
where $t=\max(0,c-b)$, $u=\max(0,b-c)$, and $Y_{s+t,s+u}^{a,b}$ is the Newcomb operator calculated in a recurrence manner \citep{hughes1981computation, murray1999solar, ellis2000disturbing}. As for the Hansen coefficient $X_c^{a,b}(e)$, its lowest power in eccentricity is ${\left| {c - b} \right|}$. It should be mentioned that the expansion given by equation (\ref{Eq21}) is convergent under the condition that the eccentricities ($e$ and $e_p$) are smaller than the critical value $e_c = 0.6627$ \citep{wintner1947analytical}.

Substituting equation (\ref{Eq21}) into equation (\ref{Eq20}), we can organize the direct part of the disturbing function as a Fourier series in terms of the classical angles $M$, $M_p$, $\Omega$, $\omega$ and ${{\varpi _p}}$ in the following form:
\begin{equation}\label{Eq22}
\begin{aligned}
{{\cal R}_D} &= \sum\limits_{n = 0}^N {\sum\limits_{m = 0}^n {\sum\limits_{k = 0}^{{k_{\max }}} {\sum\limits_{q = 0}^k {\sum\limits_{l = 0}^q {\sum\limits_{t = 0}^l {\sum\limits_{{t_1} = 0}^t {\sum\limits_{{t_2} = 0}^{l - t} {\sum\limits_{{t_3} = 0}^m {\sum\limits_{{t_4} = 0}^{n - m} {\sum\limits_{{s_1} =  - \infty }^\infty  {\sum\limits_{{s_2} =  - \infty }^\infty  {\kappa  \times \left[ {{{\rm D}_{m,n - m}}{f_q}} \right]} } } } } } } } } } } } \\
& \times {a^m}a_p^{n - m}X_{{s_1}}^{{t_3},\left( {l - 2{t_1} - 2{t_2}} \right)}\left( e \right)X_{{s_2}}^{{t_4},\left( {l - 2t + 2{t_1} - 2{t_2}} \right)}\left( {{e_p}} \right)\frac{{{{\sin }^{2l}}\left( {{I \mathord{\left/
 {\vphantom {I 2}} \right.
 \kern-\nulldelimiterspace} 2}} \right)}}{{{{\tan }^{2t}}\left( {{I \mathord{\left/
 {\vphantom {I 2}} \right.
 \kern-\nulldelimiterspace} 2}} \right)}}\\
& \times \cos \left[ {{s_1}M + {s_2}{M_p} + \left( {l - 2{t_1} - 2{t_2}} \right)\omega  + \left( {l - 2t + 2{t_1} - 2{t_2}} \right)\left( {{\varpi _p} - \Omega } \right)} \right].
\end{aligned}
\end{equation}
For the indirect part, its Fourier series form can be easily derived to be
\begin{equation}\label{Eq23}
\begin{aligned}
{{\cal R}_I} & =  - \frac{a}{{a_p^2}}\sum\limits_{{s_1} =  - \infty }^\infty  {\sum\limits_{{s_2} =  - \infty }^\infty  {X_{{s_1}}^{1,1}\left( e \right)X_{{s_2}}^{ - 2,1}\left( {{e_p}} \right)} } \\
& \times \left[ \begin{array}{l}
{\sin ^2}\frac{I}{2}\cos \left( {{s_1}M + {s_2}{M_p} + \omega  + {\varpi _p} - \Omega } \right)\\
 + {\cos ^2}\frac{I}{2}\cos \left( {{s_1}M - {s_2}{M_p} + \omega  - {\varpi _p} + \Omega } \right)
\end{array} \right].
\end{aligned}
\end{equation}

At last, we arrive at the final literal expansion of the planetary disturbing function in the following form:
\begin{equation}\label{Eq24}
\begin{aligned}
{\cal R} & = {\cal G}{m_p}\sum\limits_{n = 0}^N {\sum\limits_{m = 0}^n {\sum\limits_{k = 0}^{{k_{\max }}} {\sum\limits_{q = 0}^k {\sum\limits_{l = 0}^q {\sum\limits_{t = 0}^l {\sum\limits_{{t_1} = 0}^t {\sum\limits_{{t_2} = 0}^{l - t} {\sum\limits_{{t_3} = 0}^m {\sum\limits_{{t_4} = 0}^{n - m} {\sum\limits_{{s_1} =  - \infty }^\infty  {\sum\limits_{{s_2} =  - \infty }^\infty  \kappa  } } } } } } } } } } } \\
& \times \left[ {{{\rm D}_{m,n - m}}{f_q}} \right]{a^m}a_p^{n - m}X_{{s_1}}^{{t_3},\left( {l - 2{t_1} - 2{t_2}} \right)}\left( e \right)X_{{s_2}}^{{t_4},\left( {l - 2t + 2{t_1} - 2{t_2}} \right)}\left( {{e_p}} \right)\frac{{{{\sin }^{2l}}\left( {{I \mathord{\left/
 {\vphantom {I 2}} \right.
 \kern-\nulldelimiterspace} 2}} \right)}}{{{{\tan }^{2t}}\left( {{I \mathord{\left/
 {\vphantom {I 2}} \right.
 \kern-\nulldelimiterspace} 2}} \right)}}\\
& \times \cos \left[ \begin{array}{l}
{s_1}\lambda  + {s_2}{\lambda _p} + \left( {l - 2{t_1} - 2{t_2} - {s_1}} \right)\varpi \\
 + \left( {l - 2t + 2{t_1} - 2{t_2} - {s_2}} \right){\varpi _p} + \left( {2t + 4{t_2} - 2l} \right)\Omega
\end{array} \right]\\
& - \frac{{{\cal G}{m_p}}}{{{a_p}}}\left( {\frac{a}{{{a_p}}}} \right)\sum\limits_{{s_1} =  - \infty }^\infty  {\sum\limits_{{s_2} =  - \infty }^\infty  {X_{{s_1}}^{1,1}\left( e \right)X_{{s_2}}^{ - 2,1}\left( {{e_p}} \right)} } \\
& \times \left\{ \begin{array}{l}
{\sin ^2}\frac{I}{2}\cos \left[ {{s_1}\lambda  + {s_2}{\lambda _p} + \left( {1 - {s_1}} \right)\varpi  + \left( {1 - {s_2}} \right){\varpi _p} - 2\Omega } \right]\\
 + {\cos ^2}\frac{I}{2}\cos \left[ {{s_1}\lambda  - {s_2}{\lambda _p} + \left( {1 - {s_1}} \right)\varpi  - \left( {1 - {s_2}} \right){\varpi _p}} \right]
\end{array} \right\}
\end{aligned}
\end{equation}
where $\varpi  = \omega  + \Omega$ and $\varpi _p  = \omega _p  + \Omega _p$ are the longitudes of pericenter, and $\lambda  = M + \varpi$ and ${\lambda _p} = {M_p} + {\varpi _p}$ are the mean longitudes of the asteroid and planet, respectively. It should be mentioned that the expansion given by equation~(\ref{Eq24}) is convergent when $e < 0.6627$ and $e_p < 0.6627$ are satisfied due to the convergence of the transformation between mean and true anomalies given by equation (\ref{Eq21}).

In the final expansion, the harmonic arguments are linear combinations of the mean longitudes, longitudes of pericenter and longitudes of ascending node of the asteroid and planet, denoted by
\begin{equation*}
\varphi  = {k_1}\lambda  + {k_2}\varpi  + {k_3}\Omega  + {k_4}{\lambda _p} + {k_5}{\varpi _p} + {k_6}{\Omega _p},
\end{equation*}
where the coefficients $k_i (i=1,2,...6)$ are all integers. As a result, the expansion of disturbing function can be written in an elegant form:
\begin{equation}\label{Eq25}
{\cal R} = \sum\limits_{{\bm k}} {{{\cal C}_{{\bm k}}}\left( {a,e,I,{a_p},{e_p}} \right)\cos{\varphi}},
\end{equation}
where ${\bm k} = \left\{{k_1},{k_2},{k_3},{k_4},{k_5}\right\}$ and the explicit expressions of the force amplitudes ${{{\cal C}_{{\bm k}}}\left( {a,e,I,{a_p},{e_p}} \right)}$ can be directly derived from equation (\ref{Eq24}). Observing the literal expansion given by equation (\ref{Eq24}), we can summarize the following properties:
\begin{itemize}
  \item The expansion of the planetary disturbing function contains only cosine terms of argument $\varphi$.
  \item The coefficient $k_6$ is zero, meaning that the perturber's longitude of ascending node vanishes from the disturbing function. This is due to the choice of the reference plane as the orbit of the perturber (i.e. the inclination of the perturber's orbit is zero).
  \item The d'Alembert relation holds for the coefficients in the cosine arguments, namely $\sum\limits_i {{k_i} = 0}$.
  \item The coefficient of the longitude of ascending node $\Omega$ is always an even number.
  \item In the expression of the force amplitude, the power of $\sin(I/2)$ (or $\cos(I/2)$) is always an even number, indicating that the force amplitude is an even function with respect to the mutual inclination.
  \item The relationship between the coefficient of $\Omega$ and the power of $\sin(\frac{I}{2})$ holds $\left|2t+4t_2-2l\right| \le 2(l-t)$ (see the expression of the direct part), meaning that the power of $\sin(\frac{I}{2})$ arising in the force amplitude is greater than or equal to the absolute value of the coefficient of $\Omega$ in the cosine argument (such relationship also holds for the indirect part).
\end{itemize}

Regarding the expression of Hansen coefficient, we can determine its lowest order terms in eccentricity as $X_c^{a,b}\left( e \right) = {\cal O}\left( {{e^{\left| {b - c} \right|}}} \right)$. Applying this property to the coefficients appearing in equation (\ref{Eq24}) yields
\begin{equation*}
\begin{aligned}
& X_{{s_1}}^{{t_3},\left( {l - 2{t_1} - 2{t_2}} \right)}\left( e \right) = {\cal O}\left( {{e^{\left| {l - 2{t_1} - 2{t_2} - {s_1}} \right|}}} \right),\\
& X_{{s_2}}^{{t_4},\left( {l - 2t + 2{t_1} - 2{t_2}} \right)}\left( {{e_p}} \right) = {\cal O}\left( {{e_p^{\left| {l - 2t + 2{t_1} - 2{t_2} - {s_2}} \right|}}} \right),\\
& X_{{s_1}}^{1,1}\left( e \right) = {\cal O}\left( {{e^{\left| {1 - {s_1}} \right|}}} \right),\quad X_{{s_2}}^{ - 2,1}\left( {{e_p}} \right) = {\cal O}\left( {e_p^{\left| {1 - {s_2}} \right|}} \right).
\end{aligned}
\end{equation*}
Thus, it is not difficult to observe that, in the expansion given by equation (\ref{Eq24}), the lowest powers of $e$ and $e_p$ arising in the force amplitude are the absolute values of the coefficients of $\varpi$ and $\varpi_p$, respectively (i.e. $\left| k_2 \right|$ and $\left| k_5 \right|$). According to this property, it is possible to see that (a) the amplitude of a certain harmonic argument has positive correlation with the eccentricity $e$ ($e_p$) if the number $k_2$ ($k_5$) is not equal to zero, and (b) when the eccentricity $e$ ($e_p$) becomes zero, the terms associated with the arguments $\varphi$ with non-zero $k_2$ ($k_5$) would vanish from the disturbing function.

In particular, when the inclinations and eccentricities are small (in this situation the classical Laplace-type expansion works very well), it is possible to reserve the lowest-order terms of $e$ ($e_p$) and $\sin(I/2)$ in equation (\ref{Eq24}) to approximate the disturbing function:
\begin{equation}\label{Eq26}
{\cal R} \approx \sum\limits_{\bm{k}} {{{\tilde {\cal C}}_{\bm{k}}}\left( {a,{a_p}} \right){e^{\left| {{k_2}} \right|}}e_p^{\left| {{k_5}} \right|}{{\sin }^{\left| {{k_3}} \right|}}(I/2)\cos \left( {{k_1}\lambda  + {k_2}\varpi  + {k_3}\Omega  + {k_4}{\lambda _p} + {k_5}{\varpi _p}} \right)},
\end{equation}
which has a similar formal expression to the ones given in previous studies \citep{murray1999solar, morbidelli2002modern}.

Especially, when the planet is assumed to move around the central body in a circular orbit (i.e. $e_p = 0$), the resulting dynamical model, describing the motion of the asteroid, reduces to the circular restricted three-body problem (CRTBP) \citep{szebehely1967theory}. In this case, the Hansen coefficients $X_{{s_2}}^{{t_4},\left( {l - 2t + 2{t_1} - 2{t_2}} \right)}\left( {{e_p}} \right)$ in the expansion given by equation (\ref{Eq24}) are different from zero only when $s_2$ is equal to ${l - 2t + 2{t_1} - 2{t_2}}$, and the coefficients ${X_{{s_2}}^{ - 2,1}\left( {{e_p}} \right)}$ are different from zero only when $s_2$ is equal to unity. Consequently, the disturbing function for describing the motion of asteroids in the CRTBP can be simplified to be
\begin{equation}\label{Eq27}
\begin{aligned}
{\cal R} & = \frac{{{\cal G}{m_p}}}{{{a_p}}}\sum\limits_{n = 0}^N {\sum\limits_{k = 0}^{{k_{\max }}} {\sum\limits_{q = 0}^k {\sum\limits_{l = 0}^q {\sum\limits_{m = 0}^n {\sum\limits_{t = 0}^l {\sum\limits_{{t_1} = 0}^t {\sum\limits_{{t_2} = 0}^{l - t} {\sum\limits_{s =  - \infty }^\infty  {{\kappa _0}  X_s^{m,\left( {l - 2{t_1} - 2{t_2}} \right)}\left( e \right)\frac{{{\alpha ^n}}}{{n!}}} } } } } } } } } \\
& \times \left[ {{{\rm D}_n}{f_q}(\alpha )} \right]\frac{{{{\sin }^{2l}}\left( {\frac{I}{2}} \right)}}{{{{\tan }^{2t}}\left( {\frac{I}{2}} \right)}}\cos \left[ \begin{array}{l}
s\lambda  + \left( {l - 2t + 2{t_1} - 2{t_2}} \right){\lambda _p}\\
 + \left( {l - 2{t_1} - 2{t_2} - s} \right)\varpi  + \left( {2t - 2l + 4{t_2}} \right)\Omega
\end{array} \right]\\
& - \frac{{{\cal G}{m_p}}}{{{a_p}}}\left( {\frac{a}{{{a_p}}}} \right)\sum\limits_{s =  - \infty }^\infty  {X_s^{1,1}\left( e \right)\left\{ \begin{array}{l}
{\sin ^2}({I \mathord{\left/
 {\vphantom {I 2}} \right.
 \kern-\nulldelimiterspace} 2})\cos \left[ {s\lambda  + {\lambda _p} + \left( {1 - s} \right)\varpi  - 2\Omega } \right]\\
 + {\cos ^2}({I \mathord{\left/
 {\vphantom {I 2}} \right.
 \kern-\nulldelimiterspace} 2})\cos \left[ {s\lambda  - {\lambda _p} + \left( {1 - s} \right)\varpi } \right]
\end{array} \right\}}
\end{aligned}
\end{equation}
where $\alpha$ is the semimajor axis ratio between the asteroid and planet (i.e. $\alpha = a/a_p$) and $\kappa_0$ is
\begin{equation*}
{\kappa _0} = \frac{{{{\left( { - 1} \right)}^{n - m + k - q}}{2^{q - l}}\left( {2k - 1} \right)!!}}{{\left( {2k} \right)!!}} {k \choose q} {q \choose l} {n \choose m} {t \choose {{t_1}}} {{l - t} \choose {{t_2}}} {l \choose t} \frac{{x_c^{k - q}}}{{{{\left( {1 - {x_c}} \right)}^{1/2 + k}}}}.
\end{equation*}
In practical simulations, the upper limit of $\left|s\right|$ arising in the summations is also taken as 20. In equation (\ref{Eq27}), the functions associated with $\alpha$ are denoted by
\begin{equation*}
{f_q}(\alpha ) = \frac{{{\alpha ^q}}}{{{{\left( {1 + \alpha } \right)}^{2q + 1}}}}, \quad {{\rm D}_n}{f_q}(\alpha ) = \frac{{{{\rm d}^n}}}{{{\rm d}{\alpha ^n}}}{f_q}(\alpha ),
\end{equation*}
where the high-order derivatives are calculated by
\begin{equation*}
{{\rm D}_n}{f_q}(\alpha ) = \sum\limits_{m = 0}^n {{{\left( { - 1} \right)}^m}\frac{{\left( {2q + m} \right)!}}{{\left( {2q} \right)!}} {n \choose m} {{\left( {1 + \alpha } \right)}^{ - 2q - 1 - m}}\left[ {\frac{{{{\rm d}^{n - m}}}}{{{\rm d}{\alpha ^{n - m}}}}{\alpha ^q}} \right]}.
\end{equation*}

Remember that, in equation (\ref{Eq24}) or (\ref{Eq27}), there are two numbers characterizing the truncated orders of series expansion: $N$ and $k_{\max}$. The first number $N$ specifies the maximum order in terms of the small parameter: $\max \left(\varepsilon, {\varepsilon_p} \right)$. As discussed before, $\varepsilon$ and $\varepsilon_p$ have the same order of magnitude as the eccentricities of the asteroid and planet, thus the number $N$ corresponds to the order in terms of $\max \left(e, e_p\right)$. The second number $k_{\max}$ determines the maximum order of the Taylor expansion in the deviation of $x$ relative to its reference point, namely $\delta x = x - x_c$.

Note that the convergence of the new expansion developed in this study is not restricted by the mutual inclination and the semimajor axis ratio between the asteroid and planet because the small parameter arising in the series expansion always satisfies $\delta x \in [-0.5,0.5]$ regardless of mutual inclination and semimajor axis ratio. Thus, it becomes possible to use such a `general' expansion of disturbing function to formulate dynamical models in describing the interior ($\alpha < 1$), co-orbital ($\alpha \approx 1$) and exterior ($\alpha > 1$) mean motion resonances of minor bodies with arbitrary inclinations in planetary systems.

\section{Resonant disturbing function}
\label{Sect4}

In this section, we assume that the planet moves in a circular orbit around the central body (the model corresponds to the well-known CRTBP). In this case, the disturbing function given by equation (\ref{Eq27}) can be utilized to formulate resonant models. Regarding an asteroid located inside the $p_0$:$q_0$ mean motion resonance with respect to the planet, there is more than one critical argument (with different $k_0$), given by
\begin{equation}\label{Eq28}
\sigma _{{k_0}}^{{p_0}:{q_0}} = {q_0}\lambda  - {p_0}{\lambda _p} + \left( {{p_0} - {q_0}} \right)\Omega  + {k_0}\omega,
\end{equation}
where $k_0$ has the same parity of $p_0 + q_0$. In previous works, several special resonances associated with $p_0$:$q_0$ resonances have been discussed, such as the pure eccentricity (or inclination) resonances and retrograde resonances \citep{murray1999solar, morais2013retrograde, namouni2017disturbingb}. When $k_0 = {{p_0} - {q_0}}$, the angle defined by equation (\ref{Eq28}) becomes the critical argument of pure eccentricity resonance, denoted by $\sigma _1 = {q_0}\lambda  - {p_0}{\lambda _p} + \left( {{p_0} - {q_0}} \right)\varpi$. When $k_0 = 0$ (it requires that $p_0 + q_0$ is even), the angle defined by equation (\ref{Eq28}) represents the critical argument of pure inclination resonance, denoted by $\sigma _2 = {q_0}\lambda  - {p_0}{\lambda _p} + \left( {{p_0} - {q_0}} \right)\Omega$ (it is noted that, when $p_0 - q_0 = 1$, the pure inclination argument is defined by $\sigma _2 = 2({q_0}\lambda  - {p_0}{\lambda _p} + \left( {{p_0} - {q_0}} \right)\Omega)$). When $ k_0 = - p_0 - q_0$, the angle defined by equation (\ref{Eq28}) corresponds to the critical argument of the retrograde (pure eccentricity) resonance, denoted by $\sigma _3 = {q_0}\lambda  - {p_0}{\lambda _p} + \left( {{p_0} - {q_0}} \right)\Omega  - \left( {{p_0} + {q_0}} \right)\omega$. It is known that, for the minor bodies located on coplanar or nearly coplanar orbits, the pure eccentricity resonance with $\sigma _1$ as the critical argument has the dominant strength and, for the bodies moving in retrograde coplanar orbits, the retrograde resonance with $\sigma _3$ as the critical argument has the dominant strength. However, when the minor bodies moves on an inclined orbit, the force amplitudes of the arguments $\sigma _{{k_0}}^{{p_0}:{q_0}}$ with different $k_0$ arising in the disturbing function may have comparable amplitudes, indicating that it is not suitable to use a single critical argument in describing the resonant behavior. Please refer to Figure 16 in \citet{lei2019three} for detailed discussions on single critical argument and characteristic angle for a certain mean motion resonance.

Among all the critical arguments of the $p_0$:$q_0$ resonance, there is a common part, given by $\sigma  = {q_0}\lambda  - {p_0}{\lambda _p} + \left( {{p_0} - {q_0}} \right)\Omega$, which corresponds to the special case with $k_0 = 0$ in equation (\ref{Eq28}). In \citet{lei2019three}, the common angle $\sigma$ is defined as the characteristic angle of the $p_0$:$q_0$ resonance (notice that the expression of $\sigma$ is determined by the numbers $p_0$ and $q_0$).

According to the mathematical expression of $\sigma$, the mean longitude of the minor body can be expressed by means of $\sigma$, $\lambda_p$ and $\Omega$ as
\begin{equation*}
\lambda  = \frac{1}{{{q_0}}}\left[ {\sigma  + {p_0}{\lambda _p} - \left( {{p_0} - {q_0}} \right)\Omega } \right].
\end{equation*}
As a result, the disturbing function given by equation (\ref{Eq27}) becomes a function of the angles $\sigma$, $\omega$, $\Omega$ and $\lambda_p$. For a minor body located inside a mean motion resonance, the associated resonant angle ($\sigma$) becomes a semi-secular angle variable compared to $\lambda_p$. In addition, it is known that the angles $\omega$ and $\Omega$ are slow variables in the timescale of mean motion resonance \citep{gallardo2019strength, gallardo2020three}. Thus, among all these angles ($\sigma$, $\omega$, $\Omega$, $\lambda_p$), only the mean longitude of the planet ($\lambda_p$) is a fast variable. Usually, the terms in the disturbing function involving the fast angle $\lambda_p$ produce short-term influences upon the motion of the minor body. During the long-term evolution of the minor body, it is often to remove those short-term effects by means of averaging technique. Such an averaging process is called ``secular approximation" \citep{naoz2013secular}. Alternatively, the short-period effects can be averaged out by means of von Zeiple's method \citep{brouwer1959solution} or Hori--Deprit method \citep{hori1966theory, deprit1969canonical}. As a result, filtering out the short-term effects (i.e. averaging the disturbing function over $\lambda_p$) can be achieved by
\begin{equation}\label{Eq29}
{{\cal R}^*} \left( a,e,I, {\sigma , \omega} \right) = \frac{1}{{2 q_0 \pi }}\int\limits_0^{2 q_0 \pi } {{\cal R}\left( a,e,I,{\sigma, \omega, \Omega, {\lambda _p}} \right){\rm d}{\lambda _p}},
\end{equation}
which leads to the resonant disturbing function. After averaging the disturbing function over $q_0$ orbital periods of the planet, both the angle variables $\lambda_p$ and $\Omega$ disappear from the averaged expression, so that the number of degree of freedom is reduced by two. As a result, the resonant model determined by ${{\cal R}^*} \left( a,e,I, {\sigma , \omega} \right)$ is of two degrees of freedom. However, there is only one integral of motion (Hamiltonian of system), showing that the resulting two degree-of-freedom dynamical model determined by ${{\cal R}^*} \left( a,e,I, {\sigma , \omega} \right)$ is not integrable.

Replacing equation (\ref{Eq27}) in equation (\ref{Eq29}), we can easily obtain the resonant disturbing functions associated with the interior ($p_0 > q_0$ and $\alpha < 1$), co-orbital ($p_0 = q_0$ and $\alpha \approx 1$) and exterior ($p_0 < q_0$ and $\alpha > 1$) mean motion resonances for minor bodies with arbitrary inclinations. According to the number of $p_0$, there are two cases for the resonant disturbing function: $p_0 = 1$ and $p_0 \ne 1$. When $p_0 = 1$, the resonant disturbing function for describing the 1:$q_0$ resonances becomes
\begin{equation}\label{Eq30}
\begin{aligned}
{{\cal R}^{{*}}} & = \frac{{{\cal G}{m_p}}}{{{a_p}}}\sum\limits_{n \ge 0}^N {\sum\limits_{k \ge 0}^{{k_{\max }}} {\sum\limits_{q \ge 0}^k {\sum\limits_{l \ge 0}^q {\sum\limits_{m \ge 0}^n {\sum\limits_{t \ge 0}^l {\sum\limits_{{t_1} \ge 0}^t {\sum\limits_{{t_2} \ge 0}^{l - t} {{\kappa _0} \times X_{{q_0}\left( {2t - l - 2{t_1} + 2{t_2}} \right)}^{m,\left( {l - 2{t_1} - 2{t_2}} \right)}\left( e \right)} } } } } } } } \frac{{{\alpha ^n}}}{{n!}}\\
& \times \left[ {{{\rm D}_n}{f_q}(\alpha )} \right]\frac{{{{\sin }^{2l}}\left( {\frac{I}{2}} \right)}}{{{{\tan }^{2t}}\left( {\frac{I}{2}} \right)}}\cos \left\{ \begin{array}{l}
\left( {2t - l - 2{t_1} + 2{t_2}} \right)\sigma \\
 + \left[ {\left( {l - 2{t_1} - 2{t_2}} \right) - {q_0}\left( {2t - l - 2{t_1} + 2{t_2}} \right)} \right]\omega
\end{array} \right\}\\
& - \frac{{{\cal G}{m_p}}}{{{a_p}}}\left( {\frac{a}{{{a_p}}}} \right)\left\{ \begin{array}{l}
X_{{q_0}}^{1,1}\left( e \right){\cos ^2}({I \mathord{\left/
 {\vphantom {I 2}} \right.
 \kern-\nulldelimiterspace} 2})\cos \left[ {\sigma  + \left( {1 - {q_0}} \right)\omega } \right]\\
 + X_{ - {q_0}}^{1,1}\left( e \right){\sin ^2}({I \mathord{\left/
 {\vphantom {I 2}} \right.
 \kern-\nulldelimiterspace} 2})\cos \left[ {\sigma  - \left( {1 + {q_0}} \right)\omega } \right]
\end{array} \right\}
\end{aligned}
\end{equation}
where $\sigma  = {q_0}\lambda  - {\lambda _p} + \left( {1 - {q_0}} \right)\Omega$ and, when $p_0 \ne 1$ (in this case, the indirect part has no contribution to the resonant disturbing function), the resonant disturbing function can be organized as
\begin{equation}\label{Eq31}
\begin{aligned}
{{\cal R}^*} &= \frac{{{\cal G}{m_p}}}{{{a_p}}}\sum\limits_{n \ge 0}^N {\sum\limits_{k \ge 0}^{{k_{\max }}} {\sum\limits_{q \ge 0}^k {\sum\limits_{l \ge 0}^q {\sum\limits_{m \ge 0}^n {\sum\limits_{t \ge 0}^l {\sum\limits_{{t_1} \ge 0}^t {\sum\limits_{\scriptstyle{t_2} \ge 0\atop
\scriptstyle\bmod \left[ {\left( {2t - l - 2{t_1} + 2{t_2}} \right),{p_0}} \right] = 0}^{l - t} {{\kappa _0}} } } } } } } } \\
& \times \frac{{{\alpha ^n}}}{{n!}}\left[ {{{\rm D}_n}{f_q}(\alpha )} \right]X_{\frac{{{q_0}}}{{{p_0}}}\left( {2t - l - 2{t_1} + 2{t_2}} \right)}^{m,\left( {l - 2{t_1} - 2{t_2}} \right)}\left( e \right)\frac{{{{\sin }^{2l}}\left( {\frac{I}{2}} \right)}}{{{{\tan }^{2t}}\left( {\frac{I}{2}} \right)}}\\
&\times \cos \left\{ {\frac{1}{{{p_0}}}\left( {2t - l - 2{t_1} + 2{t_2}} \right)\sigma  + \left[ {\left( {l - 2{t_1} - 2{t_2}} \right) - \frac{{{q_0}}}{{{p_0}}}\left( {2t - l - 2{t_1} + 2{t_2}} \right)} \right]\omega } \right\}
\end{aligned}
\end{equation}
where $\bmod (A,B) = 0$ means that $A$ is divisible by $B$. It is noted that the expansion of resonant disturbing function given by equation (\ref{Eq30}) or (\ref{Eq31}) can be easily realized in computer language. By the way, direct numerical integration of equation (\ref{Eq29}) can produce numerical results of the resonant disturbing function, as adopted by \citet{gallardo2006Atlas, gallardo2019strength, gallardo2020three} in their studies. In the direct numerical integration, the original form of the disturbing function given by equation (\ref{Eq1}) can be used. In the current work, we will take the numerical integration results as references in order to validate our analytical expansion.

Obviously, the resonant disturbing function ${\cal R}^*$ given by equation (\ref{Eq30}) or (\ref{Eq31}) is a function of $a$, $e$, $I$, $\sigma$ and $\omega$. For the sake of brevity, the resonant disturbing function can be denoted by an elegant form,
\begin{equation}\label{Eq32}
{{\cal R}^*\left( a,e,I, {\sigma , \omega} \right)} = \sum\limits_{k = 0}^\infty  {\sum\limits_{{k_1}} {{\cal C}_{k,{k_1}}^{\cal R} (a,e,I) \cos \left( {k\sigma  + {k_1}\omega }\right)}},
\end{equation}
where ${k_1} \in \mathbb{Z}$ and it has the same parity of $k(p_0 + q_0)$. The coefficients ${{\cal C}_{{k},{k_1}}^{\cal R}} (a,e,I)$ are derived from the resonant disturbing function given by equation (\ref{Eq30}) or (\ref{Eq31}).

\begin{figure*}
\centering
\includegraphics[width=0.48\textwidth]{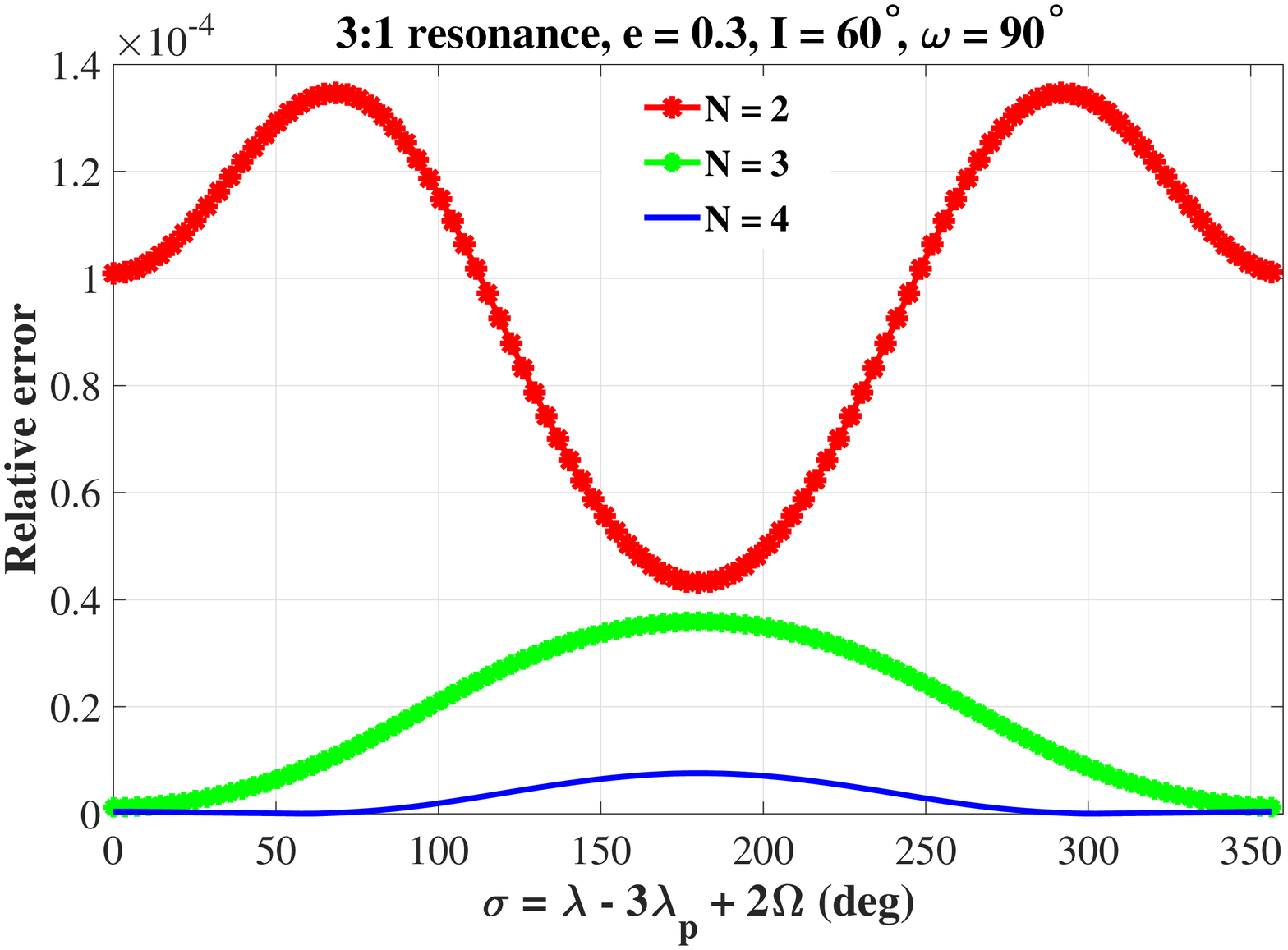}
\includegraphics[width=0.48\textwidth]{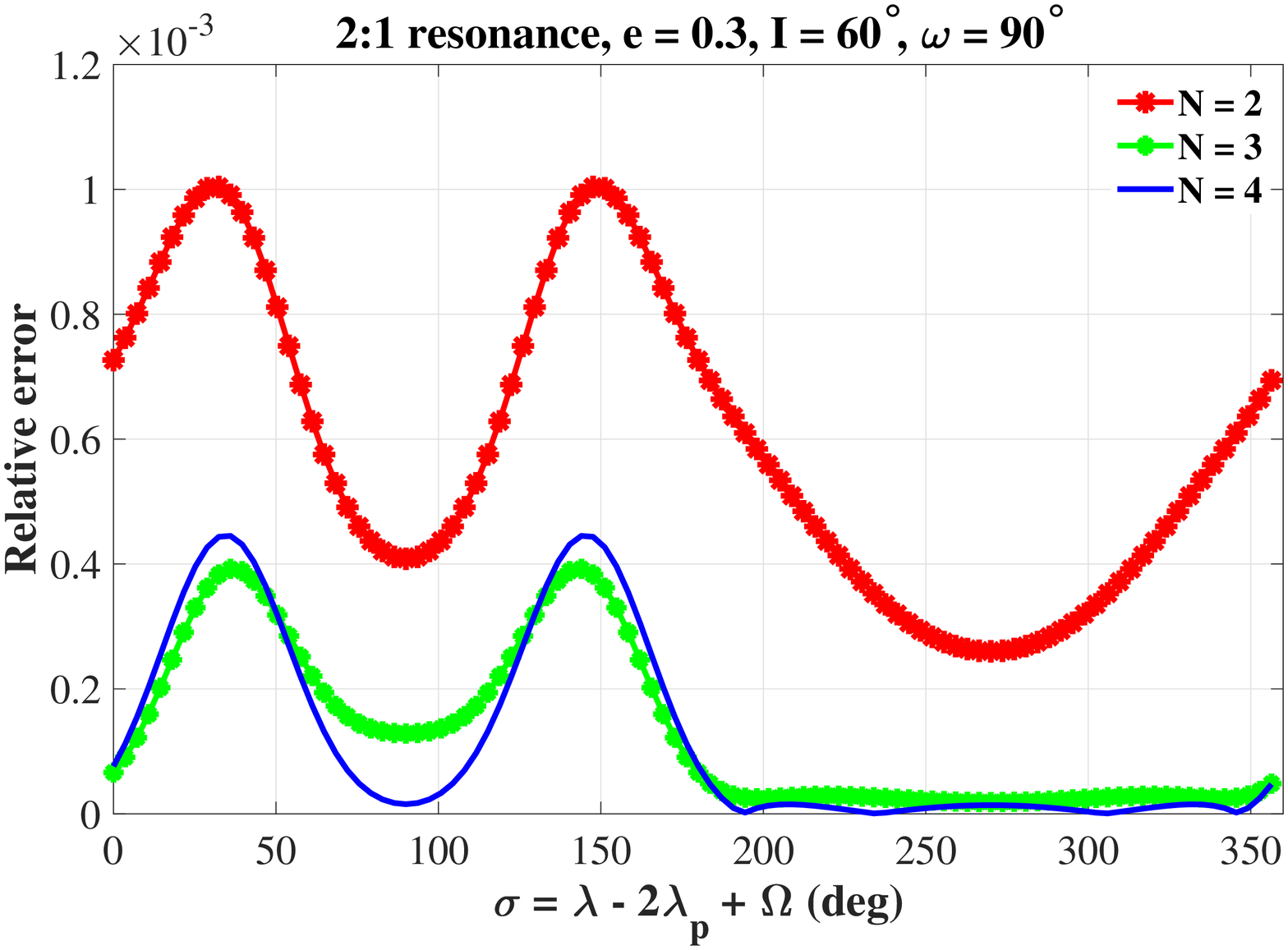}
\caption{Relative errors of the new expansion truncated at different orders in eccentricity for approximating the resonant disturbing function. Relative errors are measured by taking the numerical results produced by direct numerical integration as the accurate value of resonant disturbing function. The new expansion is truncated at orders $N = 2,3,4$ in eccentricity and, in each case, the expansion order in $\delta x$ is fixed at $k_{\max} = 30$. The left panel is for Jupiter's inner 3:1 resonance and the right one is for Jupiter's inner 2:1 resonance. In simulations, the eccentricity is fixed at $e=0.3$, the inclination at $I = 60^{\circ}$, the argument of pericenter at $\omega = 90^{\circ}$ and the semimajor axis is taken as the value at the resonant center (simulations with other parameters can be performed in a similar manner).}
\label{Fig2}
\end{figure*}

To validate the new expansion shown by Eqs. (\ref{Eq30}) and (\ref{Eq31}), it needs to discuss the precision of the analytical developments truncated at different orders. To this end, the new expansion is truncated at orders $N=2,3,4$ in eccentricity and $k_{\max} = 30$ in $\delta x$ and they are used to approximate the resonant disturbing function associated with Jupiter's interior 3:1 and 2:1 resonances. In simulations, the truncated order of $\delta x$ is assumed at $k_{\max} = 30$, the eccentricity is fixed at $e=0.3$, the inclination is taken as $I=60^{\circ}$ and the argument of pericenter is taken as $\omega = 90^{\circ}$ (these parameters are chosen artificially). The precision of $f (= {\cal R}^*)$ is measured by the relative error, defined by
\begin{equation*}
{\Delta f} = \frac{{\left| {{f_{\rm approximate}} - {f_{\rm accurate}}} \right|}}{{\left| {{f_{\rm accurate}}} \right|}}
\end{equation*}
where $f_{\rm approximate}$ and $f_{\rm accurate}$ represents the approximate and accurate values of $f$, respectively. $f_{\rm approximate}$ is computed by analytical expansion and $f_{\rm accurate}$ is produced by direct numerical integration. The relative errors of ${\cal R}^*$ are reported in Figure \ref{Fig2}, which shows that, in general, the series expansion truncated at a higher order $N$ in eccentricity has a lower relative error (or higher precision) than the lower-order expansions. This is expected by us.

\begin{figure*}
\centering
\includegraphics[width=0.45\textwidth]{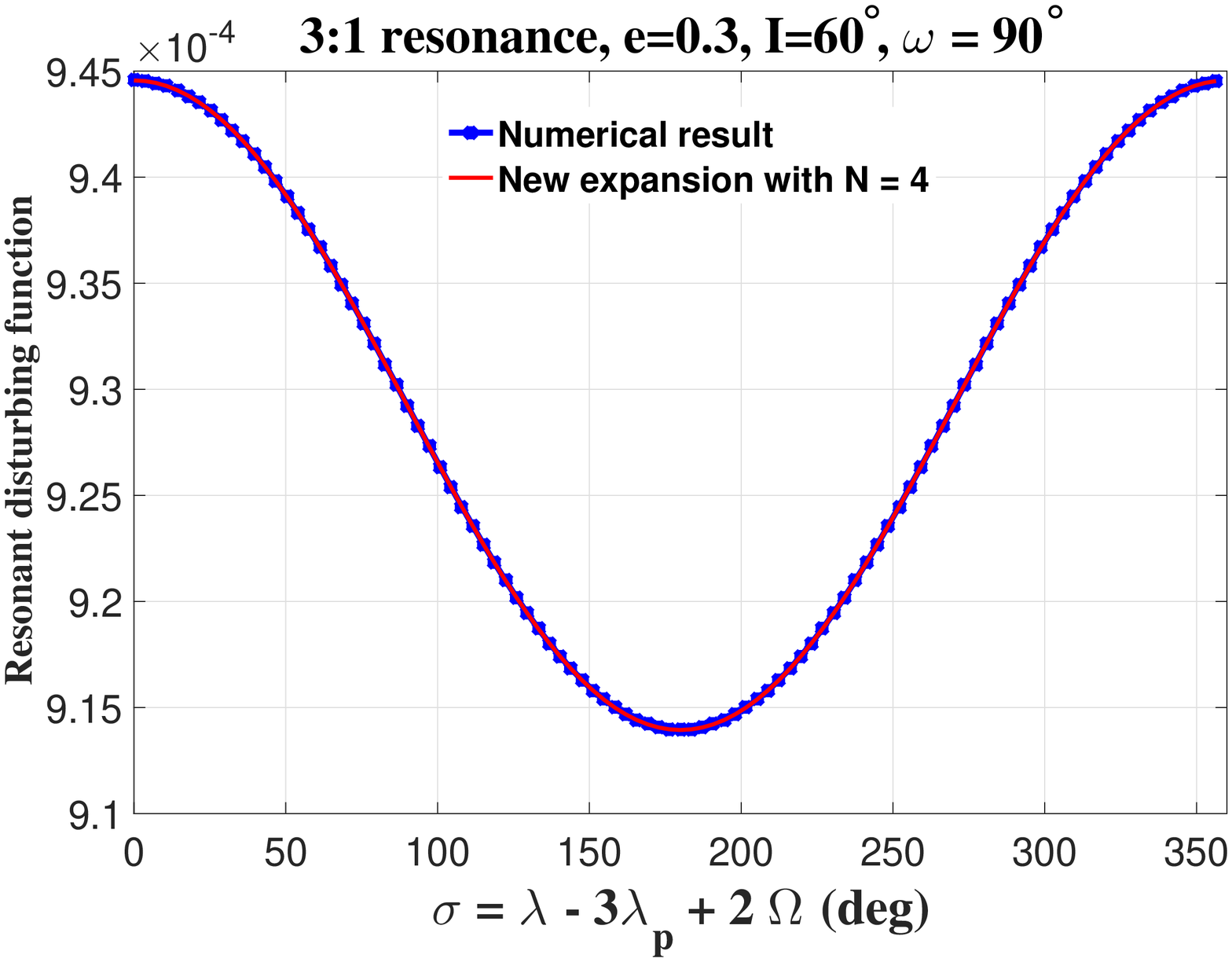}
\includegraphics[width=0.45\textwidth]{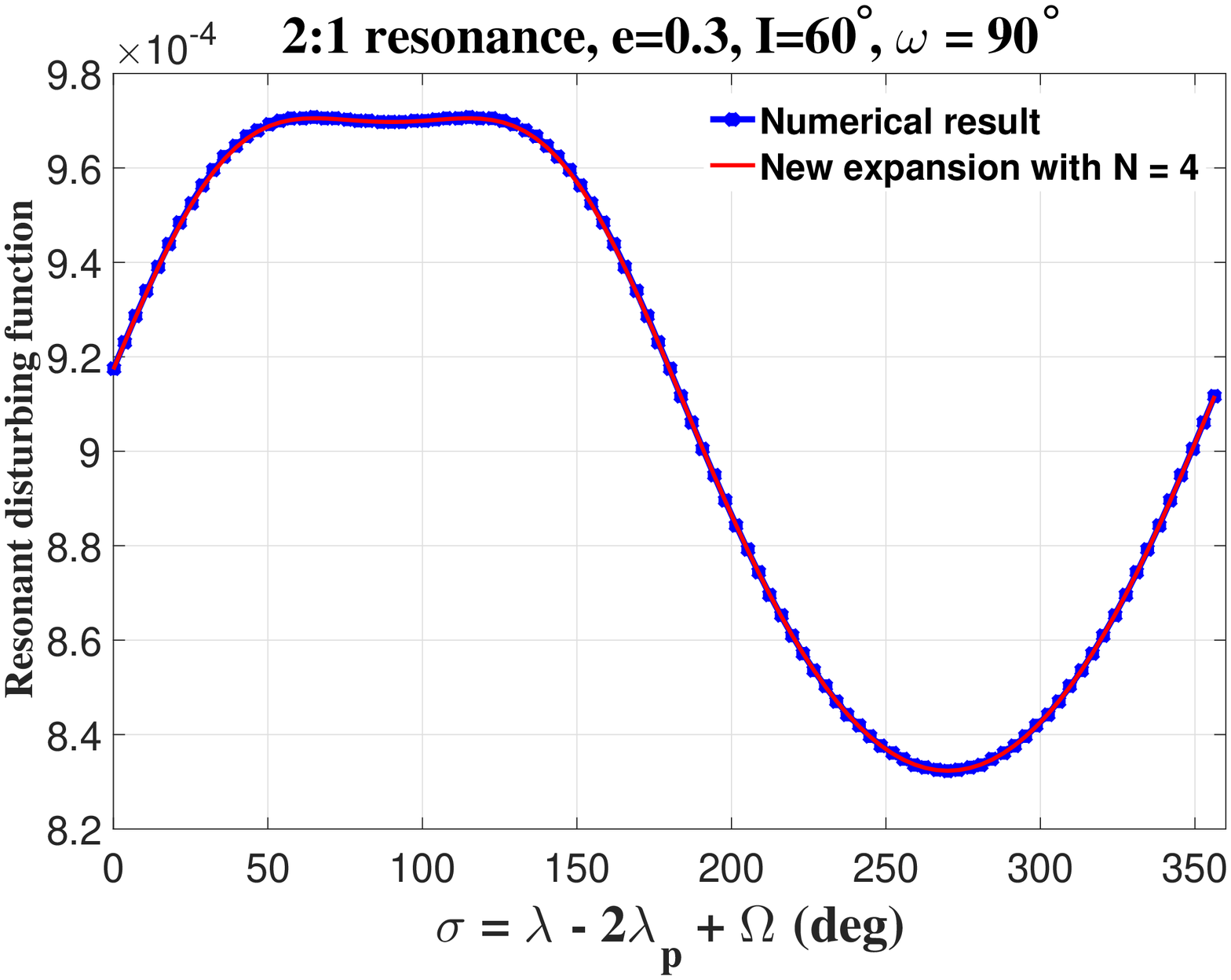}\\
\includegraphics[width=0.45\textwidth]{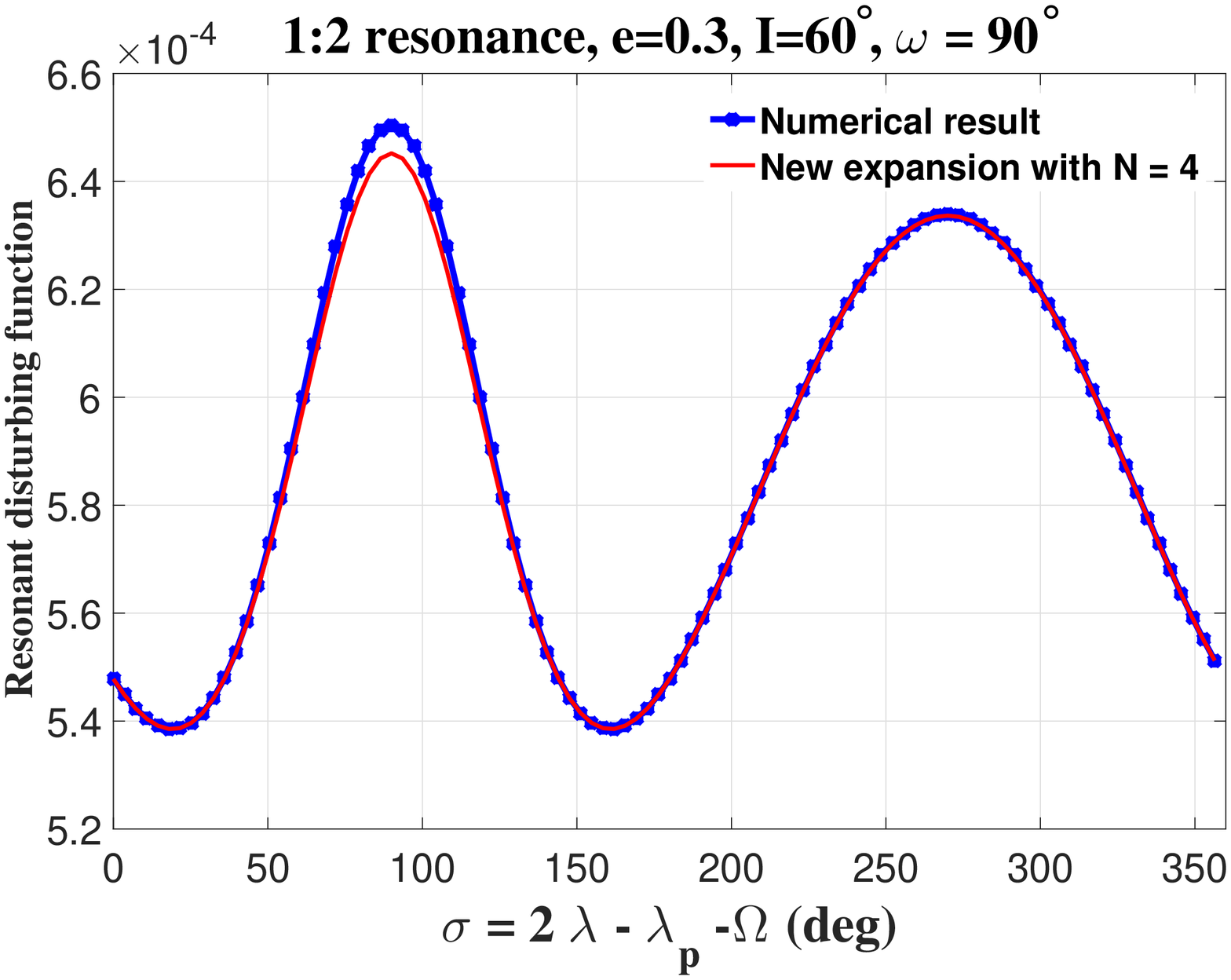}
\includegraphics[width=0.45\textwidth]{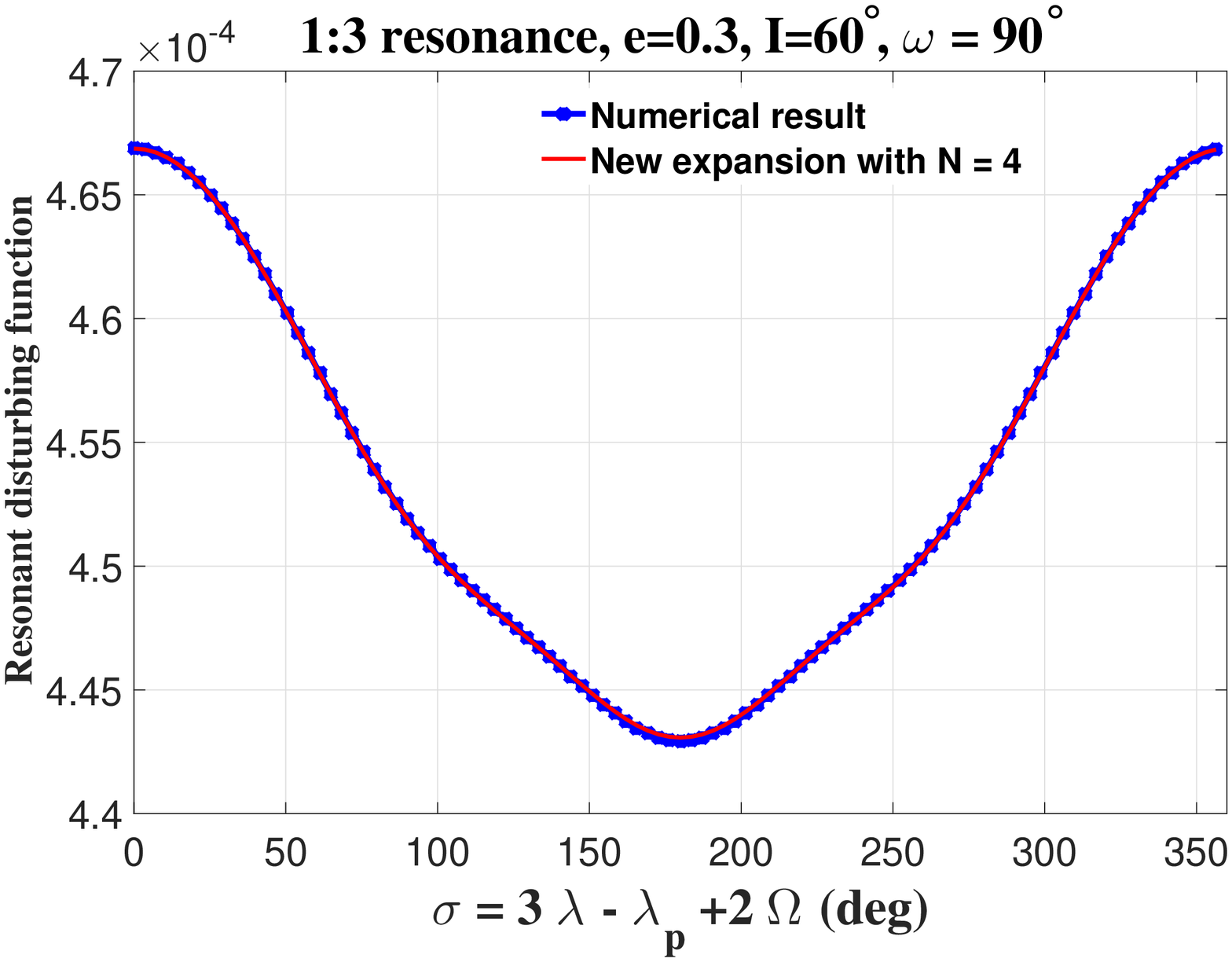}
\caption{Comparisons of the numerical results obtained by direct numerical integration and the analytical results obtained by the new expansion for computing the resonant disturbing function associated with the Jupiter's inner 3:1 and 2:1 resonances (\emph{upper panels}) and exterior 1:2 and 1:3 resonances (\emph{bottom panels}). The numerical results are taken as references for measuring the accuracy of our analytical developments. The new expansion is truncated at the fourth order in eccentricity (i.e. $N=4$) and at order $k_{\max} = 30$ in $\delta x$. In simulations, the eccentricity is assumed at $e=0.3$, the inclination is fixed at $I = 60^{\circ}$, the argument of pericenter is set as $\omega = 90^{\circ}$ and the semimajor axis is taken as the value of the resonant center.}
\label{Fig3}
\end{figure*}

Then, we compare the analytical results obtained by the new expansion truncated at order $N=4$ in eccentricity and $k_{\max}=30$ in $\delta x$ with the numerical results obtained through direct numerical integration. Concerning the resonant disturbing function associated with Jupiter's inner 3:1 and 2:1 resonances as well as Jupiter's exterior 1:2 and 1:3 resonances, in Figure \ref{Fig3} we make a direct comparison between the numerical results with the analytical results obtained by the new expansion. Please refer to the caption of Figure \ref{Fig3} for the detailed setting of other parameters. From Figure \ref{Fig3}, we can observe: (a) our new expansion could catch the peak and valley positions of resonant disturbing function accurately in comparison to the numerical results (notice that the valley and peak positions correspond to the stable and unstable equilibrium points of the resonant model and the difference of resonant disturbing function evaluated at the peak and valley positions, $\Delta {\cal R}^*$, showing the resonant strength, is positively correlated to the resonant width, as discussed in \citet{gallardo2020three}); (b) from a quantitative viewpoint, the analytical results obtained by the new expansion are in coincident with the numerical results in the entire interval of $\sigma$.

\begin{figure*}
\centering
\includegraphics[width=0.45\textwidth]{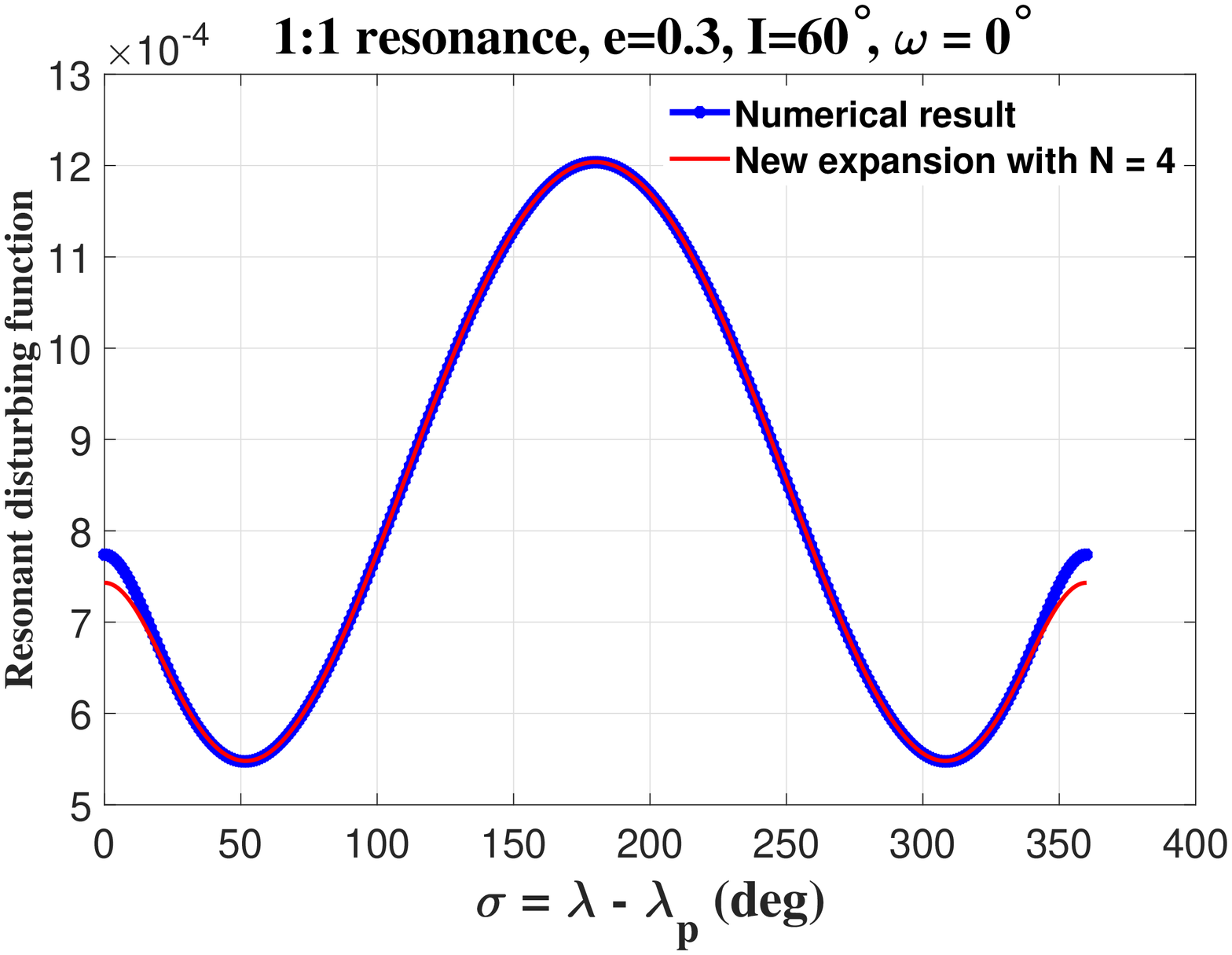}
\includegraphics[width=0.45\textwidth]{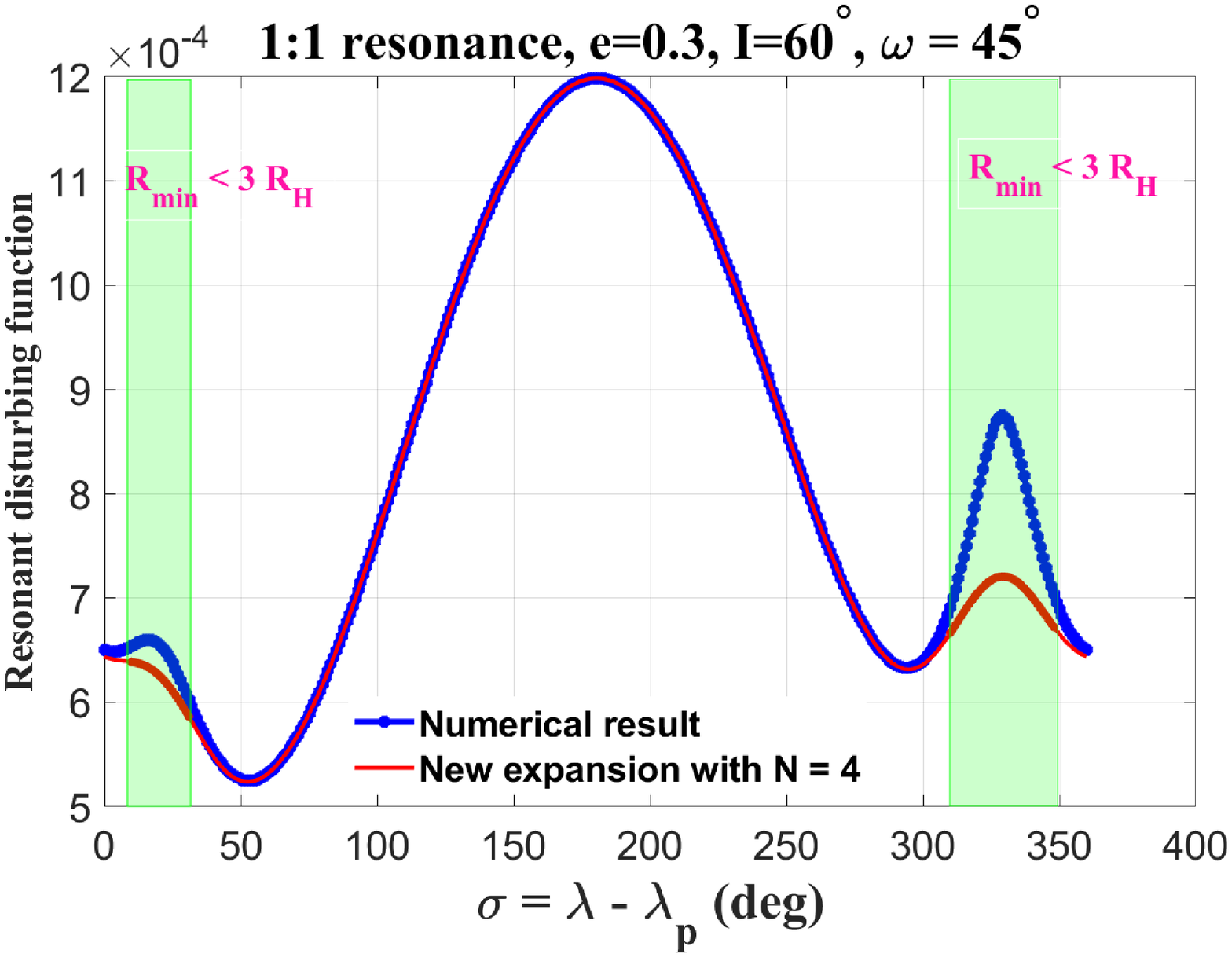}\\
\includegraphics[width=0.45\textwidth]{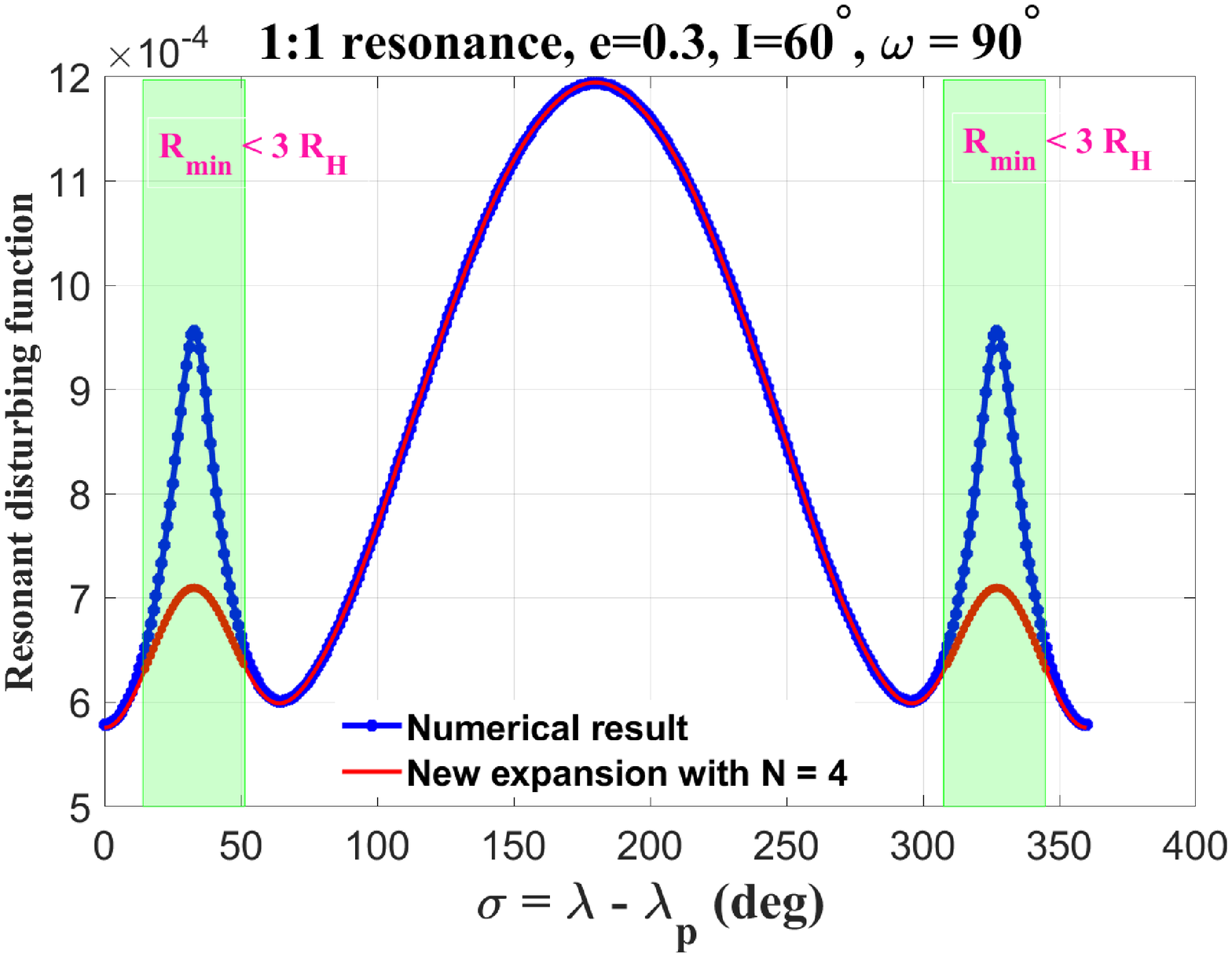}
\includegraphics[width=0.45\textwidth]{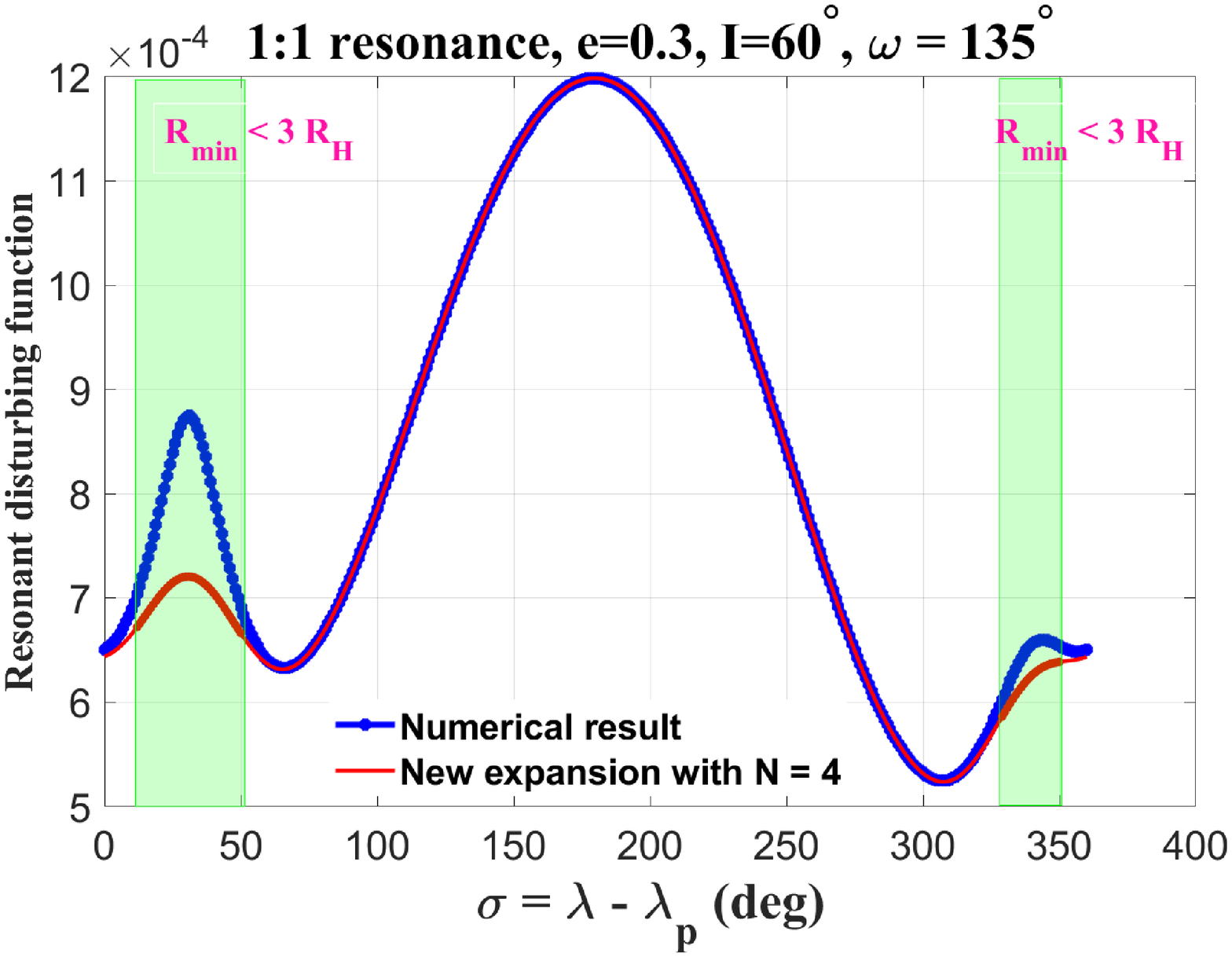}
\caption{Resonant disturbing function for co-orbital motion in the Sun--Jupiter system under different settings of $\omega$ ($\omega$ is taken as $0^{\circ}$, $45^{\circ}$, $90^{\circ}$ and $135^{\circ}$ from the top left panel to the bottom right panel), computed by means of analytical and numerical approaches. The numerical results are shown in blue dotted lines and the analytical results obtained by the new expansion truncated at order $N=4$ in eccentricity and $k_{\max} = 30$ in $\delta x$ are shown in red lines. In simulations, the eccentricity is assumed at $e=0.3$, the inclination at $I=60^{\circ}$ and the semimajor axis is taken as the value at the resonant center. The regions in shadow represent the minimum distance between the orbits of the asteroid and planet is smaller than $3R_{H}$ where $R_H$ is the Hill radius of the planet. It is observed that outside the shadow region the analytical results agree well with the numerical results.}
\label{Fig4}
\end{figure*}

At last, we apply the new expansion truncated at orders $N=4$ and $k_{\max}=30$ to the co-orbital resonances with Jupiter ($\alpha = 1$), as shown in Figure \ref{Fig4}. The numerical results are also provided for the convenience of comparison. In Figure \ref{Fig4}, the domain of $\sigma$ is marked in shadow if the minimum distance between the asteroid and planet is smaller than three times of the Jupiter's Hill radius $R_{H}$. From Figure \ref{Fig4}, it is observed that (a) in the shadow area the analytical developments underestimate the resonant disturbing function, this is because the analytical results are always smaller than the numerical results in the shadow region, and (b) outside the shadow area the analytical results produced by the new expansion are in good agreement with the numerical results. It is known that, when the asteroid is located in the shadow area, the perturbation theory usually fails to work due to the strong perturbation coming from the planet. Thus, when we are using perturbation theory to study the long-term dynamical behaviors we need to ensure that the asteroids considered are located outside the shadow area.

In summary, the comparisons made in Figures \ref{Fig3} and \ref{Fig4} strongly support the conclusion that the new expansion is valid for minor bodies located inside the interior, co-orbital and exterior resonances (for better accuracy it is required that the minimum distance between the asteroid and planet is greater than $3R_{H}$). In the coming section, we will use our analytical developments to formulate Hamiltonian model for mean motion resonances.

\section{Hamiltonian model of mean motion resonances}
\label{Sect5}

For convenience of studying the Hamiltonian dynamics, we adopt a set of modified Delaunay variables, given by \citep{morbidelli2002modern}
\begin{equation}\label{33}
\begin{aligned}
&\Lambda  = \sqrt {\mu a} ,\quad\quad\quad\quad\quad\quad\quad\quad\quad\quad \lambda  = M + \varpi,\\
&P = \sqrt {\mu a} \left( {1 - \sqrt {1 - {e^2}} } \right),\quad\quad\quad p =  - \varpi,\\
&Q = \sqrt {\mu a\left( {1 - {e^2}} \right)} \left( {1 - \cos i} \right),\quad q = - \Omega,\\
&{\Lambda _p},\quad\quad\quad\quad\quad\quad\quad\quad\quad\quad\quad\quad\quad {\lambda _p} = M_p + {\varpi}_p,
\end{aligned}
\end{equation}
where $\mu = \sqrt{{\cal G} m_0}$ is the gravitational parameter of the central mass, $\Lambda _p$ is the conjugate momentum of the mean longitude of planet $\lambda _p$. With such a set of canonical variables, the Hamiltonian of planetary system can be written as \citep{morbidelli2002modern}
\begin{equation}\label{Eq34}
{\cal H} =  - \frac{{{\mu ^2}}}{{2{\Lambda ^2}}} + {n_p}{\Lambda _p} - {\cal R}\left( {\Lambda ,P,Q,\lambda ,p,q,{\lambda _p}} \right),
\end{equation}
where $n_p$ is the mean motion of the planet and the disturbing function ${\cal R}\left( {\Lambda ,P,Q,\lambda ,p,q,{\lambda _p}} \right)$ is equal to ${\cal R}\left( a,e,i,\sigma,\omega \right)$, given by equation (\ref{Eq27}). In order to describe the dynamics associated with the $p_0$:$q_0$ resonance, we introduce a linear canonical transformation, given by
\begin{equation}\label{Eq35}
\begin{aligned}
{\Sigma _{{1}}} &= \frac{1}{q_0}\Lambda ,\quad {\sigma _{{1}}} = q_0 \lambda  - {p_0}{\lambda _p} - \left( {{p_0} - {q_0}} \right)q = \sigma,\\
{\Sigma _2} &=  - P,\quad {\sigma _2} = q - p = \omega,\\
{\Sigma _3} &=  - P - Q - \frac{{{p_0} - {q_0}}}{q_0}\Lambda ,\quad {\sigma _3} =  - q = \Omega,\\
{\Sigma _4} &= {\Lambda _p} + \frac{{{p_0}}}{q_0}\Lambda ,\quad {\sigma _4} = {\lambda _p},
\end{aligned}
\end{equation}
which can be realized by the following generating function:
\begin{equation*}
{\cal S} = \left[ {{q_0}\lambda  - \left( {{p_0} - {q_0}} \right)q - {p_0}{\lambda _p}} \right]{\Sigma _1} + \left( {q - p} \right){\Sigma _2} - q{\Sigma _3} + {\lambda _p}{\Sigma _4}.
\end{equation*}
As a result, the Hamiltonian can be written as follows:
\begin{equation}\label{Eq36}
{\cal H} =  - \frac{{{\mu ^2}}}{{2{{\left( {q_0 {\Sigma _1}} \right)}^2}}} + {n_p}\left( {{\Sigma _4} - {p_0}{\Sigma _1}} \right) - {\cal R}\left( {{\Sigma _1},{\Sigma _2},{\Sigma _3},{\Sigma _4},{\sigma _1},{\sigma _2},{\sigma _3},{\sigma _4}} \right).
\end{equation}
In the new set of canonical variables, $\sigma_1 = \sigma$ corresponds to the characteristic angle of the $p_0$:$q_0$ resonance, and $\sigma_2 = \omega$ corresponds to the argument of pericenter. It should be mentioned that the canonical transformation given by equation (\ref{Eq35}) is valid for all the interior ($p_0 > q_0$), co-orbital ($p_0 = q_0$) and exterior ($p_0 < q_0$) resonances.

In particular, when the minor body is located inside a mean motion resonance, the disturbing function is replaced by the resonant disturbing function given by equation (\ref{Eq32}). Consequently, the averaged resonant Hamiltonian becomes
\begin{equation}\label{Eq37}
{{\cal H}^*} =  - \frac{{{\mu ^2}}}{{2{{\left( {q_0 {\Sigma _1}} \right)}^2}}} - {p_0}{n_p}{\Sigma _1} - {{\cal R}^*}\left( {{\Sigma _1},{\Sigma _2},{\Sigma _3},{\Sigma _4},{\sigma _1},{\sigma _2}} \right),
\end{equation}
where the constant term $n_p \Sigma_4$ is eliminated from the Hamiltonian and the canonical equations of motion can be written as
\begin{equation}\label{Eq38}
\begin{aligned}
{\dot {\sigma _1}} = \frac{{\partial {{\cal H}^*}}}{{\partial {\Sigma _1}}},\quad {{\dot \Sigma }_1} =  - \frac{{\partial {{\cal H}^*}}}{{\partial {\sigma _1}}},\\
{\dot {\sigma _2}} = \frac{{\partial {{\cal H}^*}}}{{\partial {\Sigma _2}}},\quad {{\dot \Sigma }_2} =  - \frac{{\partial {{\cal H}^*}}}{{\partial {\sigma _2}}},
\end{aligned}
\end{equation}
which correspond to a dynamical system with two degrees of freedom ($\sigma_1$ and $\sigma_2$ are the associated angular coordinates). In the resonant model specified by equation (\ref{Eq37}), $\sigma_3$ and $\sigma_4$ are cycle coordinates, so that their conjugate momenta $\Sigma_3$ and $\Sigma_4$ become integrals of motion, given by
\begin{equation}\label{Eq39}
{\Sigma _3} =  - P - Q - \frac{{{p_0} - {q_0}}}{q_0}\Lambda  = \sqrt {\mu a} \left[ {\sqrt {1 - {e^2}} \cos I - \frac{{{p_0}}}{q_0}} \right] = {\rm const}
\end{equation}
and
\begin{equation}\label{Eq40}
{\Sigma _4} = {\Lambda _p} + \frac{{{p_0}}}{q_0}\Lambda  = {\Lambda _p} + \frac{{{p_0}}}{q_0}\sqrt {\mu a} = {\rm const}.
\end{equation}
The integral of motion given by equation (\ref{Eq39}) shows that there is coupled oscillations among semimajor axis ratio, eccentricity and inclination of the test particle, meaning that exchange between the Keplerian energy ($K=-\frac{\mu}{2a}$) and the angular momentum along the $z$ direction ($H=\sqrt{\mu a (1-e^2)}\cos{I}$) occurs in the long-term evolution.

In terms of the classical elements, the averaged resonant Hamiltonian given by equation (\ref{Eq37}) can be further expressed as
\begin{equation}\label{Eq41}
\begin{aligned}
{\cal H^*} =&  - \frac{\mu }{{2a}} - {n_p}\frac{{{p_0}}}{q_0}\sqrt {\mu a}  - {{\cal R}^*}\left( {a,e,i,\sigma ,\omega } \right)\\
=& - \frac{\mu }{{2a}} - {n_p}\frac{{{p_0}}}{q_0}\sqrt {\mu a}  - \sum\limits_{k = 0}^\infty  {\sum\limits_{{k_1}} {{\cal C}_{k,{k_1}}^{\cal R} (a,e,I) \cos \left( {k\sigma  + {k_1}\omega }\right)}}
\end{aligned}
\end{equation}
where the resonant disturbing function ${{\cal R}^*}\left( {a,e,i,\sigma ,\omega } \right)$ is given by equation (\ref{Eq32}). It should be noted that the form given by equation (\ref{Eq41}) is analogous to the one given by \citet{gallardo2020three} (see equation 5 in his work). The difference is that the resonant disturbing function in the present work is produced by analytical developments while, in \citet{gallardo2020three}, the resonant disturbing function is obtained by means of direct numerical integration.

If we are only interested in the dynamics during the timescale of mean motion resonances, the argument of pericenter $\omega$ can be assumed as a constant \citep{gallardo2006Atlas, gallardo2019strength, gallardo2020three, lei2019three}, so that the dynamical model reduces to a system with a single degree of freedom, specified by the resonant Hamiltonian
\begin{equation}\label{Eq42}
{{\cal H}^*} =  - \frac{\mu }{{2a}} - {n_p}\frac{{{p_0}}}{{{q_0}}}\sqrt {\mu a}
- \sum\limits_{k = 0}^\infty  {\left[ {{{\cal C}_k}\left( {a,e,I,\omega } \right)\cos k\sigma  + {{\cal S}_k}\left( {a,e,I,\omega } \right)\sin k\sigma } \right]},
\end{equation}
where the coefficients are given by
\begin{equation*}
\begin{aligned}
& {{\cal C}_k}\left( {a,e,I,\omega } \right) = \sum\limits_{{k_1}} {{\cal C}_{k,{k_1}}^{\cal R}\left( {a,e,I} \right)\cos \left( {{k_1}\omega } \right)},\\
& {{\cal S}_k}\left( {a,e,I,\omega } \right) =  - \sum\limits_{{k_1}} {{\cal C}_{k,{k_1}}^{\cal R}\left( {a,e,I} \right)\sin \left( {{k_1}\omega } \right)}.
\end{aligned}
\end{equation*}
In the single degree-of-freedom resonant model, the equations of motion are
\begin{equation}\label{Eq43}
{\dot \sigma _1} = \frac{{\partial {{\cal H}^*}}}{{\partial {\Sigma _1}}},\quad {\dot \Sigma _1} =  - \frac{{\partial {{\cal H}^*}}}{{\partial {\sigma _1}}},
\end{equation}
which can be written in an equivalent form:
\begin{equation}\label{Eq44}
\begin{aligned}
\frac{{{\rm d}\sigma }}{{{\rm d}t}} =& 2{q_0}\sqrt {{a \mathord{\left/
 {\vphantom {a \mu }} \right.
 \kern-\nulldelimiterspace} \mu }} \frac{{\partial {{\cal H}^*}}}{{\partial a}} = \frac{{{q_0}\sqrt {\mu a} }}{{{a^2}}} - {n_p}{p_0} - 2{q_0}\sqrt {{a \mathord{\left/
 {\vphantom {a \mu }} \right.
 \kern-\nulldelimiterspace} \mu }} \frac{{\partial {{\cal R}^*}}}{{\partial a}},\\
\frac{{{\rm d}a}}{{{\rm d}t}} =&  - 2{q_0}\sqrt {{a \mathord{\left/
 {\vphantom {a \mu }} \right.
 \kern-\nulldelimiterspace} \mu }} \frac{{\partial {{\cal H}^*}}}{{\partial \sigma }} = 2{q_0}\sqrt {{a \mathord{\left/
 {\vphantom {a \mu }} \right.
 \kern-\nulldelimiterspace} \mu }} \frac{{\partial {{\cal R}^*}}}{{\partial \sigma }}.
\end{aligned}
\end{equation}
The second equation of equation (\ref{Eq44}) indicates that the dependence of the resonant disturbing function ${\cal R}^*$ (or Hamiltonian ${\cal H}^*$) on $\sigma$ determines the dynamical behavior of semimajor axis, as pointed out in \citet{gallardo2020three}. Under the resonant model, the equilibrium points can be obtained by solving the frozen equation:
\begin{equation}\label{Eq45}
\frac{{{\rm d}\sigma }}{{{\rm d}t}} = \frac{{{\rm d}a}}{{{\rm d}t}} = 0.
\end{equation}
The first equation $\frac{{{\rm d}\sigma }}{{{\rm d}t}} \approx q_0 {\dot \lambda} - p_0 {\dot \lambda_p} =0$ shows that the equilibrium points are approximately located at
\begin{equation*}
a = {a_0} = {\left( {\frac{{q_0^2}}{{p_0^2}}\frac{\mu }{{n_p^2}}} \right)^{\frac{1}{3}}}.
\end{equation*}
The stability condition of equilibrium points requires that the second derivative of ${\cal H}^*$ with respect to $\sigma$ evaluated at the considered point is smaller than zero, expressed by
\begin{equation}\label{Eq46}
\frac{{{\partial ^2}{{\cal H}^*}}}{{\partial {\sigma ^2}}} = \sum\limits_{k = 1}^\infty  {{k^2}\left[ {{{\cal C}_k}\left( {a,e,I,\omega } \right)\cos k\sigma  + {{\cal S}_k}\left( {a,e,I,\omega } \right)\sin k\sigma } \right]} \;{{ < }}\;{{0}},
\end{equation}
meaning that the stable equilibrium points correspond to the local maxima of the resonant Hamiltonian (or local minima of the resonant disturbing function). Usually, the stable equilibrium points, denoted by $(a_0, \sigma_s)$, correspond to the resonant centers and the unstable ones, denoted by $(a_0, \sigma_u)$, correspond to saddle points. The level curves of the resonant Hamiltonian passing through the saddles points provide the dynamical separatrices, which divide the phase space into libration and circulation regions. According to the discussion in \citet{gallardo2020three}, the libration region is specified by the so-called resonance's half width $\Delta a = a_{\rm sep} - a_0$ ($a_{\rm sep}$ is defined by the separatrix), calculated by
\begin{equation}\label{Eq47}
\begin{aligned}
\Delta a &= a_{\rm sep} - a_0\\
&= \frac{{2\sqrt 6 }}{{3n}}{\left[ {{{\cal H}^*}\left( {{a_0},{e_0},{I_0},{\sigma _s},{\omega _0}} \right) - {{\cal H}^*}\left( {{a_0},{e_0},{I_0},{\sigma _u},{\omega _0}} \right)} \right]^{\frac{1}{2}}}\\
&= \frac{{2\sqrt 6 }}{{3n}}{\left[ {{{\cal R}^*}\left( {{a_0},{e_0},{I_0},{\sigma _u},{\omega _0}} \right) - {{\cal R}^*}\left( {{a_0},{e_0},{I_0},{\sigma _s},{\omega _0}} \right)} \right]^{\frac{1}{2}}}\\
&= \frac{{2\sqrt 6 }}{{3n}}{\left\{ {\sum\limits_{k = 0}^\infty  {\left[ {{{\cal C}_k}\left( {\cos k{\sigma _u} - \cos k{\sigma _s}} \right) + {{\cal S}_k}\left( {\sin k{\sigma _u} - \sin k{\sigma _s}} \right)} \right]} } \right\}^{\frac{1}{2}}}
\end{aligned}
\end{equation}
which is in agreement with the expression derived from Hamiltonian approach by \citet{gallardo2020three} (see equation 18 in his work) and in agreement with the expression of resonance's half width derived from the multi-harmonics pendulum model by \citet{lei2019three} (please refer to equation 26 in his work). Remind that the resonant disturbing function in \citet{gallardo2020three} is produced by means of direct numerical integration and, in the present work, the resonant disturbing function (or resonant Hamiltonian) can be obtained by means of both the analytical expansions and numerical integration. In Section \ref{Sect6}, we will use equation (\ref{Eq47}) to produce the analytical and numerical resonant widths in terms of $\Delta a$ and compare them in order to validate our analytical developments.

\section{Applications}
\label{Sect6}

In this section, we will apply the analytical developments performed in Sections \ref{Sect4} and \ref{Sect5} to Jupiter's inner and co-orbital resonances and to Neptune's exterior mean motion resonances. In particular, we will make direct comparisons between analytical and numerical results.

\subsection{Jupiter's inner and co-orbital resonances}
\label{Sect6-1}

The resonant Hamiltonian given by equation (\ref{Eq42}) shows that the resonant model is specified by a pair of pseudo conjugate variables $(a,\sigma)$ (notice that there is a one to one correspondence between the semimajor axis and the conjugate variable of $\sigma$). Thus, it is possible to understand the global dynamics of mean motion resonances by plotting the level curves of the resonant Hamiltonian in the space spanned by $a$ and $\sigma$, which correspond to the so-called (pseudo) phase-space structures.

\begin{figure*}
\centering
\includegraphics[width=0.48\textwidth]{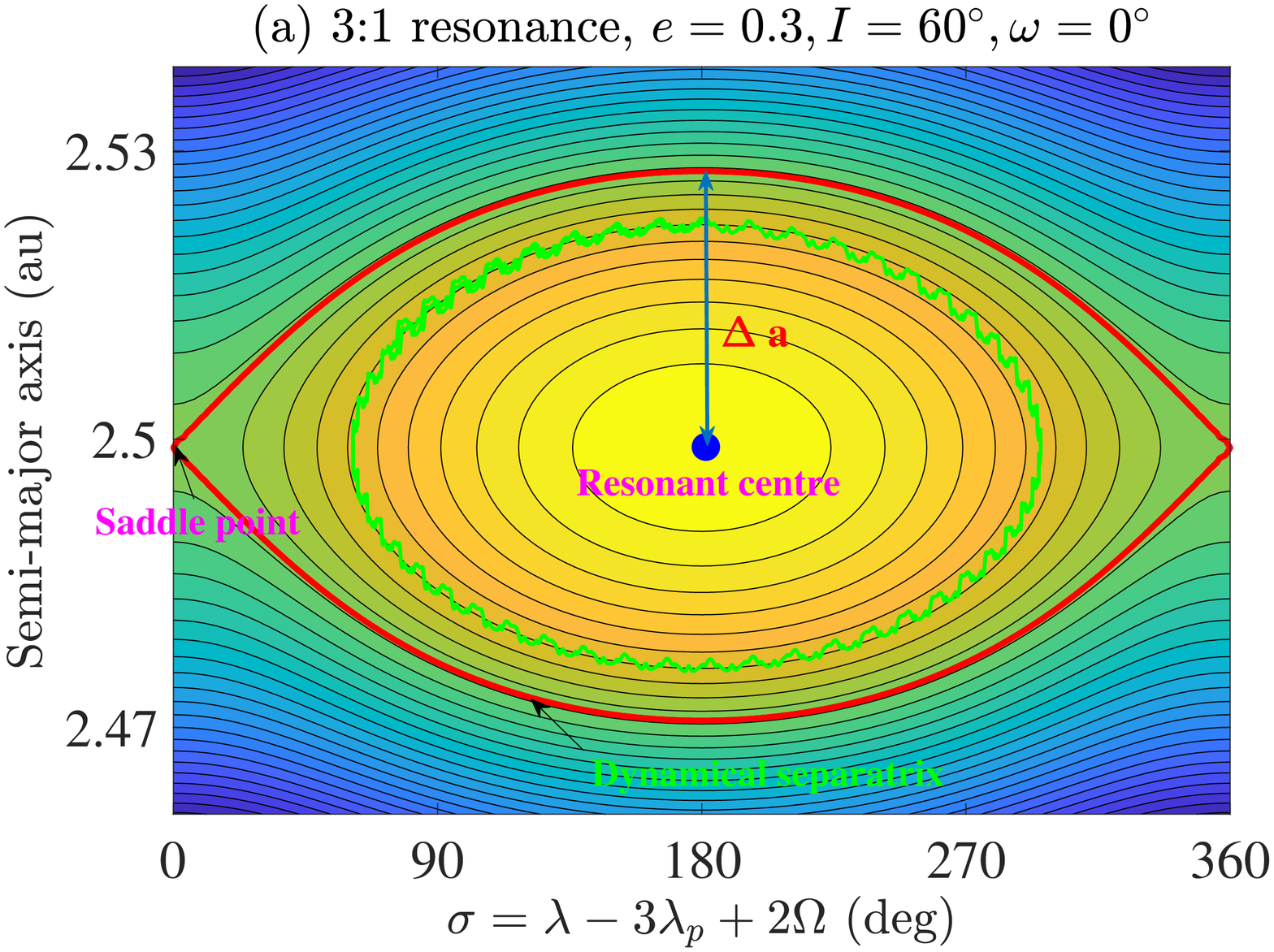}
\includegraphics[width=0.48\textwidth]{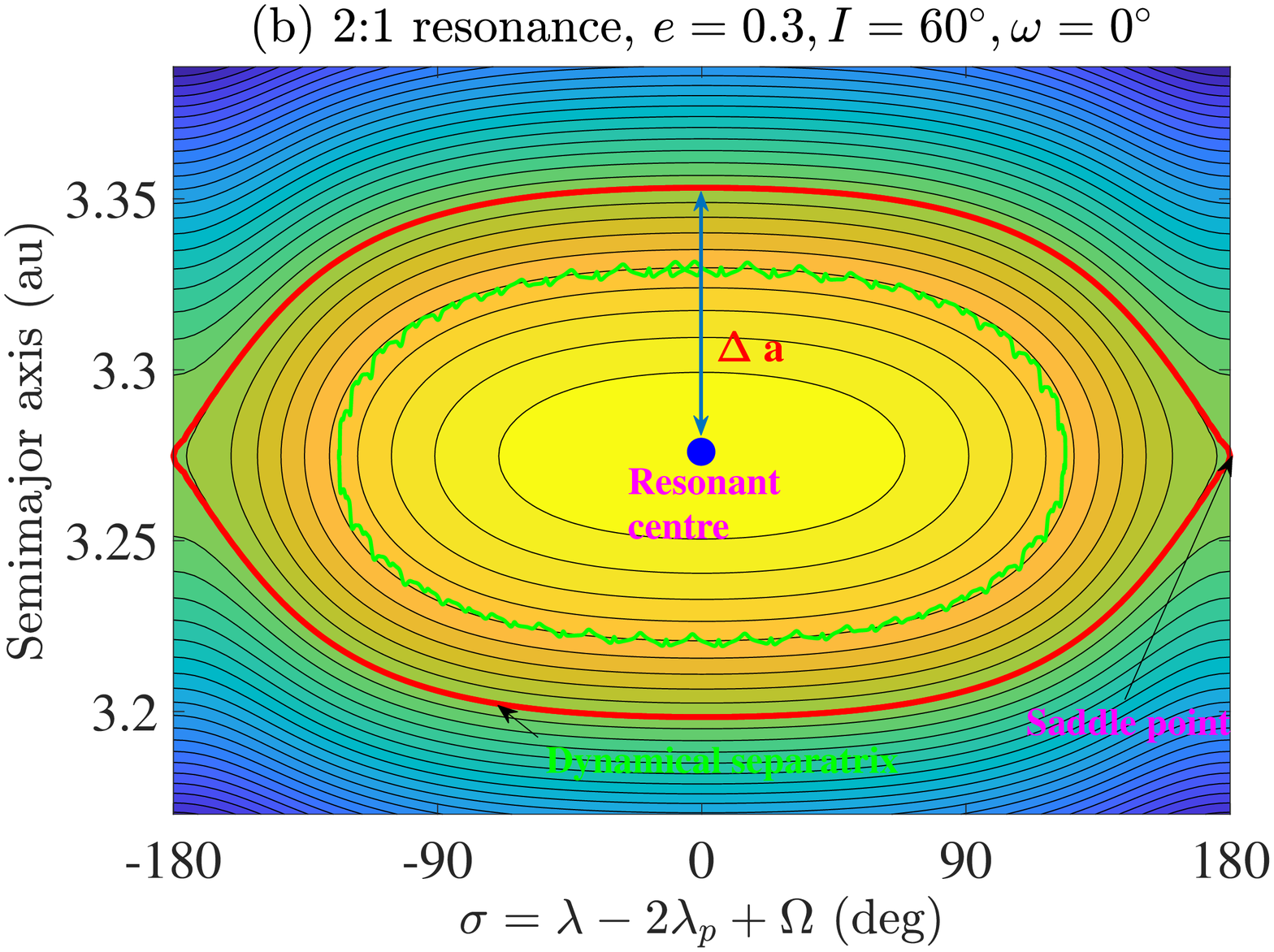}
\caption{Level curves of the resonant Hamiltonian (phase-space structures) shown in the space spanned by the characteristic angle $\sigma$ and semimajor axis $a$ for Jupiter's 3:1 and 2:1 resonances. In both plots, the eccentricity is assumed at $e=0.3$, the inclination at $I = 60^{\circ}$ and the argument of pericenter at $\omega = 0^{\circ}$. The green lines stand for the orbits numerically propagated under the Sun--Jupiter--asteroid dynamical model. The lines shown in red correspond to the dynamical separatrices passing through saddle points and the blue dots stand for the locations of resonant centers. The resonant half width, measuring the distance between the separatrix and the resonant center evaluated at $\sigma = \sigma_s$, is denoted by $\Delta a = a_{\rm sep} - a_0$. The mathematical expression of $\Delta a$ is given by equation (\ref{Eq47}).}
\label{Fig5}
\end{figure*}

In Figure \ref{Fig5}, we report the phase-space structures associated with Jupiter's inner 3:1 and 2:1 resonances. For Jupiter's inner 3:1 resonance, the characteristic argument is defined by $\sigma = \lambda - 3 \lambda_p + 2 \Omega$ and, for Jupiter's inner 2:1 resonance, the characteristic argument is defined by $\sigma = \lambda - 2 \lambda_p + \Omega$. In simulations, the eccentricity is fixed at $e=0.3$, the inclination at $60^{\circ}$ and the argument of pericenter at $\omega = 0^{\circ}$ (notice that the phase structures with other settings of elements could be produced in a similar manner). In the phase-space structures, the resonant centers (corresponding to stable equilibrium points in the resonant model) are marked in blue dots, and the level curves passing through the saddle points stand for the dynamical separatrices shown in red lines, which divide the total space into two types of regions: circulation and libration. For the 3:1 resonance, the resonant center is located at $\sigma = \lambda - 3 \lambda_p + 2 \Omega = \pi$ and, for the 2:1 resonance, the resonant center is located at $\sigma = \lambda - 2 \lambda_p + \Omega = 0$. In addition, the resonant half width measures the distance between $a_{\rm sep}$ and $a_0$, which is explicitly marked by $\Delta a$. The mathematical expression of $\Delta a$ is provided by equation (\ref{Eq47}).

To validate the phase portraits obtained from our new expansion of planetary disturbing function, orbits with the same initial conditions are numerically propagated under the Sun--Jupiter--asteroid system and they are shown in green lines, as shown in Figure \ref{Fig5}. The periodic oscillations are caused by the short-period effects arising in the $N$-body model. Evidently, the numerical trajectories could follow closely along the level curves of resonant Hamiltonian, showing that our expansion of planetary disturbing function is valid to predict resonant dynamics of asteroids.

\begin{figure*}
\centering
\includegraphics[width=0.48\textwidth]{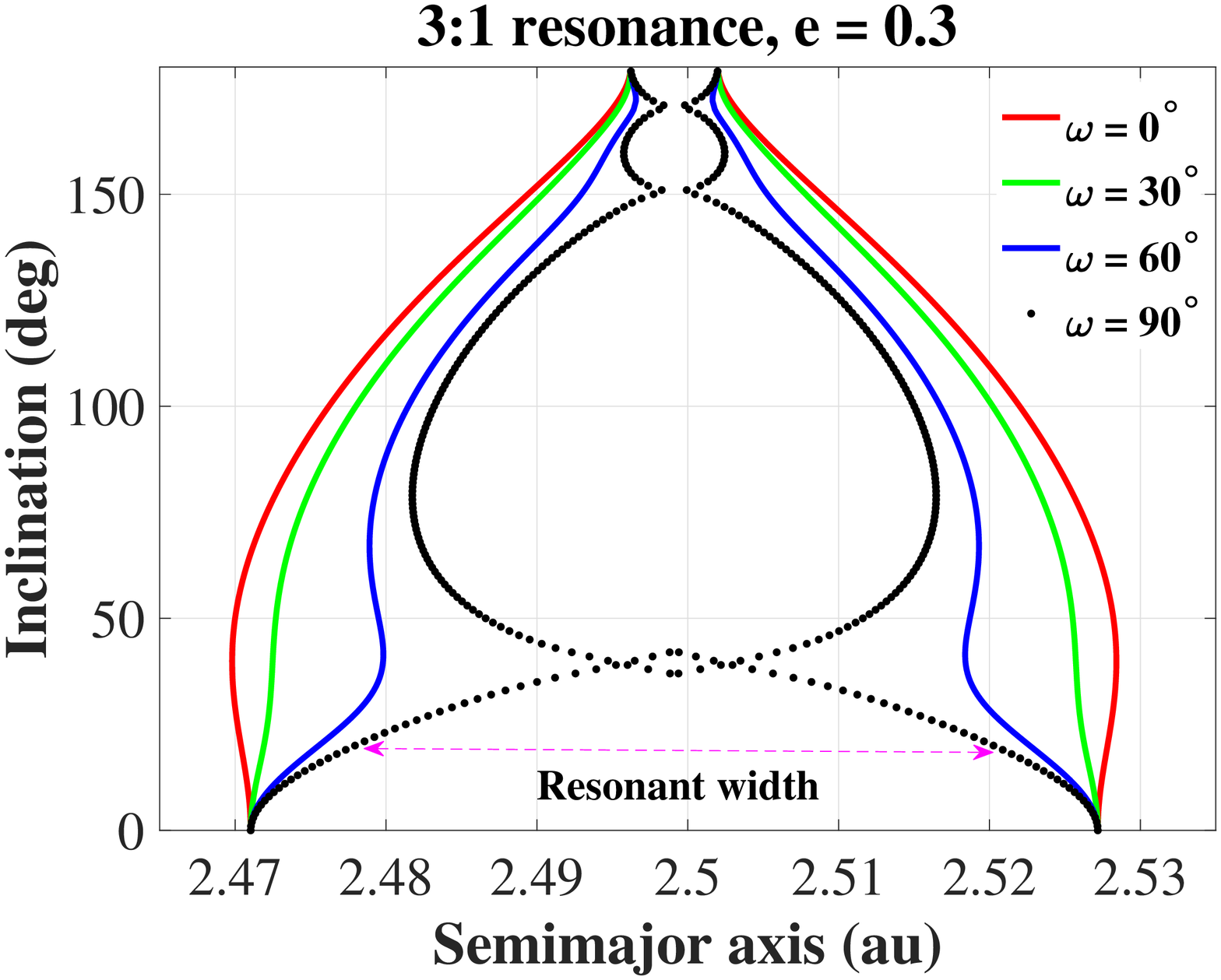}
\includegraphics[width=0.48\textwidth]{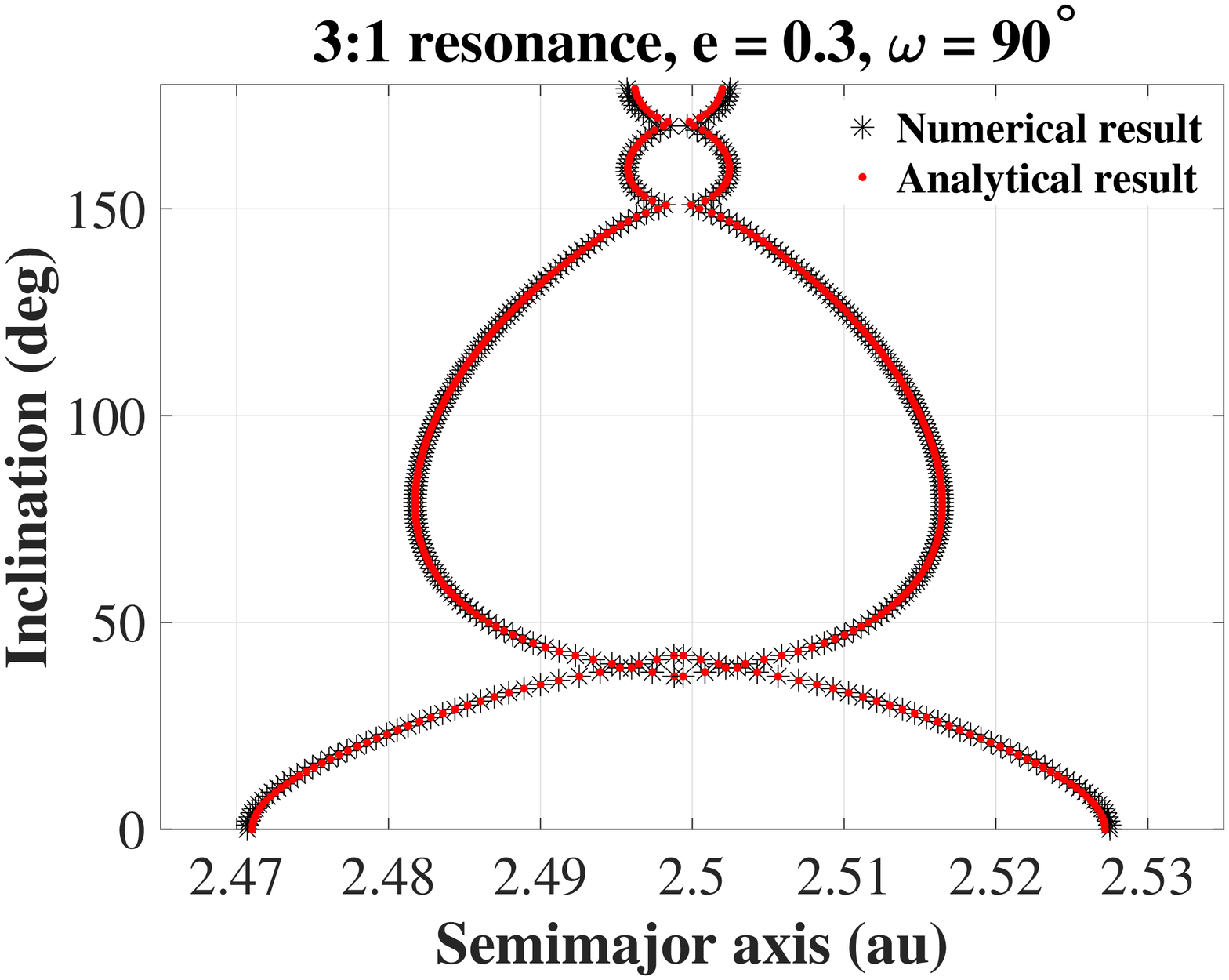}\\
\caption{Resonant widths for Jupiter's inner 3:1 resonance as functions of the mutual inclination $I$. In the left panel, analytical results of resonant width are reported for the cases of $\omega = 0^{\circ}, 30^{\circ}, 60^{\circ}, 90^{\circ}$. The resonant width in terms of the variation of semimajor axis is marked. In the right panel, a comparison is made between the analytical and numerical results of resonant width for the case of $\omega = 90^{\circ}$. In both plots, the eccentricity is fixed at $e = 0.3$.}
\label{Fig6}
\end{figure*}

As stated in Section \ref{Sect5}, the resonant disturbing function arising in equation (\ref{Eq47}) can be produced by means of direct numerical integration (the associated resonant width and center are called numerical results in the following discussions) and by means of the new expansion represented by equation (\ref{Eq30}) or (\ref{Eq30}) (the associated resonant width and center are called analytical results).

\begin{figure*}
\centering
\includegraphics[width=0.48\textwidth]{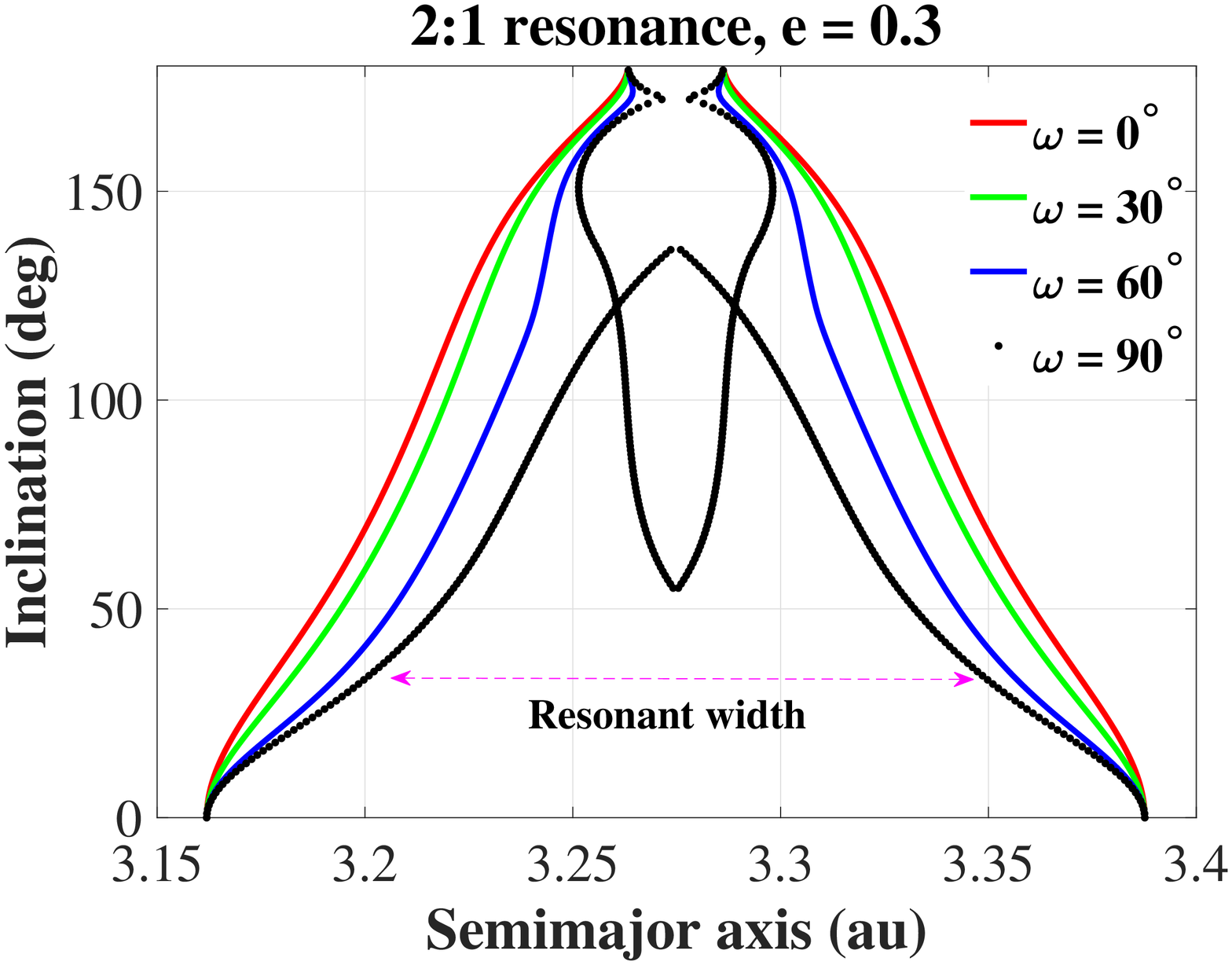}
\includegraphics[width=0.48\textwidth]{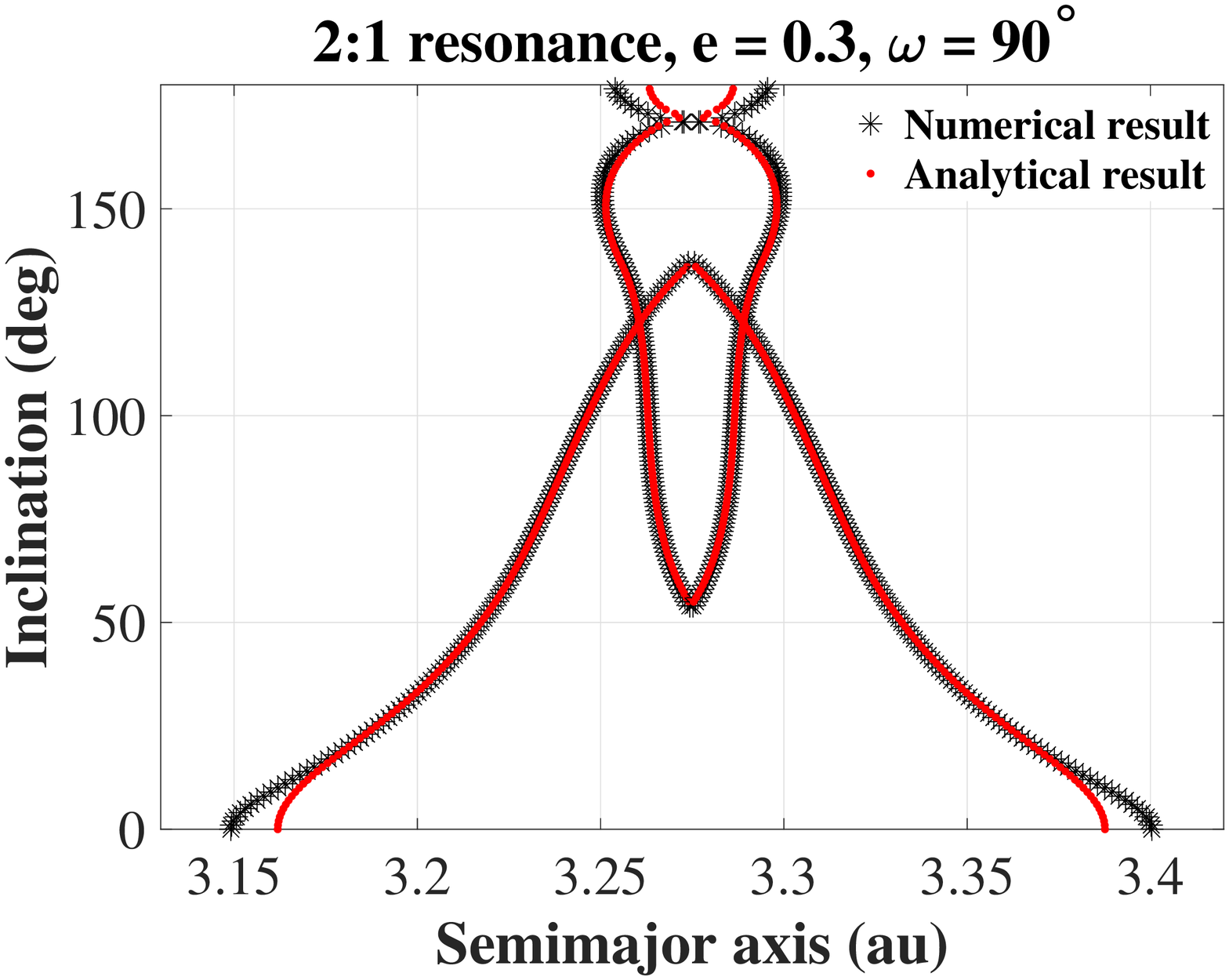}\\
\caption{Resonant widths for Jupiter's inner 2:1 resonance as functions of the mutual inclination $I$. In the left panel, analytical results of resonant width are reported for the cases of $\omega = 0^{\circ}, 30^{\circ}, 60^{\circ}, 90^{\circ}$. The resonant width in terms of the variation of semimajor axis is marked. In the right panel, a comparison is made between the analytical and numerical results of resonant width for the case of $\omega = 90^{\circ}$. In both plots, the eccentricity is fixed at $e = 0.3$.}
\label{Fig7}
\end{figure*}

Regarding Jupiter's inner 3:1 and 2:1 resonances with $e = 0.3$, Figures \ref{Fig6} and \ref{Fig7} present the analytical resonant widths as functions of the mutual inclination $I$ for the cases of $\omega = 0^{\circ}, 30^{\circ}, 60^{\circ}, 90^{\circ}$ in the left panels and provide a direct comparison between the analytical and numerical results of the resonant width for the case of $\omega = 90^{\circ}$ in the right panels. The resonant full width in terms of the variation of semimajor axis (it is equal to $2 \Delta a$) is explicitly marked.

From Figures \ref{Fig6} and \ref{Fig7}, it is observed that (a) the influence of $\omega$ upon the resonant width is significant, (b) when the mutual inclination is zero (prograde) or $\pi$ (retrograde), the resonant widths with different $\omega$ are coincident (this is because, at inclination of $I=0$ or $I=\pi$, only the prograde or retrograde pure eccentricity resonance dominates the dynamics, as discussed in Section \ref{Sect4}), and (c) the comparisons made in the right panels show that the analytical results agree well with the numerical ones. It is worth mentioning that \citet{gallardo2020three} has studied the influence of $\omega$ upon mean motion resonances by introducing the index of fragility: the index is higher, the associated resonance is more fragile (i.e. the resonant structure is more easy to be broken by changing $\omega$).

When $\omega = 90^{\circ}$, we can observe from the right panels of Figures \ref{Fig6} and \ref{Fig7} that (a) the dynamical separatrices associated with the 3:1 resonance have a gourd shape with three necks and it is found that the resonant center switches between $0$ and $\pi$ passing through each neck (it is noted that around the first neck there is a small interval of inclination $I \in [37^{\circ}, 42^{\circ}]$ at which there are two pairs of dynamical separatrices corresponding to two resonant centers located at $\sigma = 0$ and $\sigma = \pi$), (b) the dynamical separatrices associated with the 2:1 resonance have a complicated shape and it is found that, in the interval of $I \in [54^{\circ}, 137^{\circ}]$, there are two pair of dynamical separatrices bounding two different resonant centers located at $\sigma = 0$ and $\sigma = \pi$ and, in the remaining interval, there is only one resonance center at either $\sigma = 0$ or $\sigma = \pi$, and (c) for the 2:1 resonance, the resonant widths at $I = 0$ and $I=\pi$ are slightly underestimated by the analytical developments.

Regarding Jupiter's inner 3:1 and 2:1 resonances with $\omega=0^{\circ}$ and $\omega=90^{\circ}$, \citet{lei2019three} has developed a multi-harmonics pendulum model for describing the resonant dynamics and reported the curves of dynamical separatrices (see the bottom-left panels of Figures 5, 11, 18 and 20 in his work), which are in perfect agreement with the analytical (and numerical) results shown in Figures \ref{Fig6} and \ref{Fig7} (it is noted that, in \citet{lei2019three}, the Laplace-type expansions of disturbing function are used for formulating the resonant model).

\begin{figure*}
\centering
\includegraphics[width=0.48\textwidth]{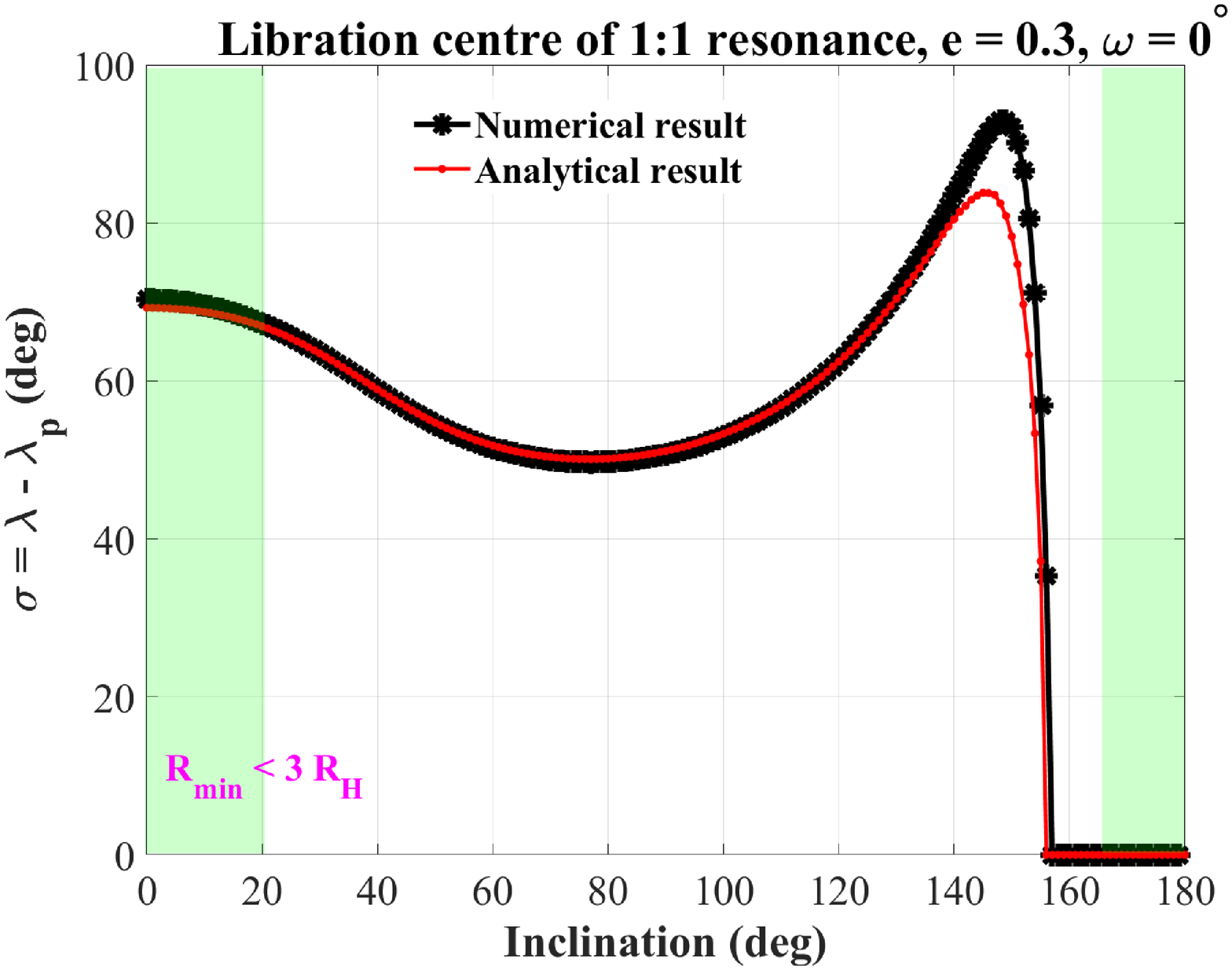}
\includegraphics[width=0.48\textwidth]{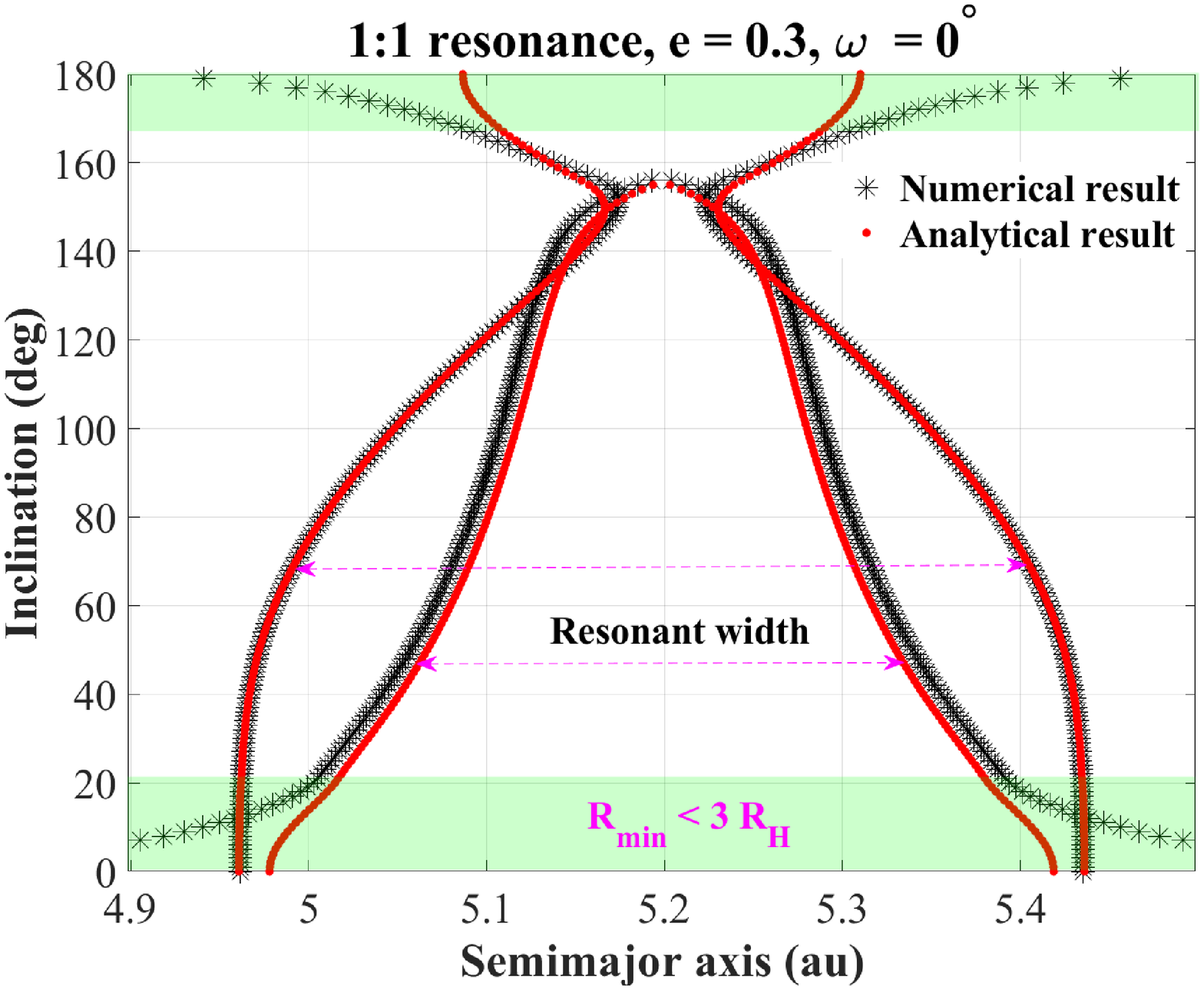}
\caption{Analytical and numerical results of the location of libration center (\emph{left panel}) and the associated resonant width (\emph{right panel}) for Jupiter's 1:1 resonance. In simulations, the eccentricity is fixed at $ e = 0.3$ and the argument of pericenter is at $\omega = 0^{\circ}$. The points inside shadow regions stand for the parameters at which the minimum distance between the asteroid and planet is smaller than three times of the planet's Hill radius (i.e. $R_{\min} < 3 R_H$). The resonant width in terms of the variation of semimajor axis is marked. It is observed from the right panel that inside the shaded area the resonant width is significantly underestimated by the analytical developments due to the underestimation of resonant disturbing function at the saddle points, as shown in Figure \ref{Fig4}.}
\label{Fig8}
\end{figure*}

For Jupiter's 1:1 resonance, the characteristic argument is defined by $\sigma = \lambda - \lambda_p$ (this argument is also called the synodic angle between the asteroid and planet in previous studies). In practical applications of our analytical developments to this case, we fix the argument of the pericenter at $\omega = 0^{\circ}$ and the eccentricity at $e = 0.3$ (simulations with other parameters can be performed in a similar manner). It is known that, for the 1:1 resonance, there are asymmetric libration centers, one of them is located around $\sigma = 60^{\circ}$ (usually called the $L_4$ point) and the other one is located around $\sigma = - 60^{\circ}$ (usually called the $L_5$ point). These asymmetric libration centers are symmetric with respect to $\sigma = \pi$, and they have the same dynamical behavior. Thus, in the practical discussions, we only focus on the one around the usual $L_4$ point. In Figure \ref{Fig8}, the location of the resonant center as a function of mutual inclination is shown in the left panel and the associated resonant width is presented in the right panel. In both plots, the analytical results are shown in red and the numerical results are shown in black. In particular, the shadow region covers those parameters at which the minimum distance between the asteroid and planet is smaller than three times of Jupiter's Hill radius (i.e. $R_{\min} < 3 R_{H}$). As discussed in Section \ref{Sect4}, the resonant disturbing function is underestimated by the analytical developments when the asteroid is located inside the shadow region, resulting in the fact that the resonant width is also underestimated by the analytical developments, as shown in the right panel of Figure \ref{Fig8}. However, when the asteroid in located outside the shadow region, the analytical results are in good agreement with the numerical ones. From the left panel of Figure \ref{Fig8}, the asymmetric libration center exists if and only if the inclination is smaller than $155^{\circ}$ and, in this region, it is observed from the right panel of Figure \ref{Fig8} that there are two pairs of dynamical separatrices bounding the asymmetric libration center. When the inclination is greater than $155^{\circ}$, the asymmetric libration center disappears and is replaced by the symmetric libration center at $\sigma = 0$ and, in this region, there is one pair of dynamical separatrices bounding the libration center, as shown in the right panel of Figure \ref{Fig8}.

\subsection{Neptune's exterior resonances}
\label{Sect6-2}

The analytical developments discussed in Section \ref{Sect5} are applied to the Sun--Neptune system for studying the resonant dynamics of exterior mean motion resonances with Neptune, including the 1:2, 1:3, 2:3, 3:5, 4:7, 3:7, 2:5 and 3:8 resonances.

\begin{figure*}
\centering
\includegraphics[width=0.48\textwidth]{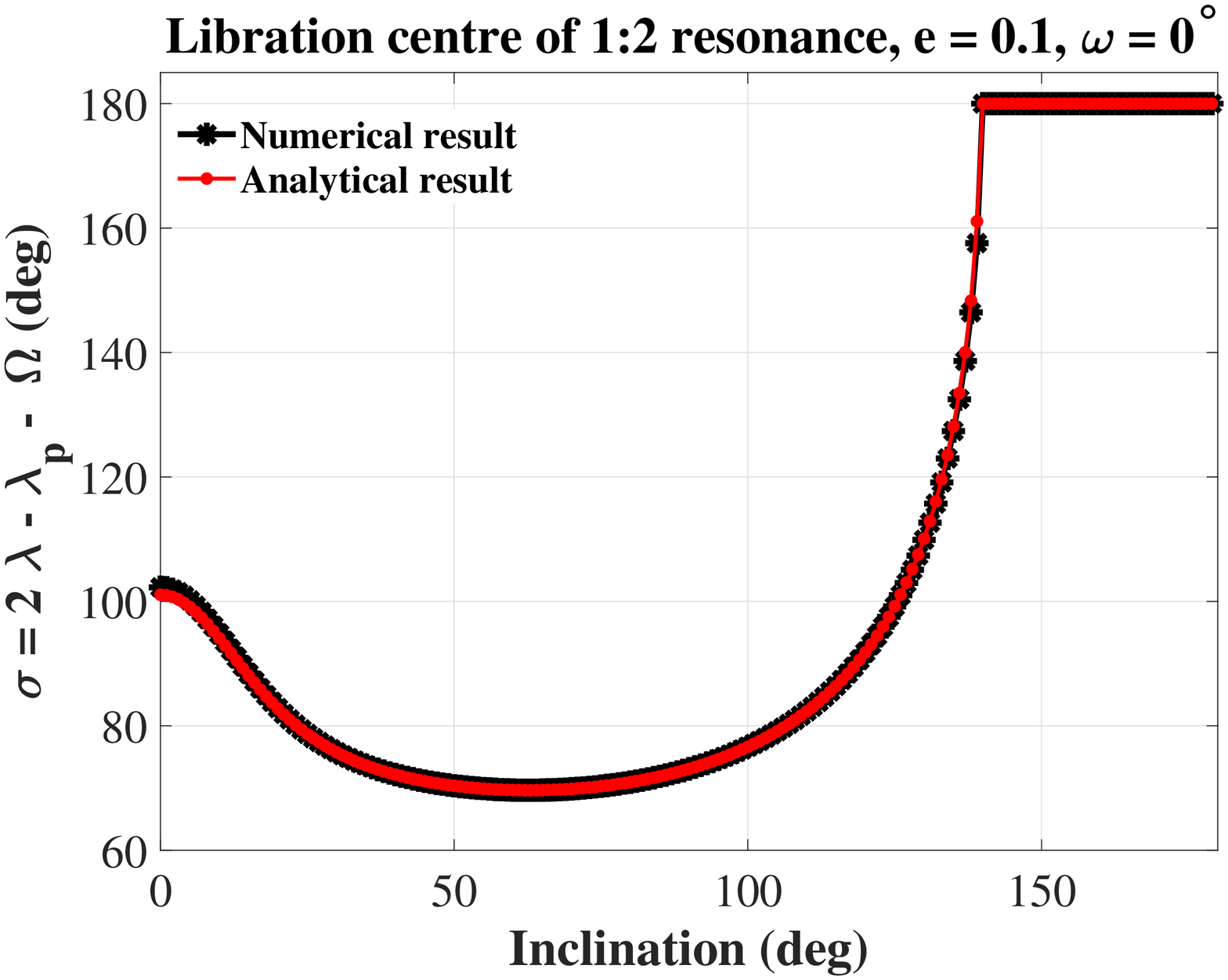}
\includegraphics[width=0.48\textwidth]{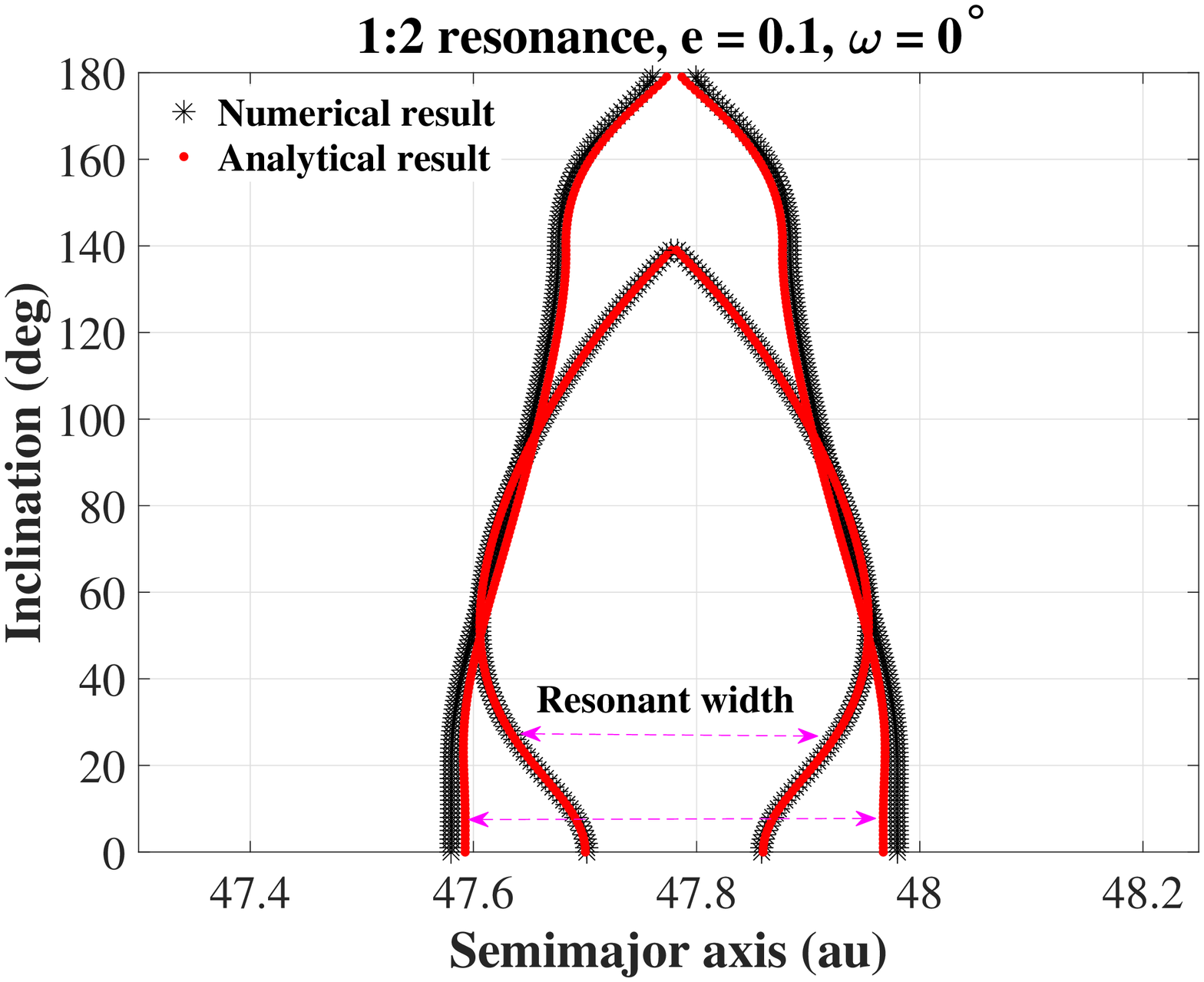}\\
\includegraphics[width=0.48\textwidth]{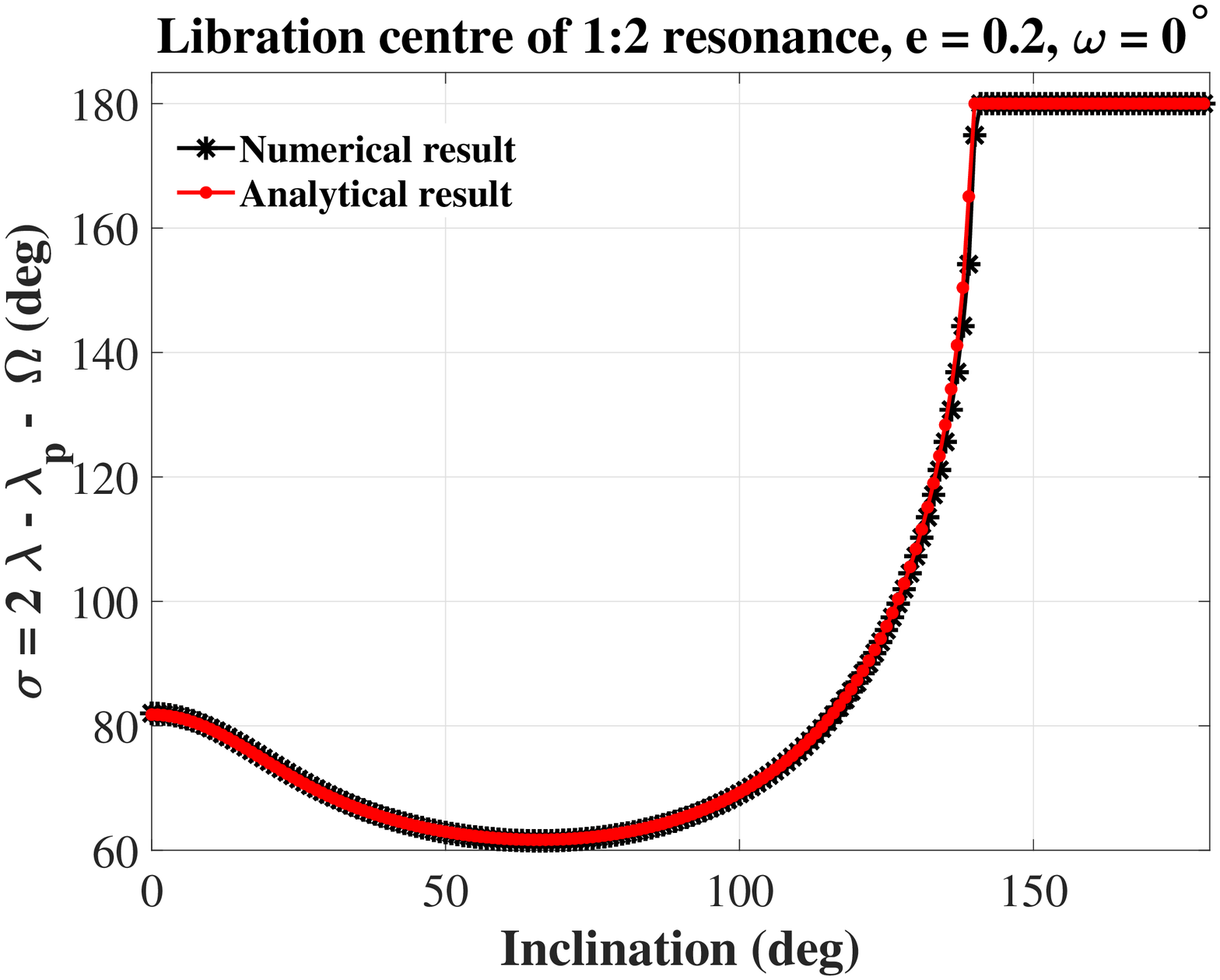}
\includegraphics[width=0.48\textwidth]{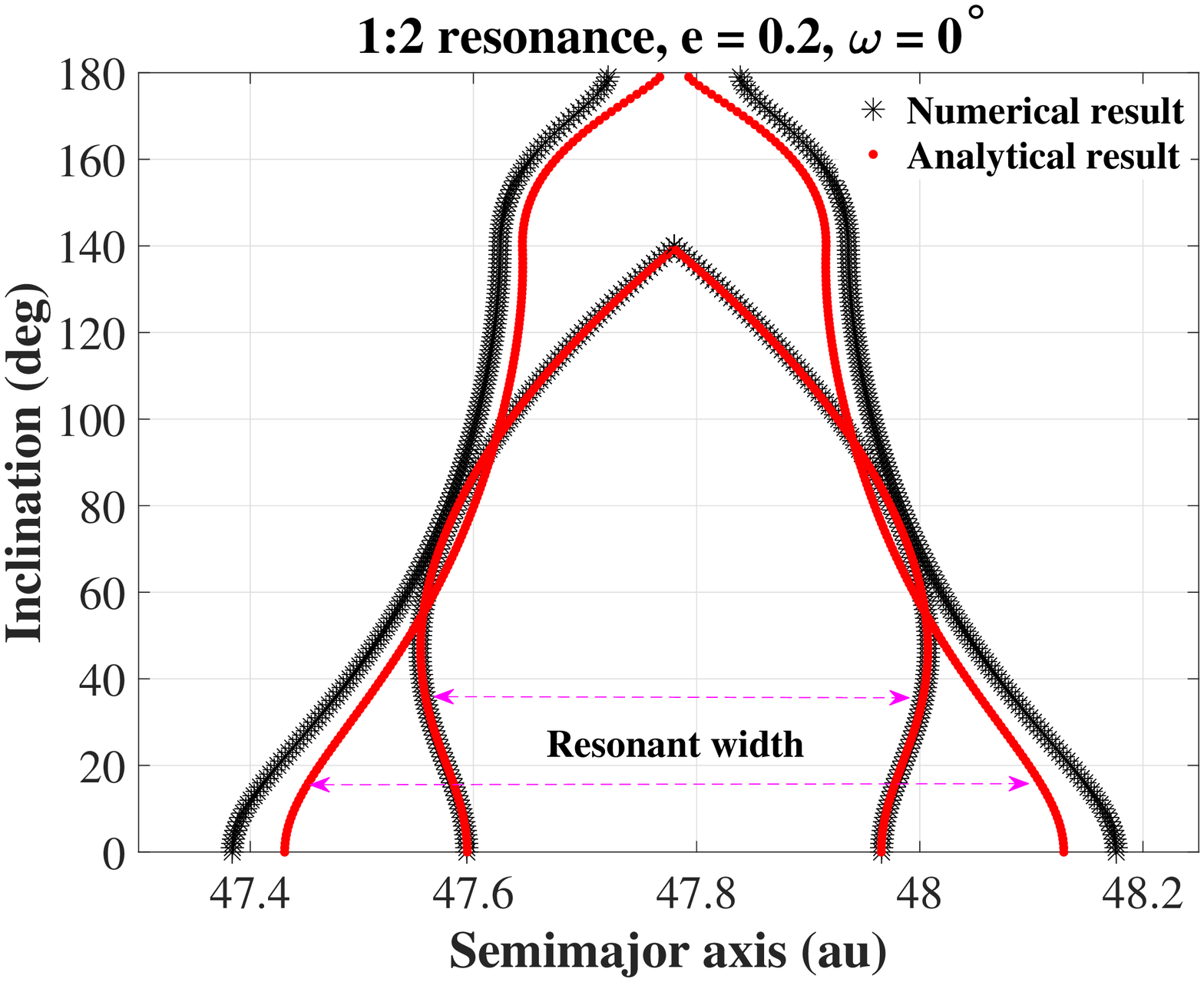}
\caption{Analytical and numerical results of the location of libration center (\emph{left panels}) and the associated resonant width (\emph{right panels}) for Neptune's exterior 1:2 resonance. In the upper panels, the eccentricity is fixed at $e=0.1$ and, in the bottom panels, the eccentricity is assumed at $e=0.2$. The resonant width in terms of the variation of semimajor axis is marked.}
\label{Fig9}
\end{figure*}

\begin{figure*}
\centering
\includegraphics[width=0.48\textwidth]{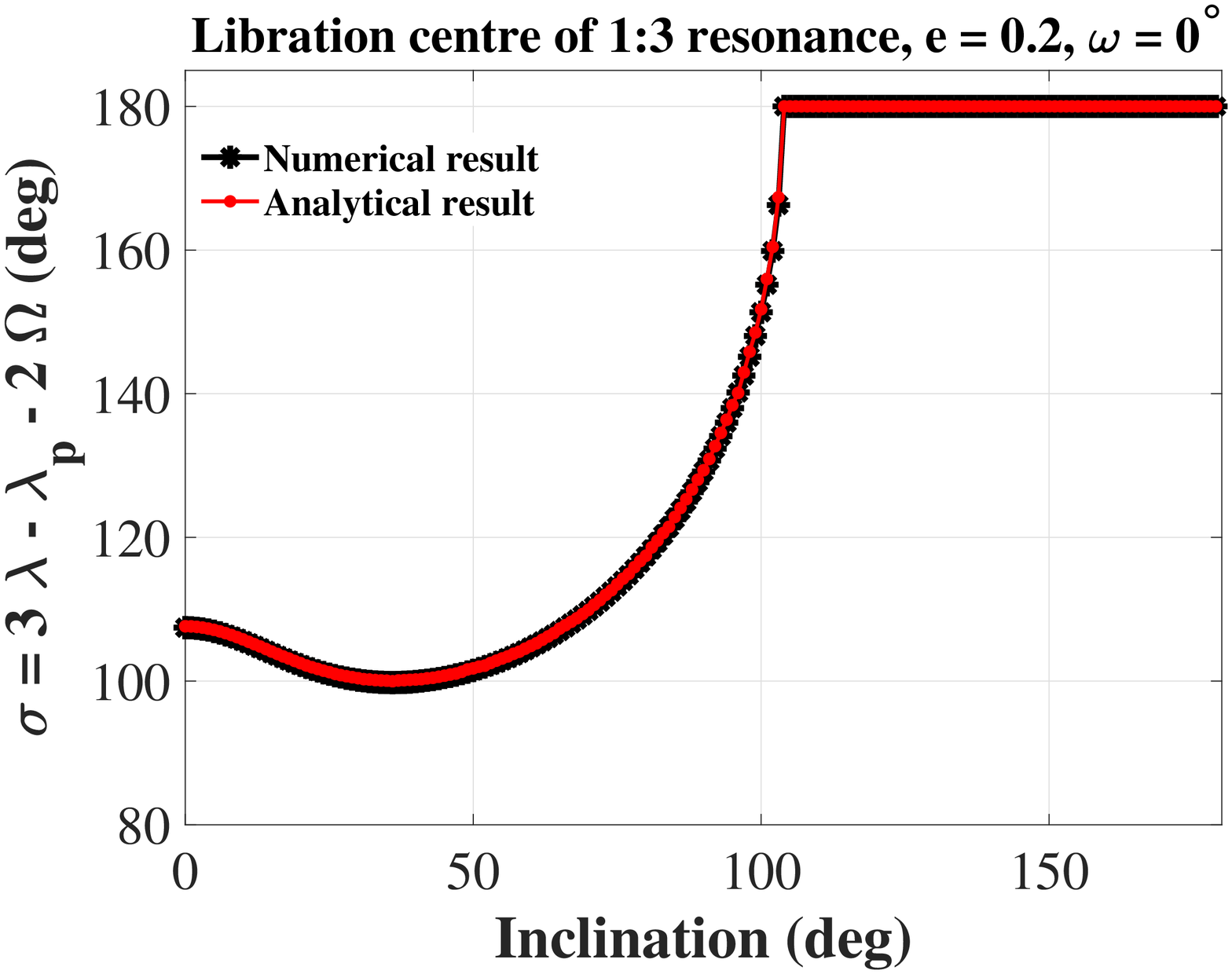}
\includegraphics[width=0.48\textwidth]{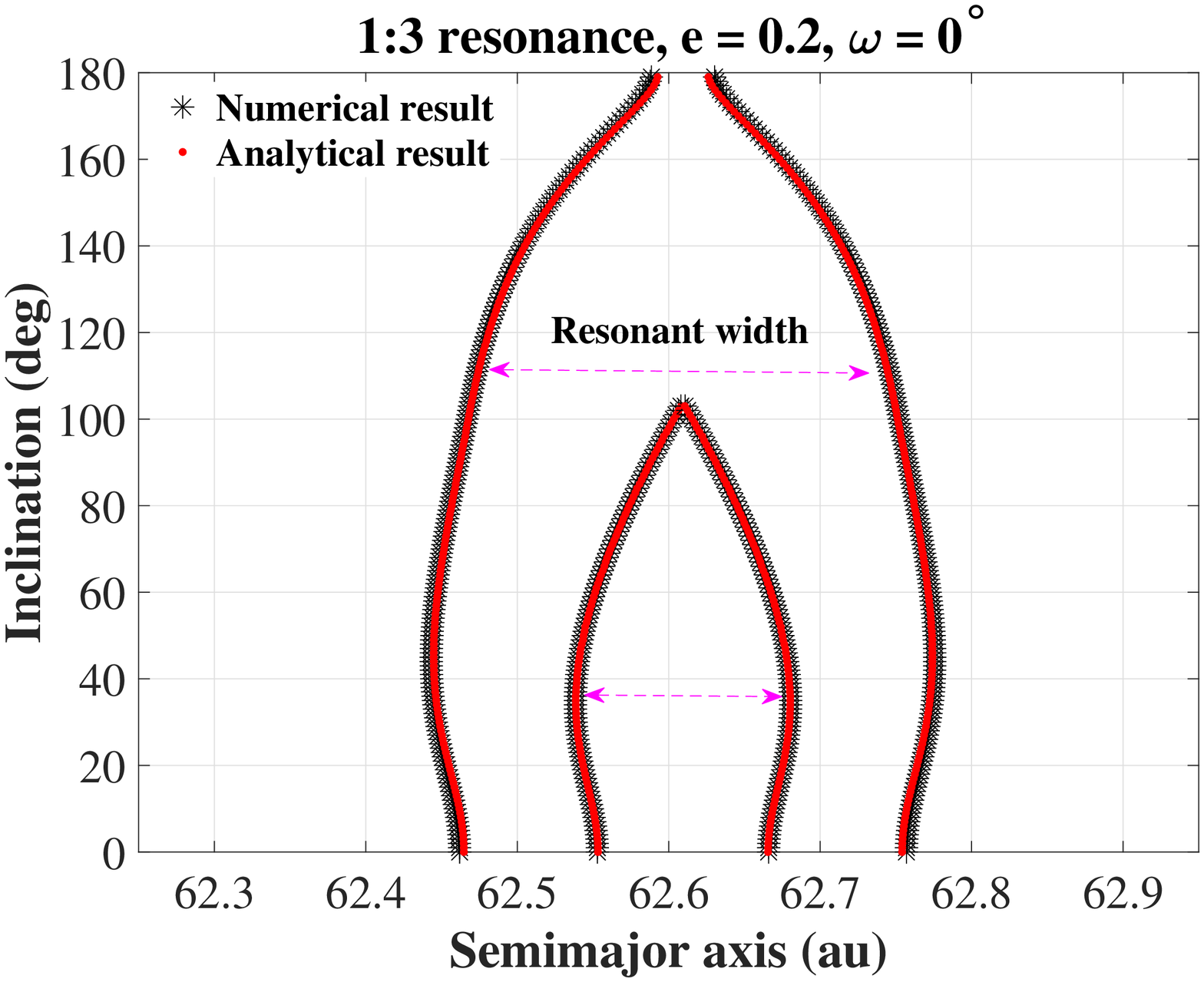}\\
\includegraphics[width=0.48\textwidth]{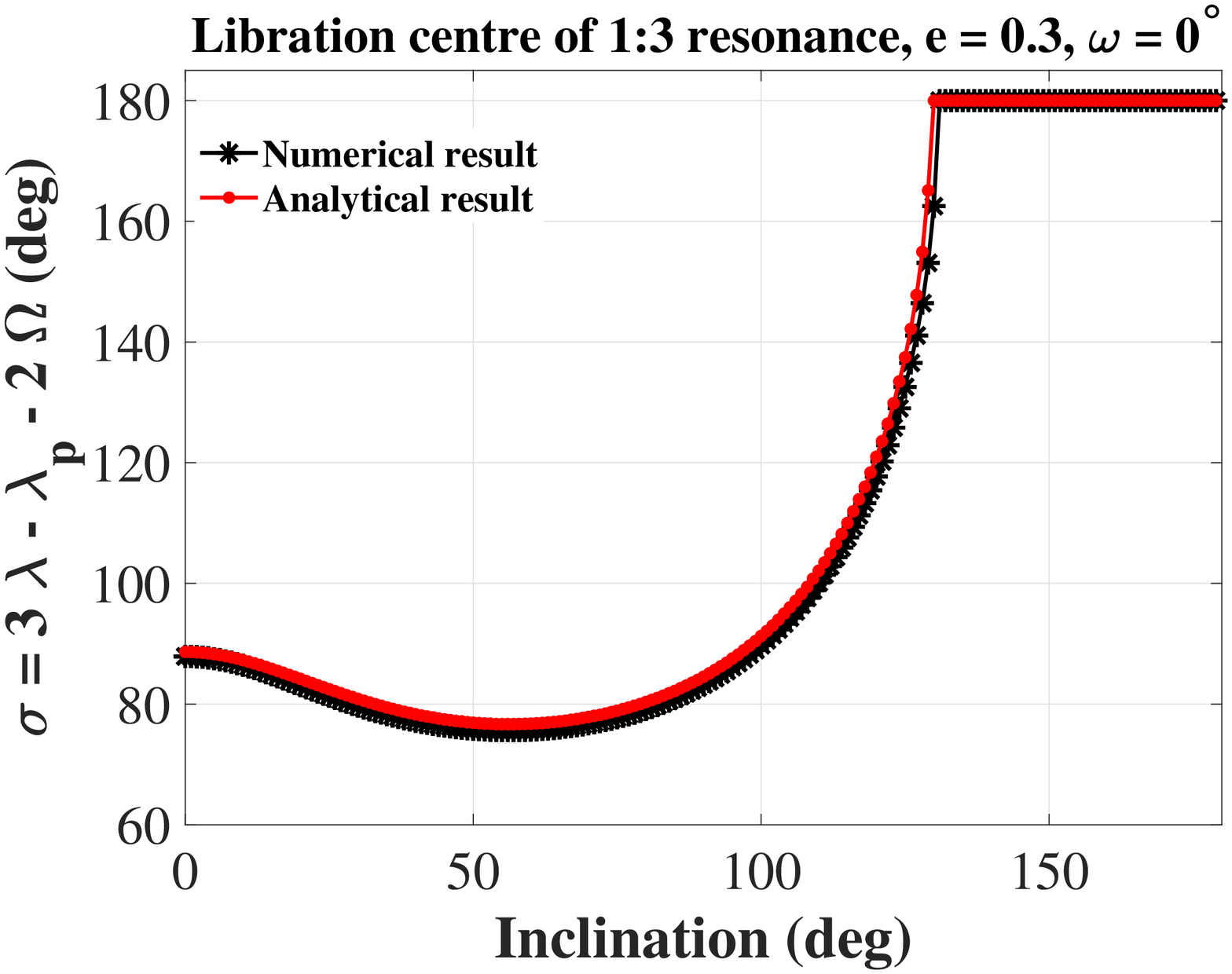}
\includegraphics[width=0.48\textwidth]{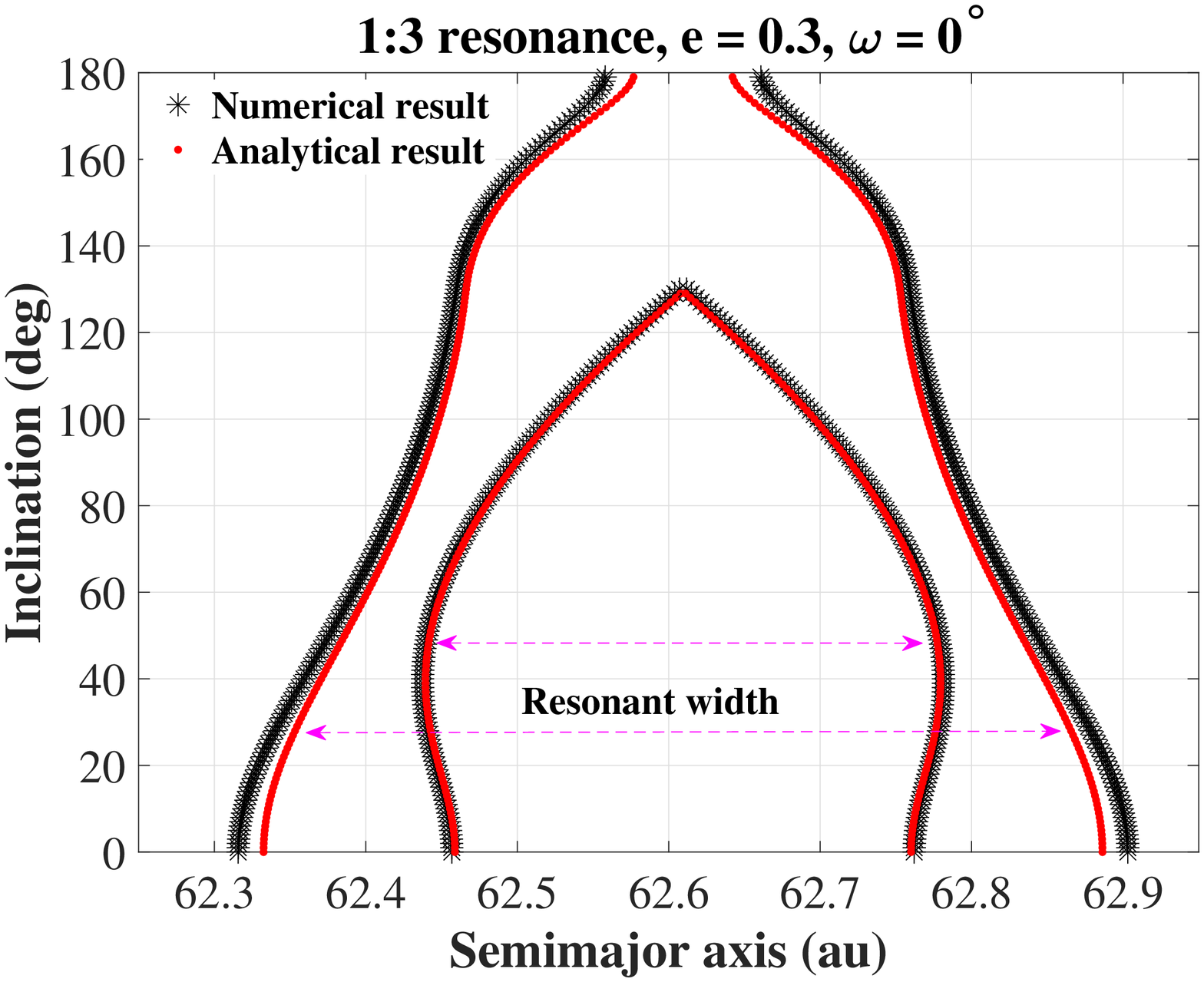}
\caption{Analytical and numerical results of the location of libration center (\emph{left panels}) and the associated resonant width (\emph{right panels}) for Neptune's exterior 1:3 resonance. In the upper panels, the eccentricity is fixed at $e=0.2$ and, in the bottom panels, the eccentricity is assumed at $e=0.3$. The resonant width in terms of the variation of semimajor axis is marked.}
\label{Fig10}
\end{figure*}

Regarding Neptune's exterior 1:2 and 1:3 resonances, their characteristic arguments are $\sigma = 2 \lambda - \lambda_p - \Omega$ and $\sigma = 3 \lambda - \lambda_p - 2 \Omega$, respectively. Figure \ref{Fig9} presents the locations of libration center and the associated resonant width as functions of the mutual inclination for the 1:2 resonance, and Figure \ref{Fig10} shows the corresponding results for the 1:3 resonance.

For the case of the 1:2 resonance as shown in Figure \ref{Fig9}, two cases with $e = 0.1$ and $e = 0.2$ are taken into account. It is observed from the left panels that the asymmetric libration centers exist when the inclination is smaller than $140^{\circ}$ and, when the inclination is greater than $140^{\circ}$, the asymmetric libration centers disappear and are replaced by symmetric libration center located at $\sigma = \pi$. It is noted that in the interval $I \in [0^{\circ}, 140^{\circ})$ there are two asymmetric libration centers which are symmetric with respect to $\pi$. In our study, we only consider one of them (smaller than $\pi$) because the other one has the same dynamical behavior. From Figure \ref{Fig9}, we can observe: (a) for both cases with $e = 0.1$ and $e = 0.2$, the analytical results could be in good agreement with the numerical results for the location of libration center and the associated resonant width, (b) with inclination changing from $0^{\circ}$ to $140^{\circ}$, the $\sigma$ of the libration center first decreases and then increases up to $\sigma = \pi$ which is the location of the symmetric libration center, (c) in the interval of $I \in [0^{\circ}, 140^{\circ})$, there are two pairs of dynamical separatrices (corresponding to the inner and outer boundaries) bounding the asymmetric libration center, (d) when the inclination is greater than $140^{\circ}$, there is only one pair of separatrices bounding the symmetric libration center, and (e) the resonant widths associated with $e = 0.2$ are greater than the ones associated with $e = 0.1$, because there is a positive correlation between the force amplitude in the expansions and the eccentricity as discussed in Section \ref{Sect3}.

For the case of the 1:3 resonance as shown in Figure \ref{Fig10}, two cases with $e = 0.2$ and $e = 0.3$ are taken into consideration. It is observed from the left panels of Figure \ref{Fig10} that, for the case of $e = 0.2$, the asymmetric libration centers disappear when the inclination is greater than $104^{\circ}$ and, for the case of $e = 0.3$, the asymmetric libration centers disappear when the inclination is greater than $130^{\circ}$. Similar to the case of the 1:2 resonance, only one of the asymmetric libration centers is considered in this study (the other one has the same dynamical behavior). From Figure \ref{Fig10}, we can observe that (a) the analytical results are in good agreement with the numerical ones, (b) resonant width in the case of $e = 0.3$ is greater than that in the case of $e = 0.2$, and (c) there are two pairs of dynamical separatrices bounding the asymmetric libration center and one pair of separatrices bounding the symmetric libration center.

\begin{figure*}
\centering
\includegraphics[width=0.48\textwidth]{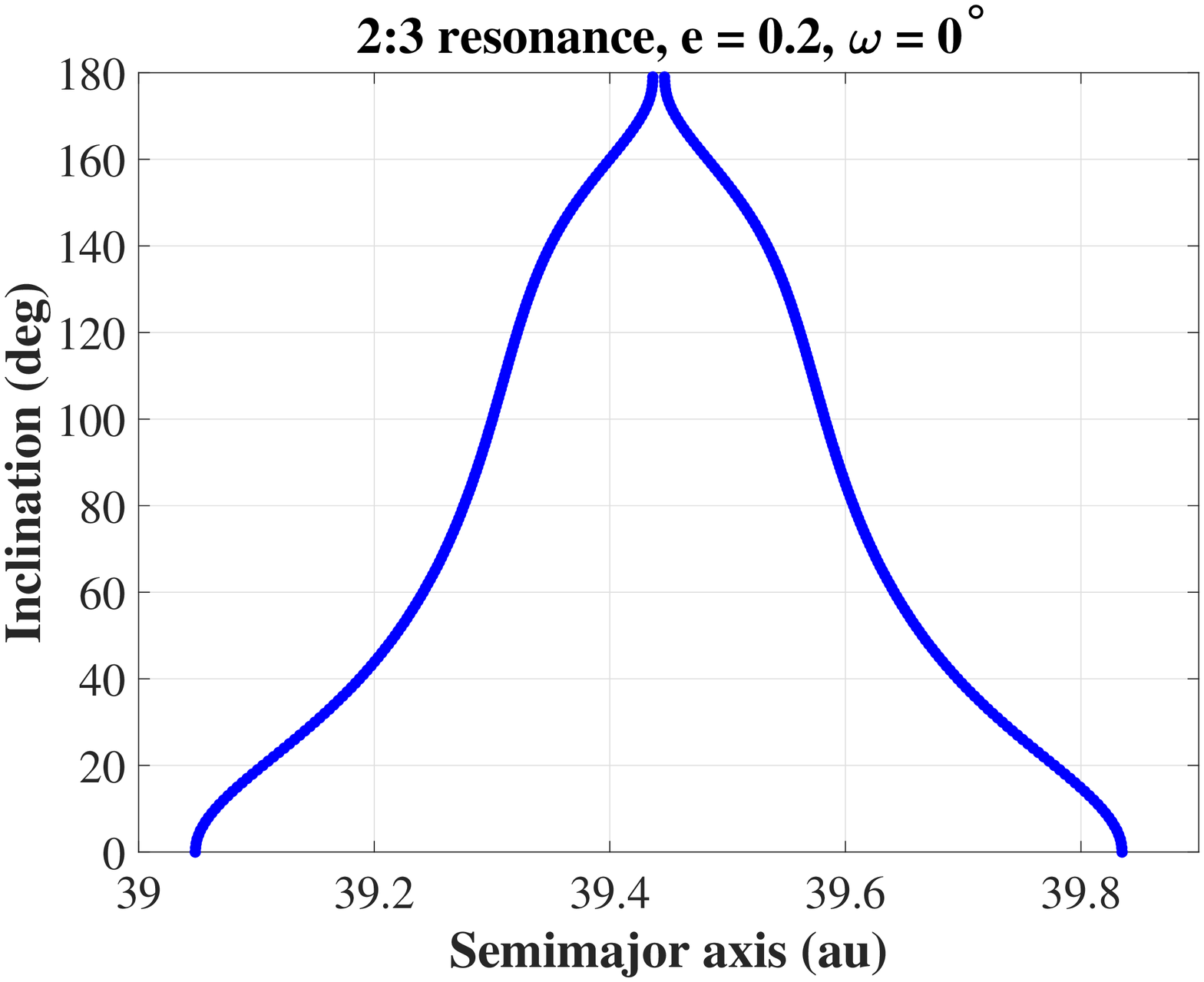}
\includegraphics[width=0.48\textwidth]{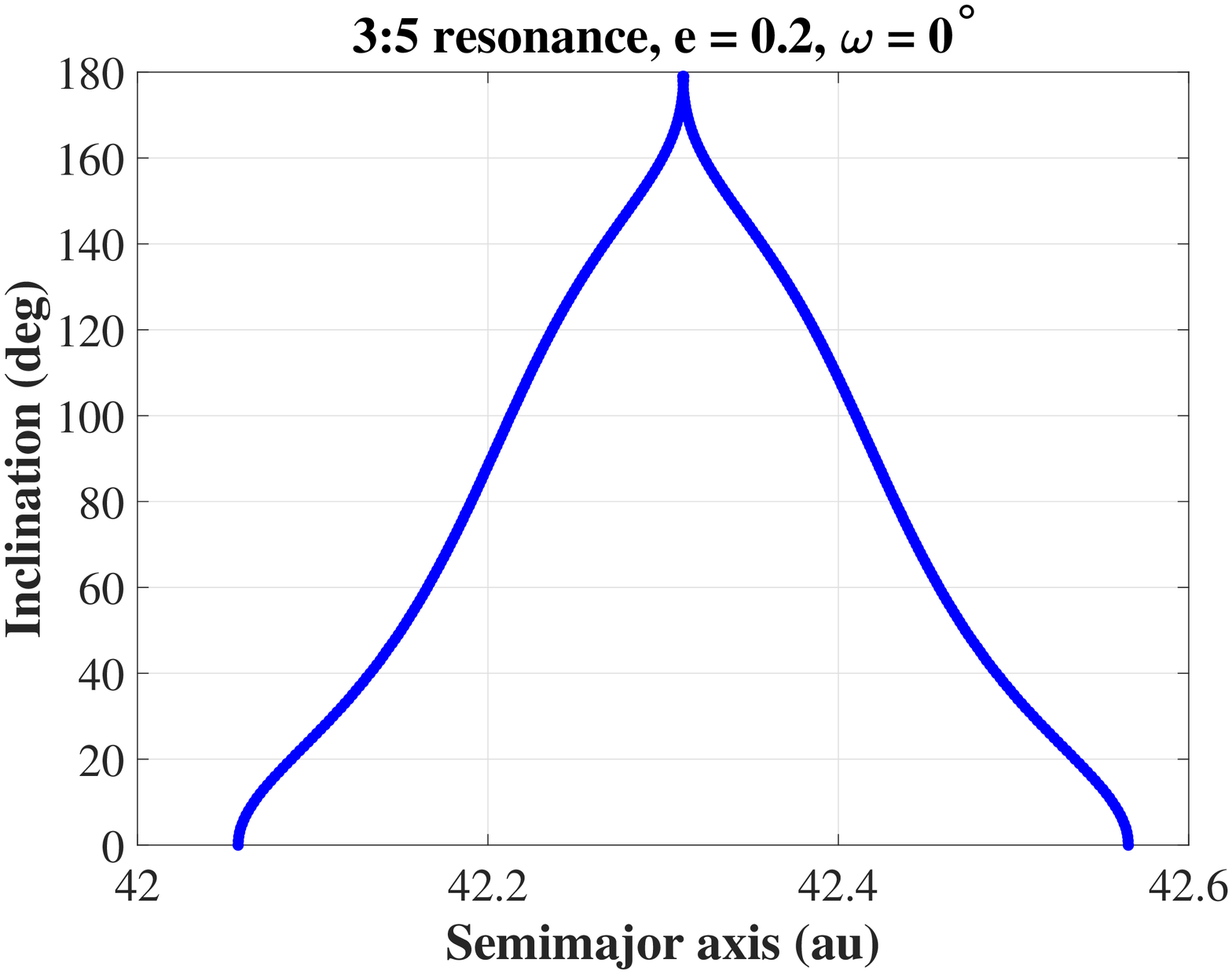}\\
\includegraphics[width=0.48\textwidth]{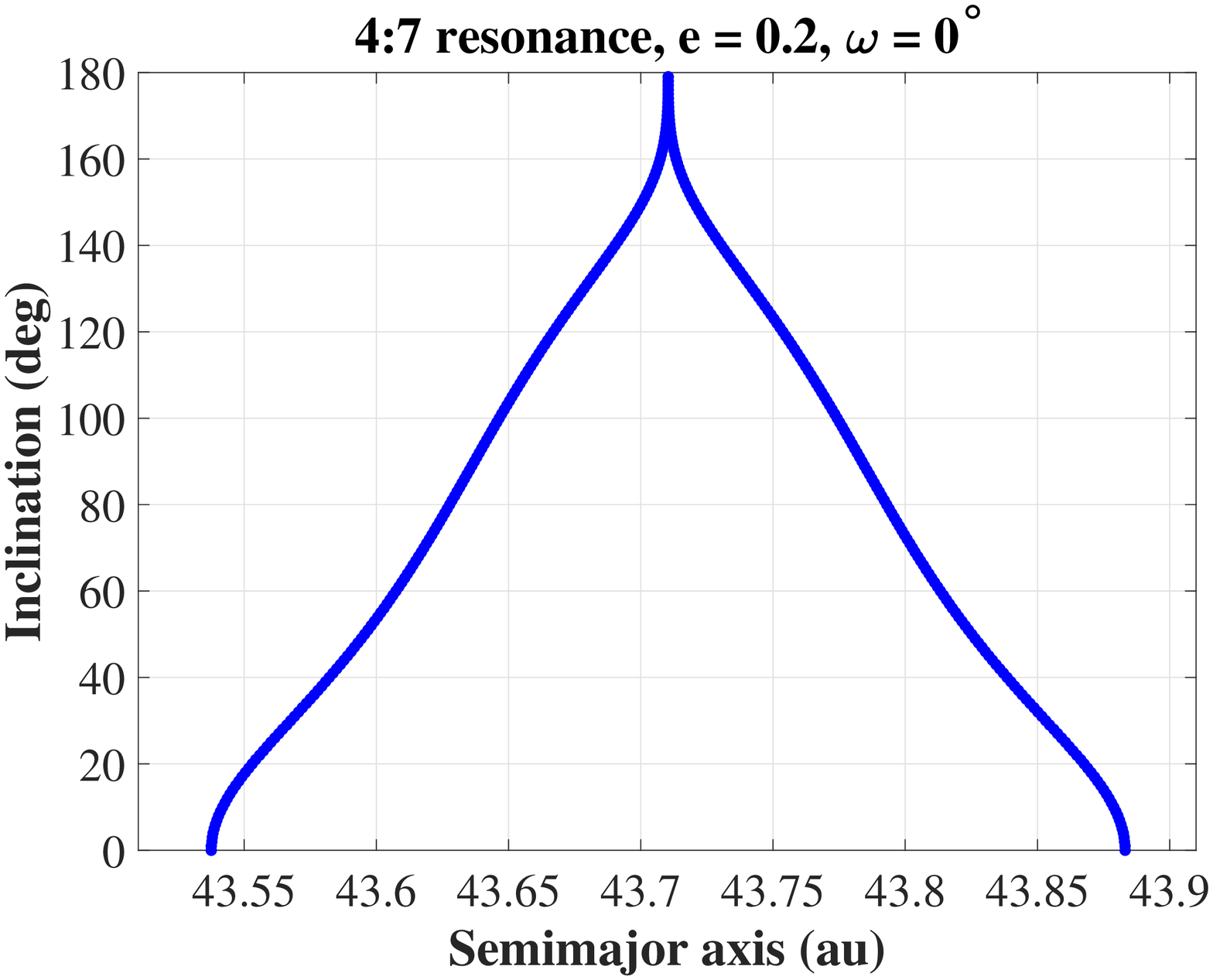}
\includegraphics[width=0.48\textwidth]{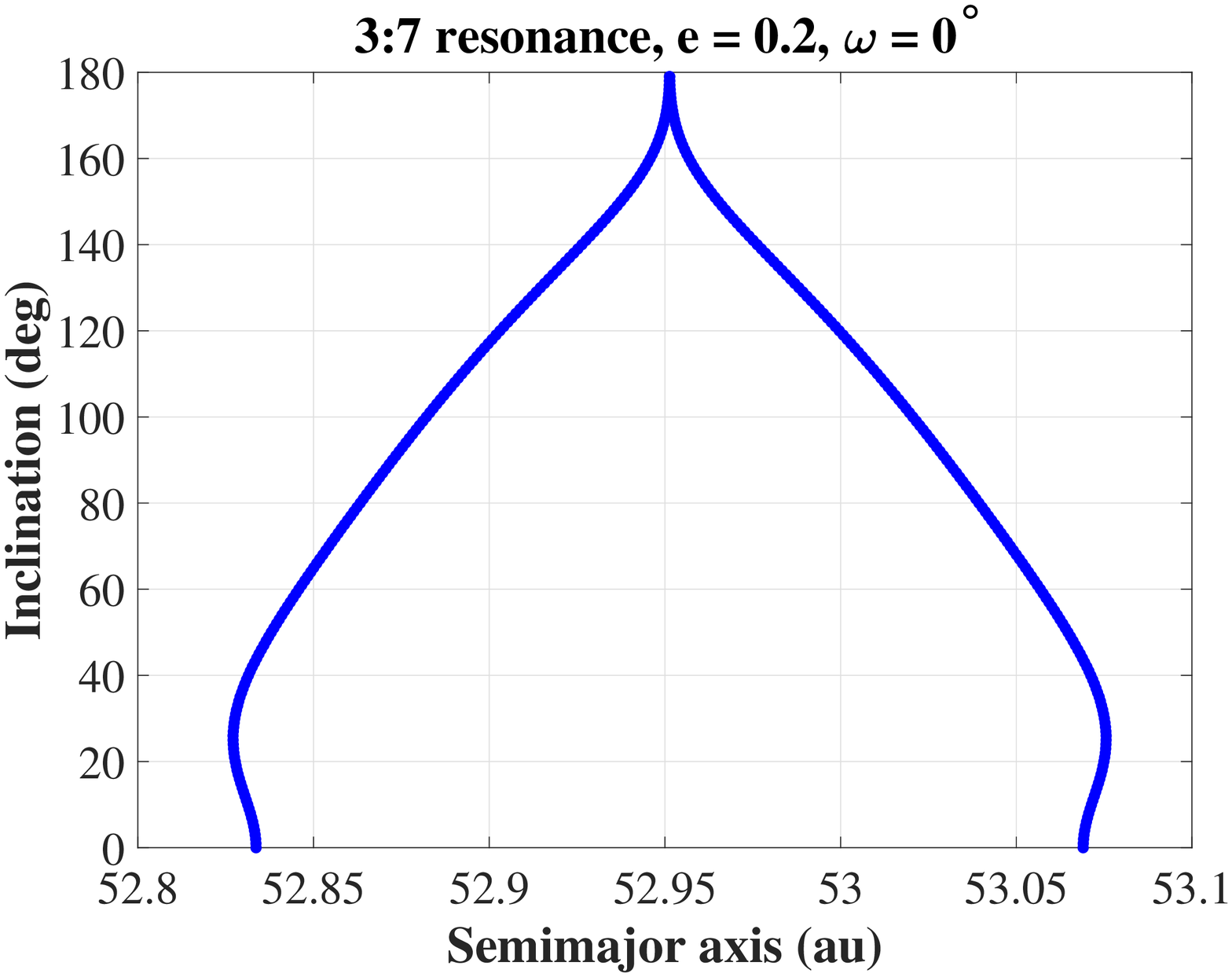}\\
\includegraphics[width=0.48\textwidth]{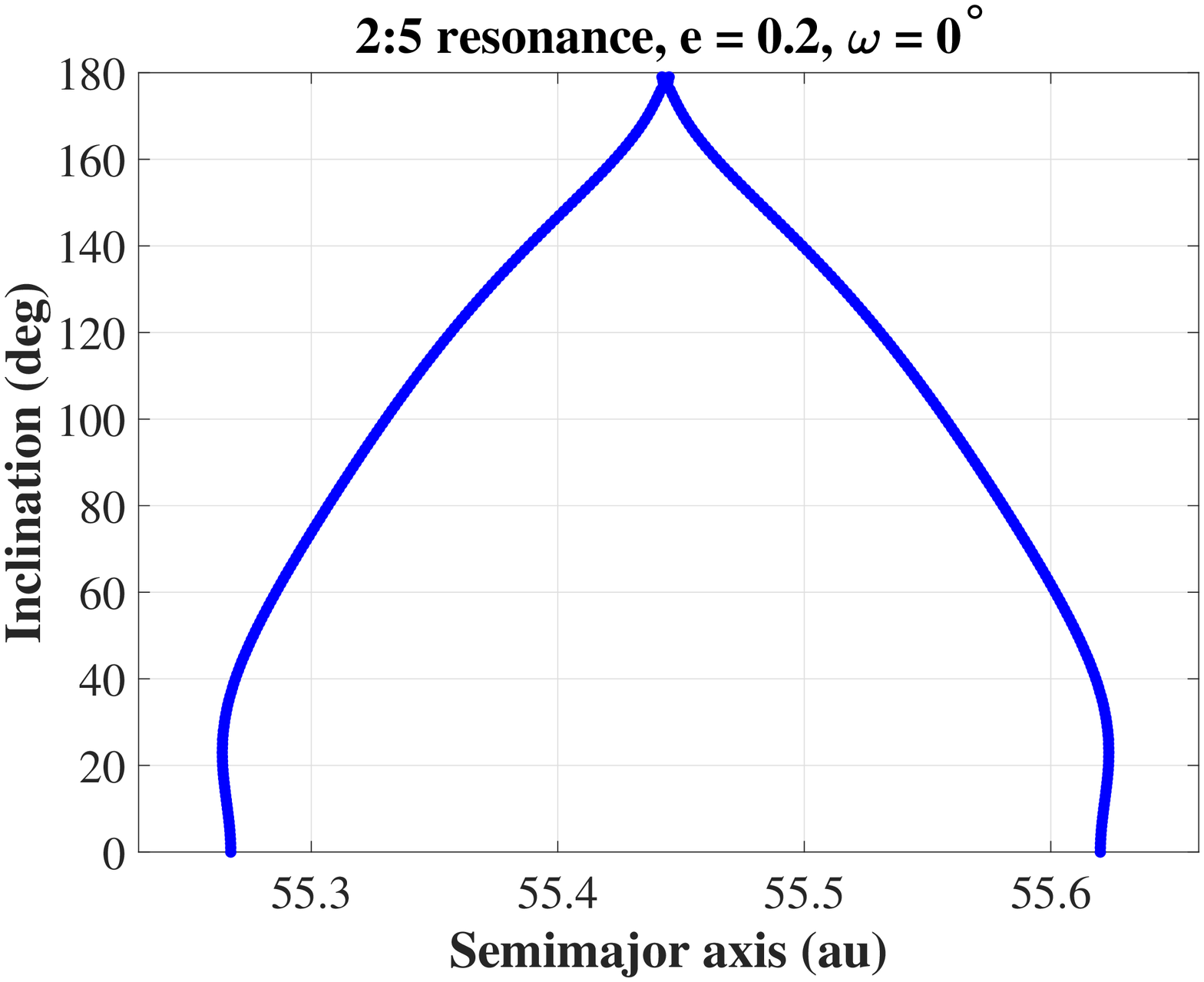}
\includegraphics[width=0.48\textwidth]{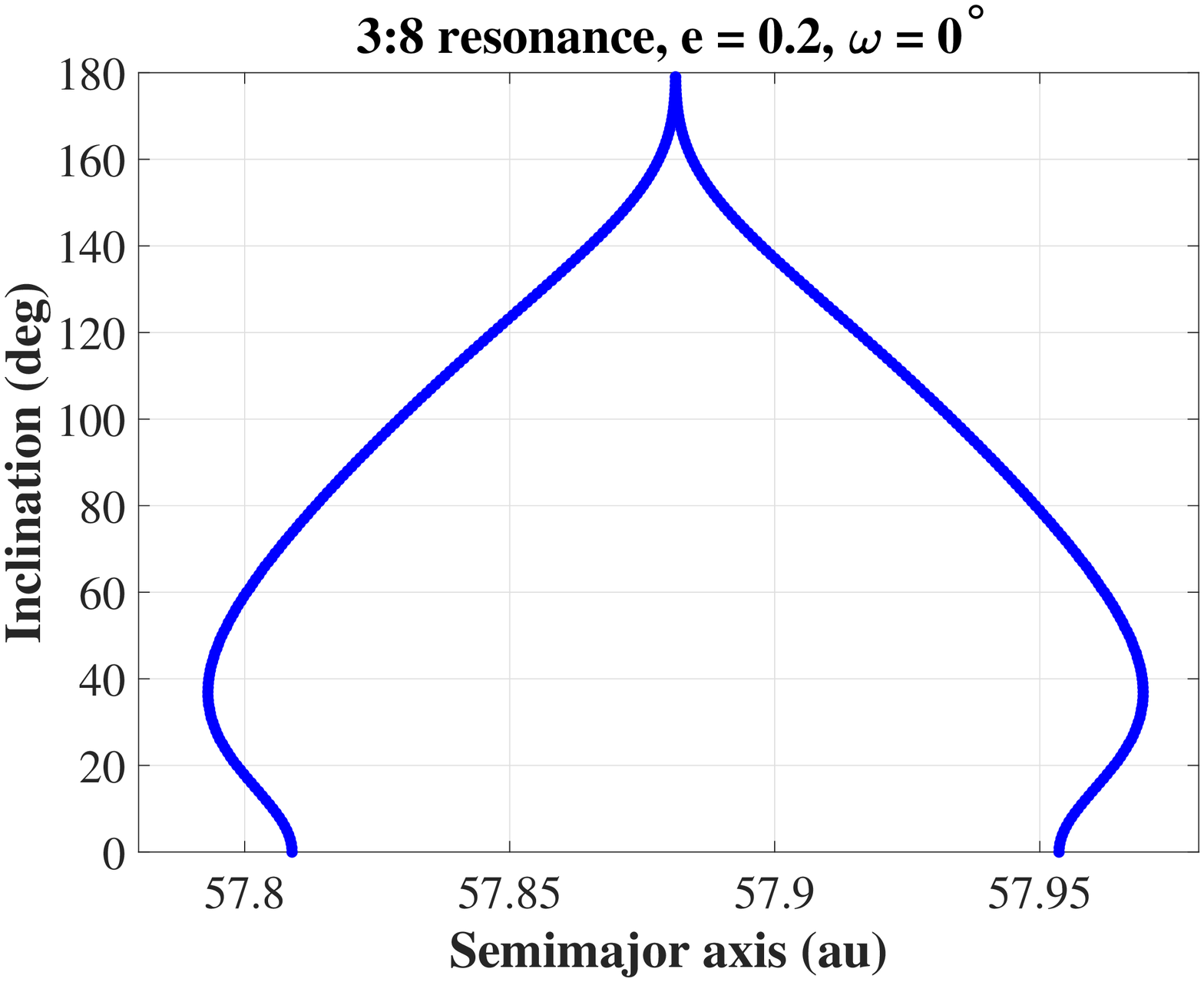}
\caption{Analytical results of the resonant width associated with Neptune's exterior 2:3, 3:5, 4:7, 3:7, 2:5 and 3:8 resonances. The eccentricity is fixed at $e=0.2$ and the argument of pericenter is fixed at $\omega = 0^{\circ}$.}
\label{Fig11}
\end{figure*}

Then, our analytical developments are applied to Neptune's exterior 2:3, 3:5, 4:7, 3:7, 2:5 and 3:8 resonances with eccentricity at $e = 0.2$ and argument of pericenter at $\omega = 0^{\circ}$. The resonant widths as functions of the mutual inclination are reported in Figure \ref{Fig11}. For these considered resonances, all of them have symmetric libration center located at $\sigma = \pi$. From Figure \ref{Fig11}, it is observed that (a) for the 2:3, 3:5 and 4:7 resonances, the resonant width is a decreasing function of the mutual inclination, (b) for the 3:7, 2:5 and 3:8 resonances, the resonant width first increases and then decreases with the inclination, and (c) when the inclination is close to $\pi$, the resonant widths of all the considered resonances are approaching zero, indicating that their retrograde resonant strengths are very weak (this point is different from that of inner resonances, as shown in the previous subsection).

\section{Conclusions}
\label{Sect7}

In this work, a new expansion of the planetary disturbing function has been developed. In the first step, the original disturbing function is expanded around circular orbits and it is expressed as a formal series in the eccentricities of the objects involved. After the first step, the disturbing function becomes a summation of the terms associated with the mutual interaction between circular orbits. In the second step, we define a parameter $x$ and expand the core function (i.e. the mutual interaction between circular orbits) around the reference point $x_c$ as Taylor series of $\delta x = x - x_c$. The disturbing function is finally organized as a Fourier series form, where the force amplitudes are related to the semimajor axis, eccentricity and inclination and the harmonic arguments are linear combinations of the mean longitude, longitude of pericenter and longitude of the ascending node of each mass. In the final expansion, there are two numbers in specifying the orders of expansion: the truncated order in eccentricities denoted by $N$ and the expansion order in $\delta x$ denoted by $k_{\max}$.

The advantages of the new expansion developed in the present work lies in the following two aspects: (a) in the process of expansion, Laplace coefficients are not used, so that the convergence problem arising from the series expansion of Laplace coefficients in semimajor axis ratio can be avoided, and (b) the new expansion is convergent regardless of the values of the mutual inclination and semimajor axis ratio, so it becomes possible to utilize the new expansion of planetary disturbing function to study the dynamics of minor bodies located inside the interior, co-orbital and exterior resonances at arbitrary inclinations.

Based on the new expansion of planetary disturbing function, the resonant Hamiltonian is formulated through the linear and canonical transformations of the modified Delaunay variables. It shows that the resonant model corresponds to a dynamical model with two degrees of freedom ($\sigma$ and $\omega$ are angular variables of the system). Considering the fact that the angle $\omega$ is much slower than $\sigma$, it is possible to make an assumption that $\omega$ is a constant during the timescale of mean motion resonances, so that the dynamical model naturally reduces to a system with a single degree of freedom (only $\sigma$ is the angle coordinate). Under this assumption, the level curves of resonant Hamiltonian in the space ($\sigma$, $a$), correspond to the (pseudo) phase-space structures, which show the global dynamics of mean motion resonances. In the Hamiltonian model of mean motion resonances, the expression of resonant half width in terms of variation of semimajor axis is provided.

The analytical developments are applied to Jupiter's inner and co-orbital resonances and Neptune's exterior resonances. As for Jupiter's inner 3:1 and 2:1 resonances, the analytical results of resonant width at different $\omega$ are produced and, in particular, a direct comparison is made between the analytical and numerical results for the case of $\omega = 90^{\circ}$, showing that the analytical and numerical results are in good agreement. Regarding Jupiter's co-orbital resonance and Neptune's exterior 1:2 and 2:3 resonances, the location of asymmetric and symmetric libration centers and the associated resonant width are produced by means of analytical and numerical approaches, and comparisons between the analytical and numerical results shows that our new expansion is valid. Furthermore, the analytical developments are applied to Neptune's exterior 2:3, 3:5, 4:7, 3:7, 2:5 and 3:8 resonances and their resonant widths as functions of mutual inclination are reported.

\begin{acknowledgements}

This work is performed with the financial support of the National Natural Science Foundation of China (No. 12073011) and the National Key R\&D Program of China (No. 2019YFA0706601).

\end{acknowledgements}

\bibliographystyle{raa}
\bibliography{mybib}

\label{lastpage}

\end{document}